\documentclass[11pt, oneside]{article}   	
\usepackage{geometry}                		
\usepackage{graphicx}				
\usepackage{amssymb}
\usepackage{amsmath}
\usepackage{subfig}
\usepackage{caption}
\usepackage{comment}

\usepackage[utf8]{inputenc}
\usepackage{amsmath}
\numberwithin{equation}{section}
\begin{document}
\title{Riemann Hypothesis, Modified Morse Potential and \\
Supersymmetric Quantum Mechanics}
\author{Michael McGuigan \\
Brookhaven National Laboratory}
\date{}
\maketitle
\begin{abstract}
In this paper we discuss various potentials related to the Riemann zeta function and the Riemann Xi function. These potentials are modified versions of Morse potentials and can also be related to modified forms of the radial harmonic oscillator and modified Coulomb potential. We use supersymmetric quantum mechanics to construct their ground state wave functions and the Fourier transform of the ground state to exhibit the Riemann zeros. This allows us to formulate the Riemann hypothesis in terms of the location of the nodes of the ground state wave function in momentum space. We also discuss the relation these potentials to one and two matrix integrals and construct a few orthogonal polynomials associated with the matrix models. We relate the Schrodinger equation in momentum space to and finite difference equation in momentum space with an infinite number of terms. We computed the uncertainty relations associated with these potentials and ground states as well as the Shannon Information entropy and compare with the unmodified Morse and harmonic oscillator potentials. Finally we discuss the extension of these methods to other functions defined by a Dirichlet series such as the the Ramanujan zeta function.
\end{abstract}
\newpage














\section{Introduction}

Recently there has been discussion of a possible future interaction between number theory and physics \cite{Witten}. An example is the  work relating quantum systems and the Riemann Hypothesis. For example recently there has been progress using the Fourier transform of the Riemann Xi function, denoted by the the $\Phi$ function, with a Gaussian modification \cite{tau} such that if the Gaussian modification is made any stronger  the Riemann hypothesis, the zeros of the Riemann zeta function lie on the axis in the complex plane with real part equal to one half, would no longer be true. This is the realization of the statement that if the Riemann Hypothesis is true it is only barely so. Perhaps this fact relates to the great difficulty in proving the Hypothesis as well as other mysteries such as the relation of the Zeta function to Matrix integrals.

In this paper we construct several potentials associated with various integral representation of the Riemann Zeta function and Riemann Xi function. This allows one to identify ground state wave functions, prepotentials, superpotentials and partner potentials in the language of supersymmetric quantum mechanics. The hope is that these potentials could be useful in elucidating the Riemann hypothesis and it's connection with physics. We identify the Riemann hypothesis in terms of the zeros of the ground state wave function in the momentum representation. We also show how to derive Matrix integral representations associated with the potentials. We show how the Riemann potentials are related to a deformation of the well studied Morse potential used to model molecules. We also compute quantities to study the uncertainty relation and Shannon information of the ground state of these potentials. We discuss modifications to the potentials from a quadratic term and how this relates to the Gaussian modification studied in \cite{tau}. Finally we discuss how these methods can be extended to other functions with a Dirichlet expansion such as the Ramanujan zeta function and state the conclusions of the paper.

Interpreting the Riemann Xi function as a ground state in momentum space allows one to study various expansions of the state in different basis which can be useful in research into the zeros of the function \cite{Griffin}\cite{Romik}\cite{Coffey}. For example if one expands the state in a simple harmonic oscillator basis we have:
$$\xi (\frac{1}{2} + ip) = \tilde \psi (p) = \left\langle {{0_R}} \right.|\left. p \right\rangle  = \sum\limits_{n = 0}^\infty  {\left\langle {{0_R}} \right.|\left. n \right\rangle \left\langle {n|} \right.\left. p \right\rangle }  $$
\begin{equation}= \sum\limits_{n = 0}^\infty  {\int_{ - \infty }^\infty  {dx} \left\langle {{0_R}} \right.|\left. x \right\rangle \left. {\left\langle x \right.|n} \right\rangle \left\langle n \right.|\left. p \right\rangle }  = \sum\limits_{n = 0}^\infty  {{a_n}\psi _n^{SHO}(p)} 
\end{equation}
with
\begin{equation}{a_n} = \int_{ - \infty }^\infty  {dx} \left\langle {{0_R}} \right.|\left. x \right\rangle \left. {\left\langle x \right.|n} \right\rangle  = \int_{ - \infty }^\infty  {dx} \psi _0^R(x)\psi _n^{SHO}(x) = \int_{ - \infty }^\infty  {dx} \Phi (x)\psi _n^{SHO}(x)\end{equation}
where the simple hamonic oscillator wave functions in position and momentum space are:
$$\psi _n^{SHO}(x) = {\left( {\frac{{m\omega }}{{\pi {2^{2n}}{{\left( {n!} \right)}^2}}}} \right)^{1/4}}{e^{ - m\omega {x^2}/2}}{H_n}({\left( {m\omega } \right)^{1/2}}x)$$
\begin{equation}\psi _n^{SHO}(p) = {\left( {\frac{{{{\left( {m\omega } \right)}^{ - 1}}}}{{\pi {2^{2n}}{{\left( {n!} \right)}^2}}}} \right)^{1/4}}{e^{ - {{\left( {m\omega } \right)}^{ - 1}}{p^2}/2}}{H_n}({\left( {m\omega } \right)^{ - 1/2}}p)\end{equation}
and the Fourier transform of the Riemann Xi function is:
\begin{equation}\Phi (q) = \psi _0^R(x) = \left\langle {{0_R}} \right.|\left. x \right\rangle  = \int_{ - \infty }^\infty  {dp} \left\langle {{0_R}} \right.|\left. p \right\rangle \left. {\left\langle p \right.|x} \right\rangle  = \int_{ - \infty }^\infty  {dp} \xi (\frac{1}{2} + ip){e^{ipx}}\end{equation}
The function $\Phi(q)$ can be explicitly written through:
\begin{align*}
&f(q) = q{\partial _q}{\theta _3}(0|q)\\
&g(q) = {q^2}\partial _q^2{\theta _3}(0|q) + q{\partial _q}{\theta _3}(0|q)\\
&\Phi (q) = 2{\pi ^2}g(q){\left( { - \log (q)/\pi } \right)^{9/4}} - 3\pi f(q){\left( { - \log (q)/\pi } \right)^{5/4}}
\end{align*}\begin{equation}\end{equation}
where ${\theta _3}(0|q)$ is the Jacobi theta function of the third kind and $q=e^{-\pi e^{-2x}}$.
Modern approaches to the Riemann hypothesis seeks even better basis expansions of the Riemann Xi function using Jensen polynomials which can approximate the Hermite polynomials at high order.

Another modern approach concerns the deformation of the $\Phi$ function by a Gaussian term \cite{tau}. Using the the relation of the ground state wave function in position space to the prepotential we are able to interpret this in terms of a quadratic deformation of the prepotential and consequently the effect on partner potentials associated with a dynamical system.

\section{Review of Supersymmetric Quantum mechanics}

Supersymmetric Quantum mechanics is reviewed in \cite{Witten:1981nf}\cite{Cooper:1994eh}\cite{Gangopadhyaya:2011wka}. Here we just recall the basic features. The fermionic annihilation operator is represented by the $2\times 2 $ matrix:
\begin{equation}b = \begin{bmatrix}
   0 & 1  \\ 
   0 & 0  \\ 
 \end{bmatrix}\end{equation}
 and the fermionic creation operator is represented as:
 \begin{equation}{b^\dag } = \begin{bmatrix}
   0 & 0  \\ 
   1 & 0  \\ 
   \end{bmatrix})\end{equation}
The supercharges $Q$ and their conjugates are defiend by:
$$Q = \begin{bmatrix}
   0 & 0  \\ 
   A & 0  \\ 

 \end{bmatrix} = A{b^\dag }$$
\begin{equation}{Q^\dag } = \begin{bmatrix}
   0 & {{A^\dag }}  \\ 
   0 & 0  \\ 

 \end{bmatrix} = {A^\dag }b\end{equation}
 
 With the operator $A$ and their conjugates defined by:
 $$A = ip + W(x)$$
 \begin{equation}{A^\dag } =  - ip + W(x)\end{equation}
 with $W(x)$ the superpotential. The two partner Hamiltonians are defined by
 $${H_ - } = {A^\dag }A$$
 \begin{equation}{H_ + } = A{A^\dag }\end{equation}
 The partner potentials are:
 $${V_ - }(x) = W{(x)^2} - W'(x)$$
 \begin{equation}{V_ + }(x) = W{(x)^2} + W'(x)\end{equation}
 The superpotential and ground state wave function of the minus partner potential are determined by the prepotential $V_0(x)$ through:
 $$W(x) = {V_0}'(x)$$
 \begin{equation}{\psi _{0 - }} = {e^{ - {V_0}(x)}}\end{equation}
 so in a sense much of the structure of supersymmetric quantum mechanics follows directly from the prepotential.

\section{Gaussian Model and quadratic potential}

To begin we can start with a simple Gaussian model. The prepotential for the Gaussian model or simple harmonic oscillator (SHO) is:
\begin{equation}{V_0}(x) = \frac{1}{4}{\omega}{x^2}\end{equation}
where we have user the convention that $2m=1$.
Defining:
\begin{equation}{\psi _0}(x) = {e^{ - {V_0}(x)}}\end{equation}
we have:
\begin{equation}{\psi _0}(x) = {e^{ - \frac{1}{4}{\omega }{x^2}}}\end{equation}
using the formalism of supersymmetric quantum mechanics one can define the superpotential:
\begin{equation}W(x) = V_0'(x)=\frac{1}{2}\omega x\end{equation}
which is related to $V_0(x)$ through:
\begin{equation}{V_0}(x) = \int\limits_0^x {W(x')dx'} \end{equation}
so that ${\psi _0}(x)$ obeys:
\begin{equation}(\frac{d}{{dx}} + W(x)){\psi _0}(x) = 0\end{equation}
Further defining:
\begin{equation}{V_ -}(x) = {W^2}(x) - \frac{d}{{dx}}W(x) = \frac{1}{4}{\omega ^2}{x^2} - \frac{1}{2}\omega\end{equation}
we have the partner Hamiltonian
\begin{equation}{H_ -} =  - \frac{{{d^2}}}{{d{x^2}}} + {V_ -}(x)\end{equation} 
The full supersymmetric Hamiltonian is written in $2 \times 2$ form as :
\begin{equation}H = \begin{bmatrix}
   p^2 + {V_ - }(q) & 0  \\ 
   0 & p^2 + {V_ + (q)}  \\ 

 \end{bmatrix}
 =
 \begin{bmatrix}
   {p^2 + {W^2}(q) - W'(q)} & 0  \\ 
   0 & p^2 + {W^2}(q) + W'(q)  \\ 
   \end{bmatrix}\end{equation}
This can be written more succinctly as:
\begin{equation}H = p^2 +{W^2}(q) + W'(q)[{b^\dag },b]\end{equation}
which can be used to calculate the eigenfunctions $ \psi_n(x)$ and can be written in terms of the well known Hermite functions:
\begin{equation}\psi_n(x) = {e^{ - \frac{1}{4}{\omega }{x^2}}}{H_n}(x\sqrt {\omega /2} )\end{equation}

\subsection{Fourier Transform of ground state}
The quantity of interest in this paper is the Fourier transform of the ground state given by:
\begin{equation}\int_{ - \infty }^\infty  {{\psi _0}(x){e^{ixp}}dx} = \int_{ - \infty }^\infty  {e^{ - {V_0}(x)}}{e^{ixp}}dx
\end{equation}
This is straightforward for the Gaussian model as it is an eigenstate under Fourier transformation meaning that it returns the same functional form under the transformation and 
\begin{equation}\hat \psi _0^ - (p) = \sqrt {2\pi } \sqrt {\frac{2}{\omega }} {e^{ - \frac{{{p^2}}}{\omega }}}\end{equation}

\subsection{Ladder operators and factorization}

Ladder operators are defined by:
$$a = \frac{d}{{dx}} + W(x) = \frac{d}{{dx}} + \frac{1}{2}\omega x$$
\begin{equation}{a^\dag } =  - \frac{d}{{dx}} + W(x)) =  - \frac{d}{{dx}} + \frac{1}{2}\omega x\end{equation}
with Hamiltonian:
\begin{equation}{H_ - } = {a^\dag }a = {p^2} + ({W^2}(x) - W'(x)) = {p^2} + \frac{1}{4}{\omega ^2}{x^2} - \frac{1}{2}\omega\end{equation}
Defining:
$$a(\omega) = {\partial _x} + \frac{1}{2}\omega x$$
\begin{equation}{a^\dag }(\omega) =  - {\partial _x} + \frac{1}{2}\omega x\end{equation}
we have:
\begin{equation}[a(\omega_1),{a^\dag }(\omega_2)] = \frac{1}{2}(\omega_1 + \omega_2)I\end{equation}
and
\begin{equation}{a^\dag }(\omega_1)a(\omega_2) + {a^\dag }(\omega_2)a(\omega_1) = 2{p^2} + \frac{1}{2}\omega_1 \omega_2 {x^2} - \frac{1}{2}(\omega_1 + \omega_2)\end{equation}
\subsection{Jacobi Matrix}

Using the recurrence relation for orthonormal functions $Q_n(x)$
\begin{equation}{Q_{n + 1}}(x) = x{Q_n}(x) - {\alpha _n}{Q_n}(x) - {\beta _n}{Q_{n - 1}}(x)
\end{equation}
 we can form the Jacobi Matrix:

\begin{equation}J_{n+1} = \begin{bmatrix}

   {{\alpha _0}} & {\sqrt {{\beta _1}} } & 0 &  \cdots  & 0  \\ 
   {\sqrt {{\beta _1}} } & {{\alpha _1}} & {\sqrt {{\beta _2}} } &  \cdots  & 0  \\ 
   0 & {\sqrt {{\beta _2}} } &  \ddots  &  \ddots  & 0  \\ 
   0 & 0 &  \ddots  & {{\alpha _{n-1}}} & {\sqrt {{\beta _n}} }  \\ 
   0 & 0 &  \cdots  & {\sqrt {{\beta _n}} } & {{\alpha _n}}  \\ 
\end{bmatrix}
 \end{equation}
 which satisfies:
 \begin{equation}\det [xI - {J_{n + 1}}] = {Q_{n + 1}}(x)\end{equation}
 So that the characteristic polynomial of the Jacobi matrix is  orthogonal polynomial itself.
 For the Hermite polynomials the recurrence relations are:
\begin{equation} {H_{n + 1}}(x) = x{H_n}(x) - n{H_{n - 1}}\end{equation}
 so the $\alpha_n=0$ and $\beta_n=n$. The Jacobi matrix then becomes:
 
 \begin{equation} 
 J_{n+1} = \begin{bmatrix}
 
   0 & {\sqrt 1 } & 0 &  \cdots  & 0  \\ 
   {\sqrt 1 } & 0 & {\sqrt 2 } &  \cdots  & 0  \\ 
   0 & {\sqrt 2 } &  \ddots  &  \ddots  & 0  \\ 
   0 & 0 &  \ddots  & 0 & {\sqrt n }  \\ 
   0 & 0 &  \cdots  & {\sqrt n } & 0  \\ 
\end{bmatrix}
  \end{equation}
 So that the characteristic polynomial of the Jacobi matrix is the Hermite polynomial. The Hermitian nature of the Jacobi matrix insure that the roots of the characteristic polynomial and hence the Hermite polynomials will always be real.
 
 \subsection{Large $N$ asymptotics}
 
 One can obtain large $N$ aymptotics of the Hermite polynomials by using saddle point methods on integral representations such as:
 \begin{equation} {\varphi _n}[x] = \frac{{{i^n}{e^{{x^2}/4}}}}{{{{(2\pi )}^{1/4}}\sqrt {n!} }}\int_{ - \infty }^\infty  {{z^n}{e^{ - {z^2}/2 + ixz}}dz}  \end{equation}
 where
  \begin{equation} {\varphi _n}(x) = \frac{{{e^{ - {x^2}/4}}}}{{{{(2\pi )}^{1/4}}\sqrt {n!} }}{H_n}(x) \end{equation}
 or
 \begin{equation}{H_n}(x) = \frac{{n!}}{{2\pi }}\oint {\frac{1}{{{z^n}}}} {e^{ - {z^2}/2 + zx}} \end{equation}
 For large $n$ the asymptotic expression for the integral takes the form:
 \begin{equation}{H_n}(2\sqrt n  + {n^{ - 1/6}}u) = \frac{{n!}}{{2\pi {n^{n/2}}{n^{1/3}}}}\exp \left( {\frac{3}{2}n + {n^{1/3}}u} \right)\int_{ - \infty }^\infty  {\exp [i} ut + i{t^3}/3 +  \ldots ]dt\end{equation}
 so that:
 \begin{equation}{H_n}(2\sqrt n  + {n^{ - 1/6}}u) \approx \frac{{n!}}{{2\pi {n^{n/2}}{n^{1/3}}}}\exp \left( {\frac{3}{2}n + {n^{1/3}}u} \right)Ai[u]\end{equation}
 where $Ai[u]$ is the Airy function.
 
 \subsection{Relation to matrix integrals}
 
 The relation to matrix integrals is through the orthogonal polynomials which can be used to compute the matrix integral. For example for the Matrix Partition function:
 \begin{equation}{Z_n} = \int {DM{e^{ - {V_0}(M)}}} \end{equation}
 then 
  \begin{equation}{Z_n} = {h_0}{h_1} \ldots {h_{n - 1}}\end{equation}
 where:
  \begin{equation}{h_k} = \int {P_k^2(z){e^{ - {V_0}(z)}}dz}\end{equation} 
 For the Gaussian model this is:
  \begin{equation}{Z_n} = {\left( {\frac{{2\pi }}{N}} \right)^{n/2}}{\prod\limits_{k = 1}^{n - 1} {\left( {\frac{k}{N}} \right)} ^{n - k}} = \frac{{{{\left( {2\pi } \right)}^{n/2}}}}{{{N^{{n^2}/2}}}}G(n + 1)\end{equation} 
 where $G(z)$ is the Barnes G-function \cite{Borot}\cite{RandomMatrix}
\subsection{Relation to the Two Matrix Model}
It is interesting that for the Gaussian potential the  prepotential can be used to define a two matrix model defined by:
\begin{equation}Z = \int {d{M_1}d{M_2}{e^{ - tr({V_0}({M_1})) - tr({M_1}{M_2})}}} \end{equation}
This type of model can be solved by using bi-orthogonal polynomials \cite{Maldacena:2004sn}\cite{Seiberg:2004at}\cite{Hashimoto:2005bf} which satisfy:
\begin{equation}\int {dadb{e^{ - {V_0}(a) - ab}}} {Q_m}(a){R_n}(b) = {h_m}{\delta _{m,n}}\end{equation}
and for the choice  $ {Q_n}(a) = {a^n}$ we find the first ten polynomials  $ {R_n}(b)$ to be
\begin{align}
&R_0(t)=1\nonumber\\
&R_0(t)=t\nonumber\\
&R_0(t)=-2 + t^2\nonumber\\
&R_0(t)=-6 t + t^3\nonumber\\
&R_0(t)=12 - 12 t^2 + t^4\nonumber\\
&R_0(t)=60 t - 20 t^3 + t^5\nonumber\\
&R_0(t)=-120 + 180 t^2 - 30 t^4 + t^6\nonumber\\
&R_0(t)=-840 t + 420 t^3 - 42 t^5 + t^7\nonumber\\
&R_0(t)=1680 - 3360 t^2 + 840 t^4 - 56 t^6 + t^8\nonumber \\
&R_0(t)=15120 t - 10080 t^3 + 1512 t^5 - 72 t^7 + t^9
\end{align}
\section{Penner Model and Morse Potential }
The Morse potential is used to describe diatomic molecules \cite{Morse:1929zz}\cite{Bordoni} as well as an example of an exactly soluble potential in supersymmetric quantum mechanics \cite{Gangopadhyaya:2011wka}. For the Morse potential model we will follow \cite{Gangopadhyaya:2011wka} and define the prepotential as:
\begin{equation}{V_0}(x) = A x + {e^{ - x}}\end{equation}
where $A$ is the Morse parameter. The vacuum state is then :
\begin{equation}{\psi _0}(x) = {e^{ - {V_0}(x)}} = {e^{ - A x - {e^{ - x}}}}\end{equation}
and one can define the superpotential:
\begin{equation}W(x) = A - {e^{ - x}}\end{equation}
related to $V_0(x)$ through:
\begin{equation}{V_0}(x) = \int\limits_0^x {W(x')dx'} -1\end{equation}
now defining:
\begin{equation}V_ - (x) = {W^2}(x) - \frac{d}{dx}W(x) =
(A - e^{-x})^2 - {e^{ - x}} = {e^{ - 2x}} - (2A+1){e^{ - x}} + A^2\end{equation}
which is the Morse potential.
The minus partner Hamiltonian is:
\begin{equation}{H_ - } =  - \frac{{{d^2}}}{{d{x^2}}} + {V_ - }(x,A) =  - \frac{{{d^2}}}{{d{x^2}}} + {e^{ - 2x}} - A{e^{ - x}} + {A^2}\end{equation}
with eigenfunctions:
\begin{equation}\psi _n^ - (x) = {e^{ - x(A - n)}}{e^{ - {e^{ - x}}}}L_n^{(2A - 2n)}(2{e^{ - x}})\end{equation}
with $L^{(k)}_n(y)$ the associated Laguerre polynomials. These can also be written as:
\begin{equation}L_n^{(\alpha )}(y) = {y^{ - \alpha }}\oint_C {\frac{{dz}}{{2\pi i}}} \frac{1}{{{z^{n + 1}}}}{(z + y)^{n + \alpha }}{e^{ - z}}\end{equation}
\subsection{Fourier transform of the ground state}
The Fourier transform of the ground state is the ground state in momentum space and can be expressed in terms of the Gamma function as:
\begin{equation}\hat \psi _0^ - (p) = \int_{ - \infty }^\infty  {{\psi _0}(x){e^{-ixp}}dx}  = \int_{ - \infty }^\infty  {{e^{ - A x - {e^{ - x}}}}{e^{-ixp}}dx}  = \Gamma \left( {A + ip} \right)\end{equation}
Note that for the Morse potential the ground state wave function in momentum space has no zeroes or nodes.

\subsection{Ladder operators and factorization}

Similar to the quadratic potential one can define ladder operators for the Morse potential \cite{Cooper}\cite{Junker} by:
$$a(A) = (\frac{d}{{dx}} + W(x)) = (\frac{d}{{dx}} + A - {e^{ - x}})$$
\begin{equation}{a^\dag }(A) = ( - \frac{d}{{dx}} + W(x)) = ( - \frac{d}{{dx}} + A - {e^{ - x}})\end{equation}
with a Hamiltonian with factorization given by:
\begin{equation}{H_ - } = {a^\dag }(A)a(A) = {p^2} + ({W^2}(x) - W'(x)) = {p^2} + ({A^2} + {e^{ - 2x}} - (2A + 1){e^{ - x}})\end{equation}
Using commutators we have:
\begin{equation}[a(A),{a^\dag }(B)] =  - 2{e^{ - x}}I\end{equation}
which can be expressed as:
\begin{equation}[a(A),{a^\dag }(B)] = a(A) + {a^\dag }(B) - \left( {A + B} \right)I\end{equation}
with product:
\begin{equation}{a^\dag }(A)a(B) + {a^\dag }(B)a(A) = 2{p^2} + 2AB + 2{e^{ - 2x}} - 2(1 + A + B){e^{ - x}}\end{equation}
and commutator:
\begin{equation}{a^\dag }(A)a(B) - {a^\dag }(B)a(A) =  - i2(A - B)p\end{equation}
Now defining the variable $y=e^{-x}$ we can write the ladder operators as:
$$a(A) = {y^{ - 1}}\left( { - y{\partial _y} + \left( {A - y} \right)} \right)$$
\begin{equation}{a^\dag }(A) = {y^{ - 1}}\left( {y{\partial _y} + \left( {A - y} \right)} \right)\end{equation}
writing the ground state as:
\begin{equation}{\psi _0}(y) = {y^A}{e^{ - y}}\end{equation}
we can use the Ladder operators to obtain the first excited state as:
\begin{equation}{\psi _1}(y) = {a^\dag }(A + 1){\psi _0}(y) = {y^{A - 1}}{e^{ - y}}(1 + 2A - 2y) = {y^{A - 1}}{e^{ - y}}L_1^{(A - 1)}(2y)\end{equation}
and expressing the Hamiltonian $H_{-}$ as:
\begin{equation}{H_ - } = ( - {y^2}\partial _y^2 - y{\partial _y} + \left( {{A^2} + {y^2} - \left( {2A + 1} \right)y} \right)\end{equation}
we have:
\begin{equation}{H_ - }{\psi _1}(y) = \left( {2A - 1} \right){\psi _1}(y)\end{equation}
which is consistent with the general eigenvalue formula:
\begin{equation}{H_ - }{\psi _n}(y) = \left( {{A^2} - {{\left( {A - n} \right)}^2}} \right){\psi _n}(y)\end{equation}

\subsection{WKB and SWKB expressions}

Similar to the simple harmonic oscillator potential the WKB and Supersymmetric WKB expression for the energies for the Morse potential is exact. We have for the WKB quantization condition \cite{Gangopadhyaya:2011wka}:
\begin{equation}n + \frac{1}{2} = \frac{1}{\pi }\int_{A + \frac{1}{2} - \sqrt {E  + A + \frac{1}{4}} }^{A + \frac{1}{2} + \sqrt {E  + A + \frac{1}{4}} } {\sqrt {E  - \left( {{V_ - }} \right)} } \frac{{dy}}{y} = \frac{1}{\pi }\int_{A + \frac{1}{2} - \sqrt {E  + A + \frac{1}{4}} }^{A + \frac{1}{2} + \sqrt {E  + A + \frac{1}{4}} } {\sqrt {E  - \left( {{A^2} + {y^2} - \left( {2A + 1} \right)y} \right)} } \frac{{dy}}{y}\end{equation}
where we have used the definition $y=e^{-x}$.

For the Supersymmetric WKB  we have the simpler expression:
\begin{equation}n = \frac{1}{\pi }\int_{A - \sqrt E  }^{A + \sqrt E  } {\sqrt {E  - {{\left( W \right)}^2}} } \frac{{dy}}{y} = \frac{1}{\pi }\int_{A - \sqrt E  }^{A + \sqrt E  } {\sqrt {E  - {{\left( {A - y} \right)}^2}} } \frac{{dy}}{y}\end{equation}
In either case one obtains the exact eigenvalue condition:
\begin{equation}{E_n} = {A^2} - {\left( {A - n} \right)^2} = 2An - {n^2}\end{equation}
or
\begin{equation}n = A - \sqrt {{A^2} - E} \end{equation}
In table 1 we list the eigenvalues for the simple case $A=5$.
\begin{table}[h]
\centering
\begin{tabular}{|l|l|l|l|}
\hline
$n$       & $E_n$ & $A$ & $  \sqrt{A^2-E} $\\ \hline
$0$   &      $0$          & $5$   & $5$    \\ \hline
$1 $            & $9$        & $5$           & $4$                    \\ \hline
$2$          & $16$                   & $5$  & $3$                     \\ \hline
$3$            & $21$                   & $5$   & $2$                    \\ \hline
$4$ & $24$                  & $5$  & $1$                      \\ \hline
$5$        & $25$                    & $5$  & $0$                     \\ \hline
\end{tabular}
\caption{\label{tab:table-name} Eigenvalues for bound states plus first unbound state for the Morse potentials for $A=5$. These states satisfy the quantization condition $n = A - \sqrt {{A^2} - E} $}
\end{table}

\subsection{Relation to two dimensional Harmonic oscillator and Coulomb potential}


It is interesting that there is a relation between the Morse potential and the two dimensional isotropic simple harmonic oscillator with potential \cite{Cooper}\cite{Kostelecky:1995fh}\cite{Kostelecky:1985fx}\cite{Levai:1998wm}:
\begin{equation}
V(r) = \frac{{{L^2}}}{{{r^2}}} + {r^2}\end{equation}
The radial wave function of the two dimensional Harmonic oscillator are:
\begin{equation}R(r) = {r^L}{e^{ - {r^2}/2}}L_{(N - L)/2}^{(L)}({r^2})\end{equation}
with the relation to the Morse potential through:
$${r^2} = 2{e^{ - x}}$$
$$N = 2A$$
\begin{equation}\frac{L}{2} = A - n\end{equation}
There is also a relation to the two dimensional Coulomb potential
\begin{equation}V({r_C}) = \frac{{{\ell ^2}}}{{r_C^2}} - \frac{1}{{{r_C}}}\end{equation}
with radial wave function given by:
\begin{equation}R({r_C}) = r_C^\ell {e^{ - {r_C}/2}}L_{{n_C} - \ell  - 1}^{(2\ell )}({r_C})\end{equation}
with:
$${r_C} = 2{e^{ - x}}$$
$${n_C} = A + 1$$
\begin{equation}\ell  = A - n\end{equation}

\subsection{Complete basis}

The Morse bound states do not form a complete basis of Hilbert space because they don't include the unbound states. One can form the set \cite{Bordoni}:
\begin{equation}\sqrt {\frac{{\alpha n!}}{{\Gamma (2\sigma  + 1)}}} {y^\sigma }{e^{ - y/2}}L_n^{(2\sigma  - 1)}(y)\end{equation}
which does form a complete basis. These complete basis can be useful for example to perform a basis expansion of a modification to the Morse potential.

\subsection{Jacobi Matrix}
The Jacobi Matrix associated with the Morse potential is determined by the recurrence relations for the the Laguerre polynomials which are:
\begin{equation}(n + 1)L_{n + 1}^{(\alpha)} (y) = (1 + 2n + \alpha  - y)L_n^{(\alpha) }(y) - (n + \alpha )L_{n - 1}^{(\alpha)} (y)\end{equation}
Now defining monic polynomials by:
        \begin{equation}L_n^{(\alpha )}(y) = \frac{{{{\left( { - 1} \right)}^n}}}{{n!}}p_n^\alpha (y)\end{equation}
the recurrence relations become:
\begin{equation}p_{n + 1}^\alpha (y) = yp_n^\alpha  - (2n + 1 + \alpha )p_n^\alpha (y) - n(n + \alpha )p_{n - 1}^\alpha \end{equation}
So that 
$${\alpha _n} = 2n + 1 + \alpha $$
\begin{equation}{\beta _n} = n(n + \alpha )\end{equation}
and the Jacobi matrix is:
\begin{equation}
 J_{n+1} = \begin{bmatrix}

   {1 + \alpha } & {\sqrt {1(1 + \alpha )} } & 0 &  \cdots  & 0  \\ 
   {\sqrt {1(1 + \alpha )} } & {3 + \alpha } & {\sqrt {2(2 + \alpha )} } &  \cdots  & 0  \\ 
   0 & {\sqrt {2(2 + \alpha )} } &  \ddots  &  \ddots  & 0  \\ 
   0 & 0 &  \ddots  & {2n - `1 + \alpha } & {\sqrt {n(n + \alpha )} }  \\ 
   0 & 0 &  \cdots  & {\sqrt {n(n + \alpha )} } & {2n + 1 + \alpha }  \\ 
\end{bmatrix}
 \end{equation}

\subsection{Large $N$ asymptotics}


Asymtotically can use a saddle point approximation to the integral definition of the Laguerre polynomial to determine the relation:
\begin{equation}L_n^{(\alpha )}(\frac{y}{n}) \approx {\left( {\frac{y}{n}} \right)^{ - \alpha /2}}{e^{ - y/2n}}{J_\alpha }\left( {2\sqrt y } \right)\end{equation}
Thus one can study the zeros of the Bessel function as the limit of an infinitely large characteristic polynomial associated with the Hermitean Jacobi matrix (.).

\subsection{Relation to matrix integrals}

The Penner Matrix integral is given by:

\begin{equation}
Z=\int {DM{e^{ - \frac{N}{\gamma }trM - \frac{N}{\gamma }tr(\log(M))}}} \end{equation}
It is related to the Morse potential prepotential in the $y=e^{-x}$ coordinate represntaion through $V_0(y)=-A\log y + y$. The Penner matrix model can be solved by using the associated Laguerre orthogonal polynomials \cite{Penner:1988cza}\cite{Distler:1990pg}\cite{Chaudhuri:1991hv}\cite{Alvarez}\cite{Deo}. The polynomials are normalized so that:
\begin{equation}\int_0^\infty  {dy{e^{ - \alpha y}}{y^\alpha }P_n^\alpha (y)P_m^\alpha (y) = {\delta _{n,m}}} {\alpha ^{ - 2n - \alpha  - 1}}n!\Gamma (n + \alpha  + 1)\end{equation}
so that 
\begin{equation}h_n = {\alpha ^{ - 2n - \alpha  - 1}}n!\Gamma (n + \alpha  + 1)\end{equation}
and 
\begin{equation}Z = N!\prod\limits_{n = 1}^N {{h_n}}  = N!\prod\limits_{n = 1}^N {{\alpha ^{ - 2n - \alpha  - 1}}n!\Gamma (n + \alpha  + 1)} \end{equation}
We aslo have:
\begin{equation}Z = {e^{ - \frac{{{N^2}}}{\gamma }}}\int_0^\infty  {\prod\limits_{i = 1}^N {d\lambda _i^\alpha {e^{ - \alpha {\lambda _i}}}\det (P_i^\alpha ({\lambda _j}))} } \end{equation}
with  $  \alpha  =  - \frac{N}{\gamma } > 0$.
which can be written:
\begin{equation}{Z_n} = {\left( {\frac{{\Gamma (N + 1)}}{{{N^{N + 1}}}}} \right)^n}{\prod\limits_{k = 1}^{n - 1} {\left( {\frac{k}{N} + \frac{{{k^2}}}{{{N^2}}}} \right)} ^{n - k}} = \frac{1}{{{N^{n(n + N)}}}}\frac{{G(n + 1)G(N + n + 1)}}{{G(N + 1)}}\end{equation}

\section{Riemann potential I }

Similar to the treatment of the Morse potential one can develop a potential associated with the Riemann zeta function which we call the Riemann potential.  To discuss the Riemann potential model our starting point is the potential
\begin{equation}{V_0}(x) = A x + \log (1 + {\exp(e^{ - x}}))\end{equation}
so that the vacuum state is:
\begin{equation}{\psi _0}(x) = {e^{ - {V_0}(x)}} = {e^{ - A x - \log(1+{\exp(e^{ - x}}))}}
={e^{ - A x/}}\frac{1}{{1 + \exp({e^{ - x}})}}
\end{equation}
Note that unlike the Gaussian model this vacuum state is not the vacuum state of $V_0(x)$. Instead we form the superpotential:
\begin{equation}W(x) = V_0'(x)= A - {e^{ - x}}\frac{{\exp ({e^{ - x}})}}{{1 + \exp ({e^{ - x}})}}\end{equation}
This can be written as:
\begin{equation}W(x) = A - {e^{ - x}} + \frac{{{e^{ - x}}}}{{1 + \exp ({e^{ - x}})}} = A - {e^{ - x}} + f_1(x)\end{equation}
with:
\begin{equation}f_1(x) = \frac{{{e^{ - x}}}}{{1 + \exp ({e^{ - x}})}}\end{equation}
so that 
\begin{equation}{\psi _0}(x) = {e^{ - \int_0^x {W(z)dz} }}\end{equation}
as is usual in supersymmetric quantum mechanics.
We can then define the Riemann potential as:
\begin{equation}{V_{ - }}(x,A) = {W^2}(x) - W'[x] \end{equation}
This can also be written:
\begin{equation}{V_ - }(x,A) = {A^2} + {e^{ - 2x}} - (2A + 1){e^{ - x}} + {e^{ - 2x}}\frac{{\left( { - 1 - 3\exp ({e^{ - x}})} \right)}}{{{{\left( {1 + \exp ({e^{ - x}})} \right)}^2}}} + (2A + 1){e^{ - x}}\frac{1}{{1 + \exp ({e^{ - x}})}}\end{equation}
and:
\begin{equation}{V_ + }(x,A) = {A^2} + {e^{ - 2x}} - (2A - 1){e^{ - x}} + {e^{ - 2x}}\frac{{\left( { - 1 - 1\exp ({e^{ - x}})} \right)}}{{{{\left( {1 + \exp ({e^{ - x}})} \right)}^2}}} + (2A - 1){e^{ - x}}\frac{1}{{1 + \exp ({e^{ - x}})}}\end{equation}
or
\begin{equation}{V_ - }(x,A) = {A^2} + {e^{ - 2x}} - (2A + 1){e^{ - x}} + {e^{ - 2x}}\frac{{\left( { - 1 - (2 + 1)\exp ({e^{ - x}})} \right)}}{{{{\left( {1 + \exp ({e^{ - x}})} \right)}^2}}} + (2A + 1){e^{ - x}}\frac{1}{{1 + \exp ({e^{ - x}})}}\end{equation}
and
\begin{equation}{V_ + }(x,A) = {A^2} + {e^{ - 2x}} - (2A - 1){e^{ - x}} + {e^{ - 2x}}\frac{{\left( { - 1 - (2 - 1)\exp ({e^{ - x}})} \right)}}{{{{\left( {1 + \exp ({e^{ - x}})} \right)}^2}}} + (2A - 1){e^{ - x}}\frac{1}{{1 + \exp ({e^{ - x}})}}\end{equation}
Note:
\begin{equation}{V_ + }(x,A + 1) = {V_ - }(x,A) + f_2(x)\end{equation}
with:
\begin{equation}f_2(x) = 2\frac{{{e^{ - 2x}}\exp ({e^{ - x}})}}{{{{\left( {1 + \exp ({e^{ - x}})} \right)}^2}}}\end{equation}
So there is a generalization of shape invariance for the Riemann potential.
The relation between the two potentials can written symmetrically as:
\begin{equation}{V_ + }(x,A + 1) - {\left( {A + 1} \right)^2} - {e^{ - 2x}}\frac{{\exp ({e^{ - x}})}}{{{{\left( {1 + \exp ({e^{ - x}})} \right)}^2}}} = {V_ - }(x,A) - {A^2} + {e^{ - 2x}}\frac{{\exp ({e^{ - x}})}}{{{{\left( {1 + \exp ({e^{ - x}})} \right)}^2}}}\end{equation}
Defining $y=e^{-x}$ we can form the ladder operators in the $y$ representation as:
$$\alpha(A) = {y^{ - 1}}\left( { - y{\partial _y} + (A - y) + \frac{y}{{1 + {e^y}}}} \right)$$
\begin{equation}\alpha^\dagger(A) = {y^{ - 1}}\left( {y{\partial _y} + (A - y) + \frac{y}{{1 + {e^y}}}} \right)\end{equation}
We plot the Riemann potential as well as the Morse potential in Figure 4. WE can see that they agree for large and small $x$ but the Riemann potential is deeper with a minimum shifted to the left with respect to the Morse potential. The maximum of the ground state wave function is also shifted to the left with respect for the Riemann function with respect to the Morse potential.

Given the Riemann potential we can for the Riemann Hamiltonian from:
\begin{equation}{H_ - } =  - {y^2}\partial _y^2 - y{\partial _y} + {A^2} + {y^2} - \left( {2A + 1} \right)y + (2A + 1)y\frac{1}{{{e^y} + 1}} + {y^2}\frac{{\left( { - 1 - 3{e^y}} \right)}}{{{{\left( {{e^y} + 1} \right)}^2}}}\end{equation}
and creation and annihilation operators:
$$a(A) =  - y{\partial _y} + A - y + y\frac{1}{{1 + {e^y}}}$$
\begin{equation}{a^\dag }(A) =  - y{\partial _y} + A - y + y\frac{1}{{1 + {e^y}}}\end{equation}
we can readily verify that:
\begin{equation}{H_ - } = {a^\dag }(A)a(A)\end{equation}
so that
\begin{equation}{H_ - }{\psi _0} =0\end{equation}
and
\begin{equation}a(A){\psi _0}=0\end{equation}
where
\begin{equation}{\psi _0} = {y^A}\frac{1}{{{e^y} + 1}}\end{equation}
This represents the ground state of the Riemann potential.

\subsection{Fourier transform of the ground state}
The quantity of interest is the Fourier transform of the ground state given by:
\begin{equation}{{\hat \psi }_0}(p) = \int_{ - \infty }^\infty  {{\psi _0}(x){e^{ - ipx}}dx = } \int_{ - \infty }^\infty  {\frac{{{e^{ - A x}}}}{{{e^{{e^{ - x}}}} + 1}}{e^{ - ipx}}dx = } \Gamma (A + ip)\eta (A + ip)\end{equation}
where $\eta(z)$ is the Dirichlet eta function related to the Riemann zeta function through:
\begin{equation}\eta (z) = (1 - {2^{1 - z}})\zeta (z)\end{equation}
In many physical representations of the Riemann zeta function one has a physical way of expressing the Riemann hypothesis \cite{Sierra:2016rgn}\cite{Sierra:2014wua}\cite{Khuri:2001yd}\cite{Das:2018dit}\cite{Lagarias}\cite{Rahm}. In the representation in terms of the Riemann potential one can state the hypothesis so that only for a Morse-like parameter $A=1/2$ does the ground state wave function in the momentum presentation have nodes or zeros. Note that the representation in momentum space is important as there is a theorem that the ground state wave function in position space has no nodes or zeros. Although the connection to supersymmetric quantum mechanics to the Riemann Zeta function and the Morse potential has been made in \cite{Das:2018dit}\cite{Lagarias}\cite{Rahm} our approach is somewhat different in that we concentrate on the ground state in momentum space to form the hypothesis.

\subsection{Ladder operators and factorization}
Similar to the Morse potential one can form ladder operators in $x$ space and their commutators. Defining:
$$a(A) = (\frac{d}{{dx}} + W(x)) = \frac{d}{{dx}} + 
A - {e^{ - x}}\frac{{\exp ({e^{ - x}})}}{{1 + \exp ({e^{ - x}})}}$$
\begin{equation}{a^\dag }(A) = ( - \frac{d}{{dx}} + W(x)) = - \frac{d}{{dx}}
+ A - {e^{ - x}}\frac{{\exp ({e^{ - x}})}}{{1 + \exp ({e^{ - x}})}}\end{equation}

with
$${H_ - } = {a^\dag }(A)a(A) = {p^2} + {W^2}(x) - W'(x) = $$
\begin{equation}{p^2} + {A^2} + {e^{ - 2x}}\frac{{\exp (2{e^{ - x}}) - \exp ({e^{ - x}})}}{{{{\left( {1 + \exp ({e^{ - x}})} \right)}^2}}} - (2A + 1){e^{ - x}}\frac{{\exp ({e^{ - x}})}}{{1 + \exp ({e^{ - x}})}}\end{equation}
Taking the commutation relation we have:
\begin{equation}[{a^\dag }(B),a(A)] = 2{e^{ - 2x}}\frac{{exp({e^{ - x}})}}{{{{\left( {1 + \exp ({e^{ - x}})} \right)}^2}}} + AB( - 1 + {e^{ - x}})\frac{{exp(2{e^{ - x}})}}{{{{\left( {1 + \exp ({e^{ - x}})} \right)}^2}}}\end{equation}

\subsection{Gram-Schmidt Orthogonalization Process}

Orthogonal polynomials for the Riemann Potential can be determined by the Gram-Schmidt process associated with weight function. One starts with:
\begin{equation}R_0^{(\alpha )}(y) = 1\end{equation}
and the weight function
\begin{equation}w(y) = {y^\alpha }{e^{ - y}}\frac{1}{{1 + {e^{ - y}}}}\end{equation}
to determine
\begin{equation}{B_1} = \frac{{\int_0^\infty  {yw(y){{\left( {R_0^{(\alpha }(y)} \right)}^2}dy} }}{{\int_0^\infty  {w(y){{\left( {R_0^{(\alpha }(y)} \right)}^2}dy} }}\end{equation}
and 
\begin{equation}R_1^{(\alpha )}(y) = y - {B_1}\end{equation}
The rest of the polynomials are determined from:
\begin{equation}R_k^{(\alpha )}(y) = (y - {B_k})R_{k - 1}^{(\alpha )}(y) - {C_k}R_{k - 2}^{(\alpha )}(y)\end{equation}
with
\begin{equation}{B_k} = \frac{{\int_0^\infty  {yw(y){{\left( {R_{k - 1}^{(\alpha }(y)} \right)}^2}dy} }}{{\int_0^\infty  {w(y){{\left( {R_{k - 1}^{(\alpha }(y)} \right)}^2}dy} }}\end{equation}
and
\begin{equation}{C_k} = \frac{{\int_0^\infty  {yw(y)R_{k - 1}^{(\alpha }(y)R_{k - 2}^{(\alpha }(y)dy} }}{{\int_0^\infty  {w(y){{\left( {R_{k - 2}^{(\alpha }(y)} \right)}^2}dy} }}\end{equation}
Using the Gram-Schmidt process as well as integrals of the form:
\begin{equation}\int_0^\infty  {yw(y){y^m}{y^n}dy}  = (1 - {2^{ - m - n - \alpha }})\Gamma (1 + m + n + \alpha )\zeta (1 + m + n + \alpha )\end{equation}
we determine the first few polynomials as:
\begin{align}
  & R_0^{(1)}(y) = 1  \nonumber\\ 
  & R_1^{(1)}(y) = {\text{ - 2.19229  +  y}}  \nonumber\\ 
  & R_2^{(1)}(y) = 6.87631 - 6.28796y + {y^2}  \nonumber\\
  & R_3^{(1)}(y) =  - 28.2686 + 38.905y - 12.3597{y^2} + {y^3} 
  \end{align}
To determine the Jacobi Matrix we can use:
\begin{equation}R_{k + 1}^{(\alpha )}(y) = (y - {B_{k + 1}})R_k^{(\alpha )}(y) - {C_{k + 1}}R_{k - 1}^{(\alpha )}(y) = yR_k^{(\alpha )}(y) - {\alpha _k}R_k^{(\alpha )}(y) - {\beta _k}R_{k - 1}^{(\alpha )}(y)\end{equation}
so that
$${\alpha _k} = {B_{k + 1}}$$
\begin{equation}{\beta _k} = {C_{k + 1}}\end{equation}
and for the first few coefficients we have:
\newpage
$$B_1=2.19229$$
$$B_2=4.09567$$
$$B_3=6.07169$$
$$C_2=2.10259$$
\begin{equation}C_3=6.14983\end{equation}

\subsection{Orthogonal polynomials for the Matrix intregral}

We also determine the first few orthonomal polynomials associated with the Matrix integral (2).
\begin{equation}Z(A,B,C) = \int {dM{e^{ATr\left( {LogM} \right) - BTr\left( M \right) - CTr\left( {Log\left( {1 + Exp( - M/C} \right)} \right)}})} \end{equation}
For $A=0$, $B=1$ and $C=2$ this is:
\begin{equation}Z(0,1,2) = \int {dM{e^{- Tr\left( M \right) - 2Tr\left( {Log\left( {1 + Exp( - M/2} \right)} \right)}})} \end{equation}
the first few orthogonal polynomials associated with the this matrix integral and measure:
\begin{equation}w(y) ={e^{ - y}}\frac{1}{{{{\left( {1 + {e^{ - y/2}}} \right)}^2}}}\end{equation}
are:
\begin{align}
  & R_0^{(0)}(y) = 1 \nonumber \\ 
  & R_1^{(0)}(y) = {{ - 1.33908  +  y}} \nonumber \\ 
  & R_2^{(0)}(y) = 2.97619 - 4.66845y + {y^2} \nonumber \\
  & R_3^{(0)}(y) =  -9.40578 + 22.8139y - 9.90732{y^2} + {y^3} 
  \end{align}
To determine the Jacobi matrix we use the recursion coefficients:
$$B_1=1.33908$$
$$B_2=3.32937$$
$$B_3=5.23886$$
$$C_2=1.48211$$
\begin{equation}C_3=4.61963\end{equation}
The value of the Matrix integral is determined by the product quantities $h_n$ which are given by 
\begin{equation}{h_n} = \int_0^\infty  {w(y){{\left( {R_n^{(\alpha )}(y)} \right)}^2}dy}  = \int_0^\infty  {{e^{ - y}}\frac{1}{{{{\left( {1 + {e^{ - y/2}}} \right)}^2}}}{{\left( {R_n^{(0 )}(y)} \right)}^2}dy} \end{equation}
and take the values:
$(h_0, h_1, h_2, h_3) = (.386294,.52531,2.64488, 25.5684)$
one can define orthonormal functions by:
\begin{equation}\psi _n^{(0)}(y) = \frac{1}{{\sqrt {{h_n}} }}{e^{ - V(y)/2}}R_n^{(0)}(y) = \frac{1}{{\sqrt {{h_n}} }}\sqrt {w(y)} R_n^{(0)}(y) = \frac{1}{{\sqrt {{h_n}} }}{e^{ - y/2}}\frac{1}{{\left( {1 + {e^{ - y/2}}} \right)}}R_n^{(0)}(y)\end{equation}
Finally the ground state for the Riemann potential I can be expressed as:
\begin{equation}\psi _0^{(RI)}(x) = \sqrt {{e^{ - (x - \log (2))}}} \psi _0^{(0)}\left( {{e^{ - (x - \log (2))}}} \right)\end{equation}





%



\section{Interpolation between the Riemann and  Morse potential}

Isospectral deformation of a potential is a one parameter deformation of the ground state, superpotential and partner potentials that preserve the energy eigenvalues and reflection or transmission coefficients despite a potentially large variation is the function form of the potentials.

For the Morse potential the isospectral deformation of the Morse potential ground state is \cite{Gangopadhyaya:2011wka}:
\begin{equation}{\psi _0}(x,\lambda ) = \frac{{\sqrt {\lambda (\lambda  + 1)} \sqrt 2 {e^{ - x/2}}}}{{{e^{ - {e^{ - x}}}} + \lambda {e^{{e^{ - x}}}}}} = \frac{{\sqrt {\lambda (\lambda  + 1)} }}{\lambda }\frac{{\sqrt 2 {e^{ - x/2}}}}{{\left( {\frac{1}{\lambda }{e^{ - {e^{ - x}}}} + {e^{{e^{ - x}}}}} \right)}}\end{equation}
Despite the similarity it is not possible to reach the Riemann potential ground state by isospectral deformation which is given by:
\begin{equation}{\psi _0}(x) = \frac{{{e^{ - x/2}}}}{{1 + {e^{{e^{ - x}}}}}} = \frac{{{e^{ - x/2}}{e^{ - {e^{ - x}}}}}}{{\left( {{e^{ - {e^{ - x}}}} + 1} \right)}}\end{equation}
Instead we consider a different deformation of the Morse potential as:
\begin{equation}{V_0}(x,T) = Ax + {e^{ - x}} + T\log (1 + {e^{ - {e^{ - x}}/T}})\end{equation}
so that the ground state wave function is:
\begin{equation}{\psi _0}(x,T) = {e^{ - {V_0}(x,T)}} = {e^{ - Ax}}{e^{ - {e^{ - x}}}}{\left( {\frac{1}{{1 + {e^{ - {e^{ - x}}/T}}}}} \right)^T}\end{equation}
The superpotential is then:
\begin{equation}W(x,T) = {\partial _x}{V_0}(x,T) = A - {e^{ - x}} + \frac{{{e^{ - x}}}}{{1 + {e^{{e^{ - x}}/T}}}}\end{equation}
The minus partner potential take a complicated form given by:
\begin{equation}{V_ - }(x,T) = {A^2} + {e^{ - 2x}} - (2A + 1){e^{ - x}} + (2A + 1)\frac{{{e^{ - x}}}}{{1 + {e^{{e^{ - x}}/T}}}} + {e^{ - 2x}}\left( { - 1 - \left( {2 + \frac{1}{T}} \right){e^{{e^{ - x}}}}} \right)\frac{1}{{{{\left( {1 + {e^{{e^{ - x}}/T}}} \right)}^2}}}\end{equation}
This agrees with the Morse potential for $T=0$ and the Riemann potential for $T=1$. For  $T = \infty $ this become the Morse potential shifted to the left by $   \log (2)$. The relation between these three cases are shown in figure 1 for $A = 1/2$. Note that as $T$ varies from zero to infinity the deepest potentials is obtained for the Riemann potential at $T=1/\beta =1$. As such minima are also obtained in compactified theories one might also refer to this type of deformation as compactified radius deformation with radius $R=2 \pi/T$.

\begin{figure}
  \includegraphics[width =  \linewidth]{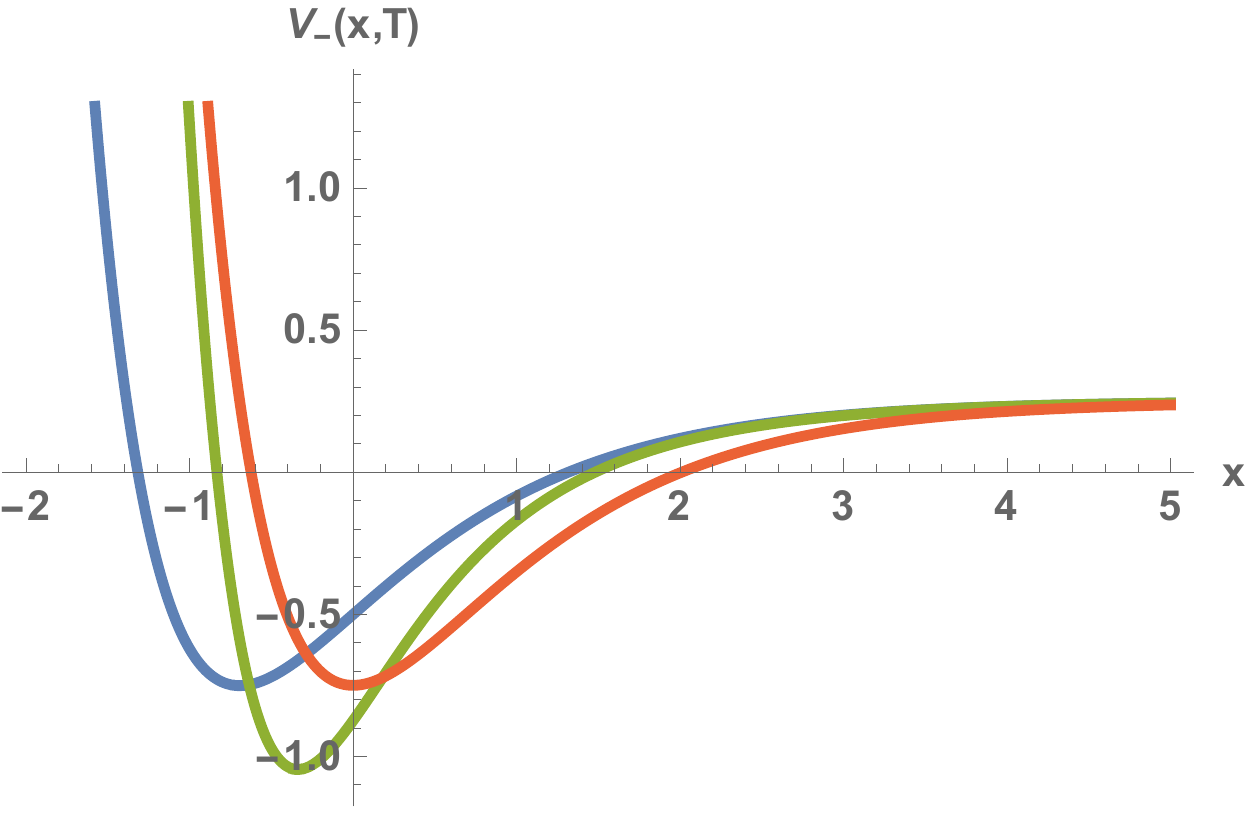}
  \caption{Riemann Potential I (green) for $T=1$ and $A=1/2$. For $T=0$ this agrees with the Morse potential (red) and for $T=\infty$ this agrees with the shifted Morse Potential.}
  \label{fig:Radion Potential}
\end{figure}

\section{ Finite difference equation in momentum space }

It is known the momentum eigenfunctions of the Morse potential obey a finite difference equation involving translations in imaginary momentum \cite{Dahl2}. For example the position space equation:
\begin{equation}\left( {\frac{d}{{dx}} + A - {e^{ - x}}} \right){\psi _0}(x) = 0\end{equation}
which becomes is momentum space:
\begin{equation}(A + ip){{\tilde \psi }_0}(p) - {{\tilde \psi }_0}(p - i) = 0\end{equation}
For the Riemann potential we have the more complicated equation:
\begin{equation}\left( {\frac{d}{{dx}} + A - {e^{ - x}} + \frac{{{e^{ - x}}}}{{1 + {e^{{e^{ - x}}}}}}} \right){\psi _0}(x) = 0\end{equation}
However we can use generating function for the Euler numbers given by:
\begin{equation}\frac{2}{{{e^t} + 1}} = \sum\limits_{n = 0}^\infty  {{E_n}(0)\frac{{{t^n}}}{{n!}}} \end{equation}
to represent:
\begin{equation}\frac{{{e^{ - x}}}}{{1 + {e^{{e^{ - x}}}}}} = \frac{1}{2}\sum\limits_{n = 0}^\infty  {{E_n}(0)\frac{{{e^{ - x(n + 1)}}}}{{n!}}} \end{equation}
Then the momentum eigenfunction will obey the following difference equation involving discrete translations in imaginary momentum:
\begin{equation}(A + ip){{\tilde \psi }_0}(p) - {{\tilde \psi }_0}(p - i) + \frac{1}{2}\sum\limits_{n = 0}^\infty  {{E_n}(0)\frac{{{{\tilde \psi }_0}(p - i(n + 1))}}{{n!}}}  = 0\end{equation}

\section{Relation to Morse, radial harmonic oscillator and Coulomb potential}

Just as for the ordinary Morse potential there is a relation of the Riemann potential to the two dimensional modified simple harmonic oscillator and two dimensional modified Coulomb potential  in radial coordinates.

\subsection*{Relation to Morse potential}
The usual Morse potential is given by:
\begin{equation}V_M(x) = {A^2} + {e^{ - 2x}} - (2A + 1){e^{ - x}} \end{equation}
While the Riemann potential is:
\begin{equation}V_R(x) = {A^2} + {e^{ - 2x}} - (2A + 1){e^{ - x}} + {e^{ - 2x}}\frac{{\left( { - 1 - 3\exp ({e^{ - x}})} \right)}}{{{{\left( {1 + \exp ({e^{ - x}})} \right)}^2}}} + (2A + 1){e^{ - x}}\frac{1}{{1 + \exp ({e^{ - x}})}}\end{equation}
The ground state of the Riemann potential is:
\begin{equation}{\psi _{0R}}(x) = {e^{ - xA}}{e^{ - {e^{ - x}}}}\frac{1}{{1 + {e^{ - {e^{ - x}}}}}}\end{equation}
We plot these two functions for $A=1/2$ in Figure 2.
\begin{figure}
  \includegraphics[width =  \linewidth]{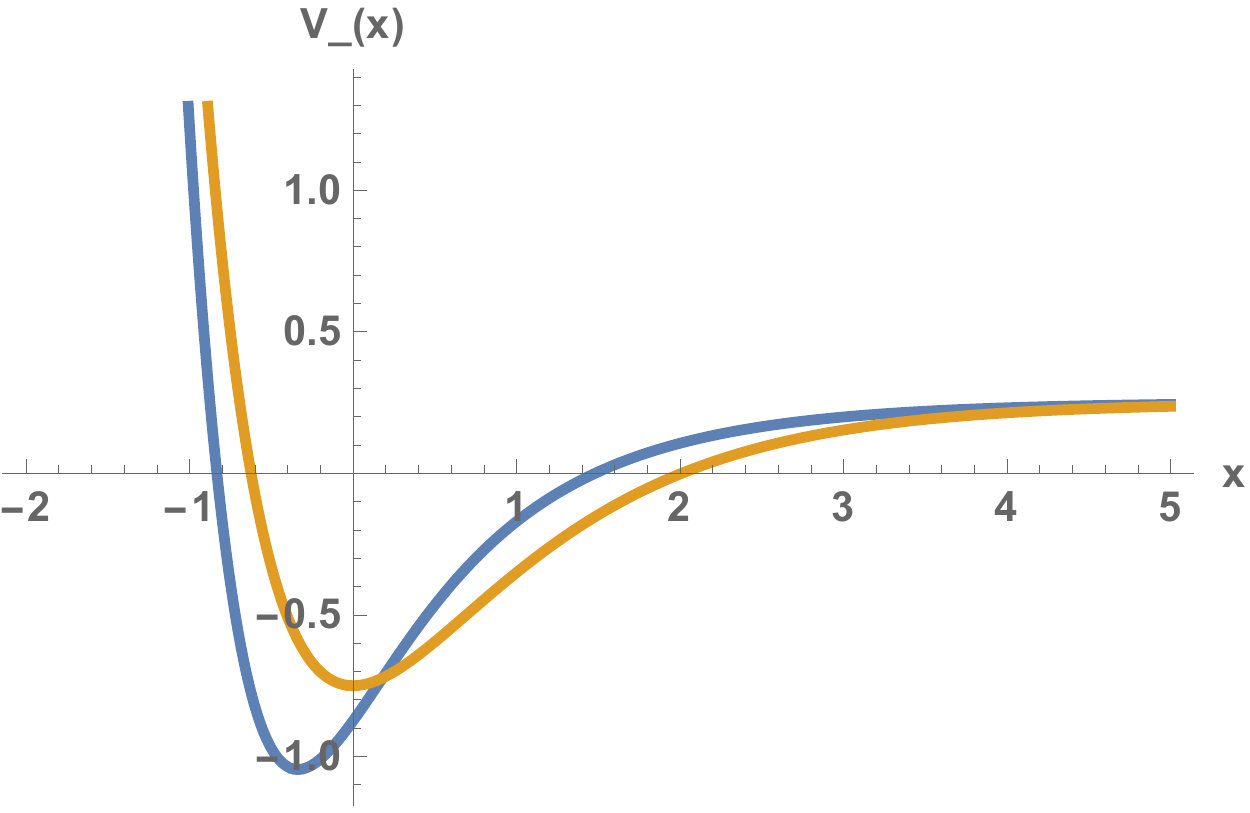}
  \caption{Ordinary Morse potential orange, Riemann modified potential blue.}
  \label{fig:Radion Potential}
\end{figure}

\subsection*{Relation to radial harmonic oscillator potential}

For the Riemann potential there is a relation to a central force potential in two dimensions with a complicated potential given by:
\begin{equation}V(r) = \frac{{{r^2}}}{2} + \frac{{{r^2}}}{2}\frac{{\left( { - 1 - 3{e^{{r^2}/2}}} \right)}}{{{{\left( {1 + {e^{{r^2}/2}}} \right)}^2}}} + (2A + 1)\frac{1}{{\left( {1 + {e^{{r^2}/2}}} \right)}}\end{equation}
The state:
\begin{equation}\psi (r) = {r^{2A}}\frac{1}{{1 + {e^{{r^2}/2}}}}\end{equation}
corresponds to the ground state of the Riemann potential through the relation:
\begin{equation}r^2 = 2 e^{-x}\end{equation}
Hamiltonian:
\begin{equation}{H_ - } =  - \frac{{{r^2}}}{4}\partial _r^2 - \frac{r}{4}{\partial _r} + {A^2} + \frac{{{r^4}}}{4} - (2A + 1)\frac{{{r^2}}}{2} + (2A + 1)\frac{{{r^2}}}{2}\frac{1}{{{e^{{r^2}/2}} + 1}} + \frac{{{r^4}}}{4}\frac{{\left( { - 1 - 3{e^{{r^2}/2}}} \right)}}{{{{\left( {{e^{{r^2}/2}} + 1} \right)}^2}}}\end{equation}
or:
\begin{equation}{r^{ - 2}}({H_ - } - {E_n}) =  - \frac{1}{4}\partial _r^2 - \frac{1}{{4r}}{\partial _r} + \frac{{{A^2} - {E_n}}}{{{r^2}}} + \frac{{{r^2}}}{4} - (2A + 1)\frac{1}{2} + (2A + 1)\frac{1}{2}\frac{1}{{{e^{{r^2}/2}} + 1}} + \frac{{{r^2}}}{4}\frac{{\left( { - 1 - 3{e^{{r^2}/2}}} \right)}}{{{{\left( {{e^{{r^2}/2}} + 1} \right)}^2}}}\end{equation}
For the radial two dimensional isotropic harmonic oscillator coordinates
\begin{equation}r^2=2e^{-x}\end{equation}
the Riemann potential simple harmonic central potential is:
\begin{equation}V(r) =  4\frac{{{A^2} - {E_n}}}{{{r^2}}} + r^2 - 2(2A + 1) + 2(2A + 1)\frac{1}{{{e^{{r^2}/2}} + 1}} + r^2\frac{{\left( { - 1 - 3{e^{{r^2}/2}}} \right)}}{{{{\left( {{e^{{r^2}/2}} + 1} \right)}^2}}}\end{equation}
We plot this potential in figure 3.
\begin{figure}
  \includegraphics[width =  \linewidth]{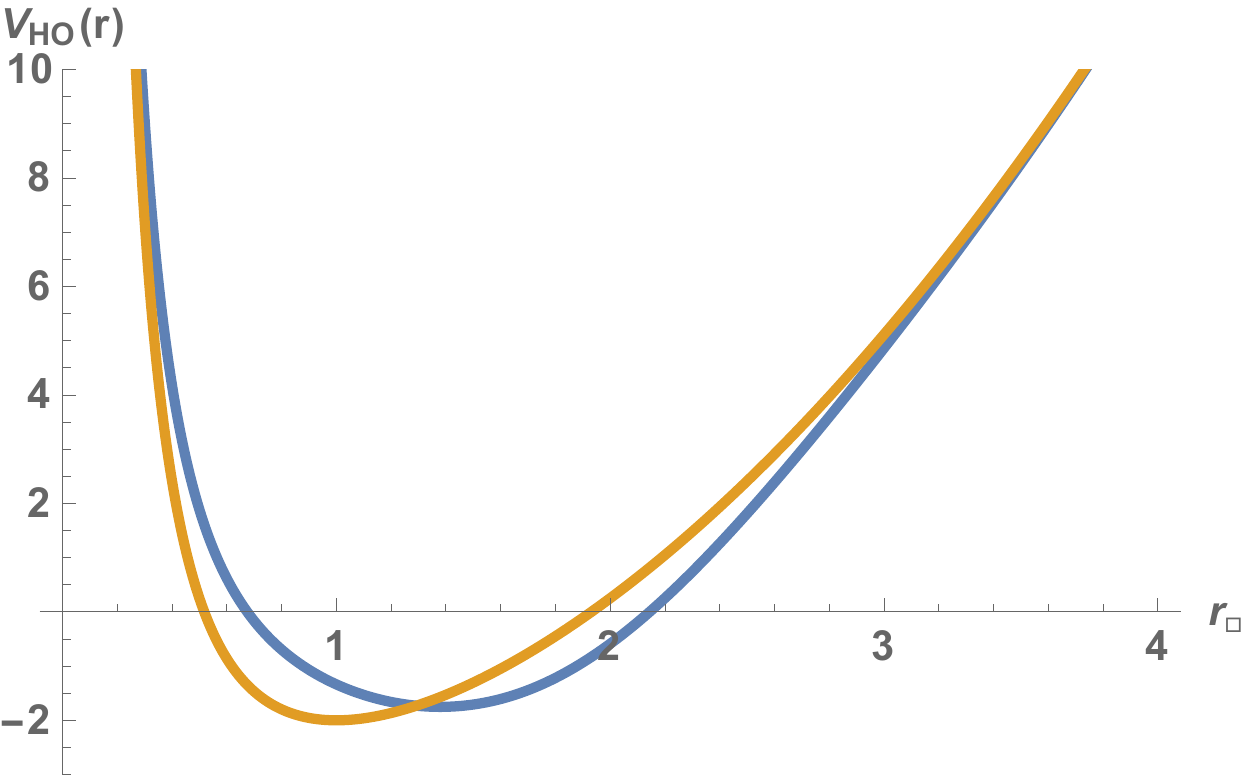}
  \caption{Ordinary two dimensional simple harmonic oscillator potential orange, Riemann harmonic oscillator potential blue.}
  \label{fig:Radion Potential}
\end{figure}

\subsection*{Relation to the Coulomb potential}

For the Coulomb potential in two spatial dimensions we have for Coulomb coordinates
\begin{equation}
r_C=2e^{-x}
\end{equation}
the central potential
\begin{equation}{V_C}({r_C}) = \frac{1}{4} - \frac{{2A + 1}}{{2{r_C}}} + \frac{{{A^2} - {E_n}}}{{r_C^2}} + \frac{{2A + 1}}{{2{r_C}}}\frac{1}{{{e^{{r_C}/2}} + 1}} + \frac{1}{4}\frac{{\left( { - 1 - 3{e^{{r_C}/2}}} \right)}}{{{{\left( {{e^{{r_C}/2}} + 1} \right)}^2}}}\end{equation}
and the ground state wave function is of the form:
\begin{equation}{\psi _{0C}}({r_C}) = r_C^A\frac{1}{{{e^{{r_C}/2}} + 1}}\end{equation}
\begin{figure}
  \includegraphics[width =  \linewidth]{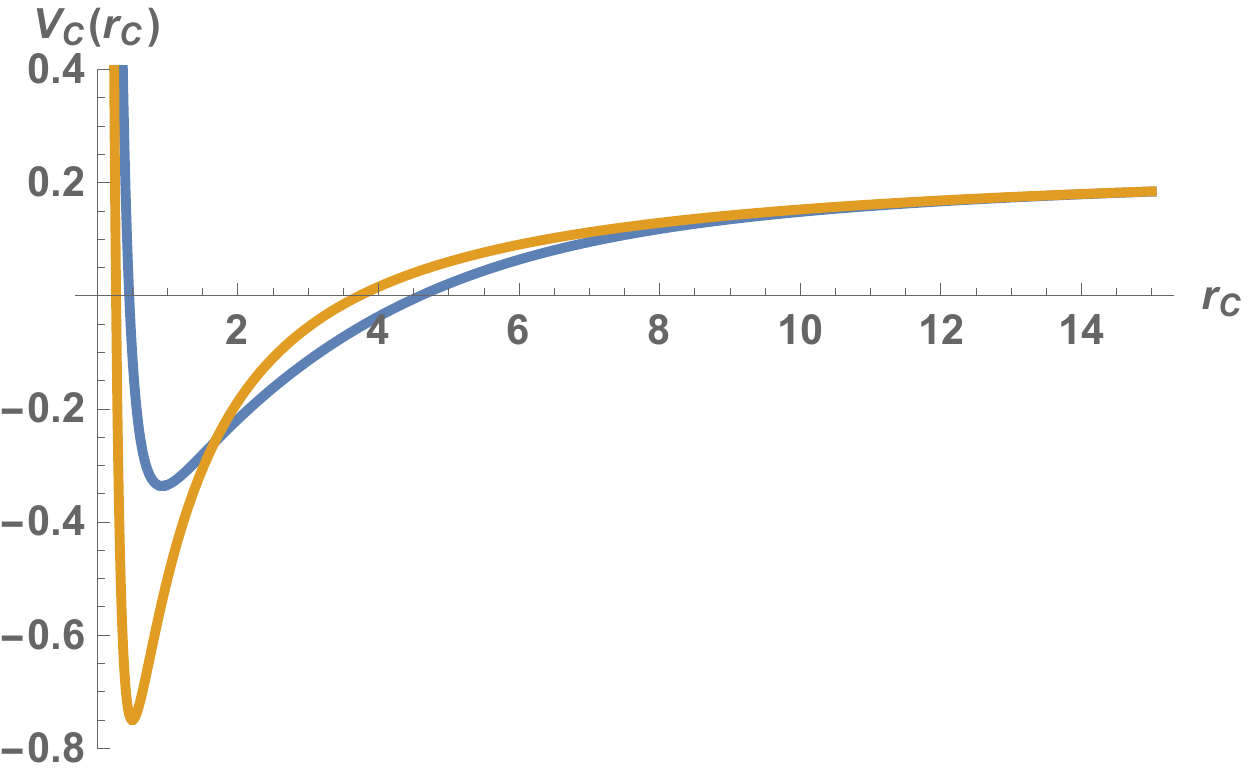}
  \caption{Ordinary two dimensional Coulomb potential orange, Riemann Coulomb potential blue. The Riemann potential in Coulomb coordinates $r_C$ is shallower and broader than the ordinary two dimensional Coulomb potential}
  \label{fig:Radion Potential}
\end{figure}
We plot this potential in figure 4.

\section{Other representations of the Riemann potential}

\subsection*{Riemann Potential II}

We can build another representation for Riemann potential based on the intergal representation:
\begin{equation}{2^{1 - A - ip}}\left( {A + ip} \right)\Gamma (A + ip)\eta (A + ip) = \int_{ - \infty }^\infty  {\frac{{{e^{ - x(A + 1)}}{e^{ - ipx}}}}{{{{\cosh }^2}({e^{ - x}})}}} dx\end{equation}
So that we can define the prepotential:
\begin{equation}{V_0}(x) = (A + 1)x + 2\log (\cosh ({e^{ - x}}))\end{equation}
and ground state wave function in position space:
\begin{equation}{\psi _0}(x) = \frac{1}{\sqrt{N_0}}\frac{{{e^{ - x(A + 1)}}}}{{{{\cosh }^2}({e^{ - x}})}}\end{equation}
The ground state wave function in momentum space is:
\begin{equation}{{\tilde \psi }_0}(p) =  \frac{1}{{\sqrt {2\pi } }}\frac{1}{{\sqrt {{N_0}} }}{2^{1 - A - ip}}\left( {A + ip} \right)\Gamma (A + ip)\eta (A + ip)\end{equation}
with ${N_0} = \frac{1}{{18}}\left( { - 6 + {\pi ^2}} \right)$. From $V_0(x)$ we obtain the superpotential
\begin{equation}W(x) = (A + 1) - 2\tanh ({e^{ - x}})\end{equation}
\begin{figure}%
\centering
\subfloat[]{%
\label{fig:first}%
\includegraphics[height=1.5in]{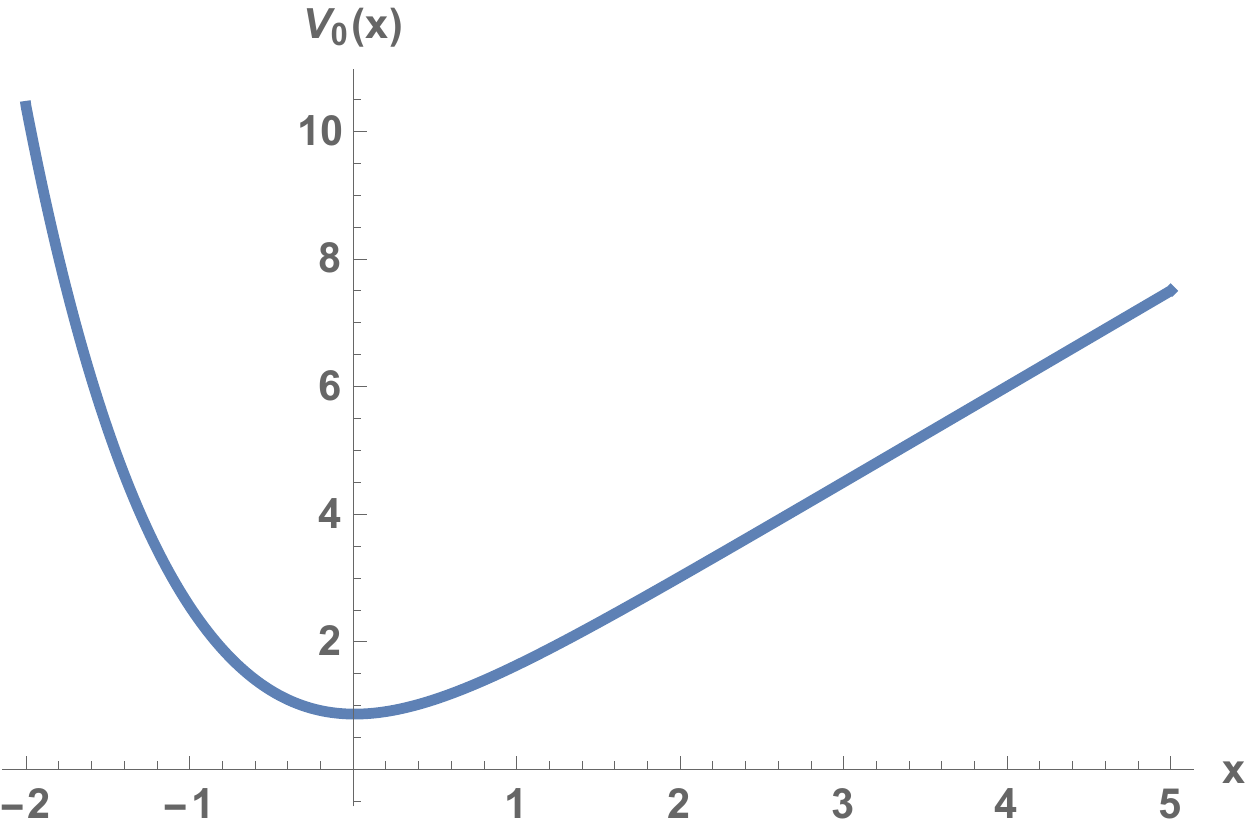}}%
\qquad
\subfloat[]{%
\label{fig:second}%
\includegraphics[height=1.5in]{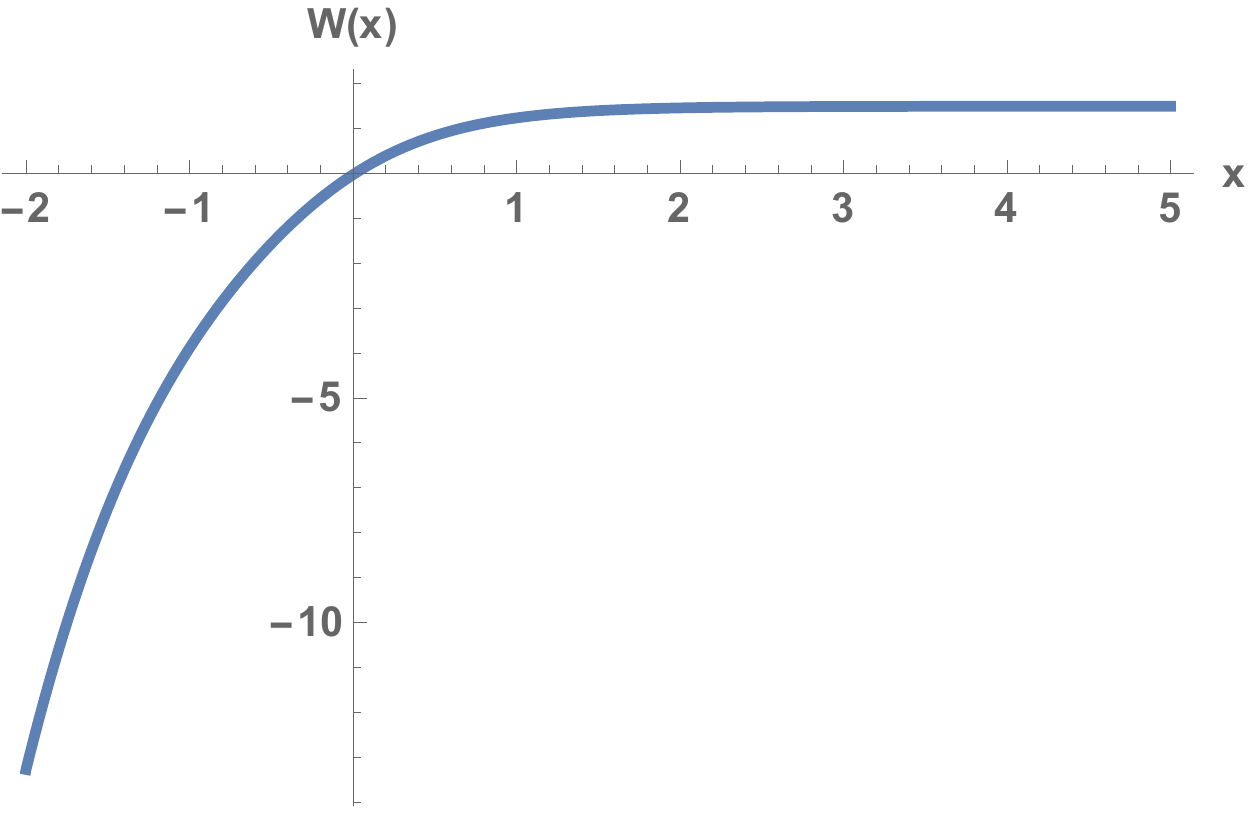}}%
\caption{(a) Prepotential for Riemann Potential II for $A=1/2$. (b) Superpotential for Riemann Pontential II for $A=1/2$.}
\end{figure}
These are plotted in figure 5.
\begin{figure}%
\centering
\subfloat[]{%
\label{fig:first}%
\includegraphics[height=1.5in]{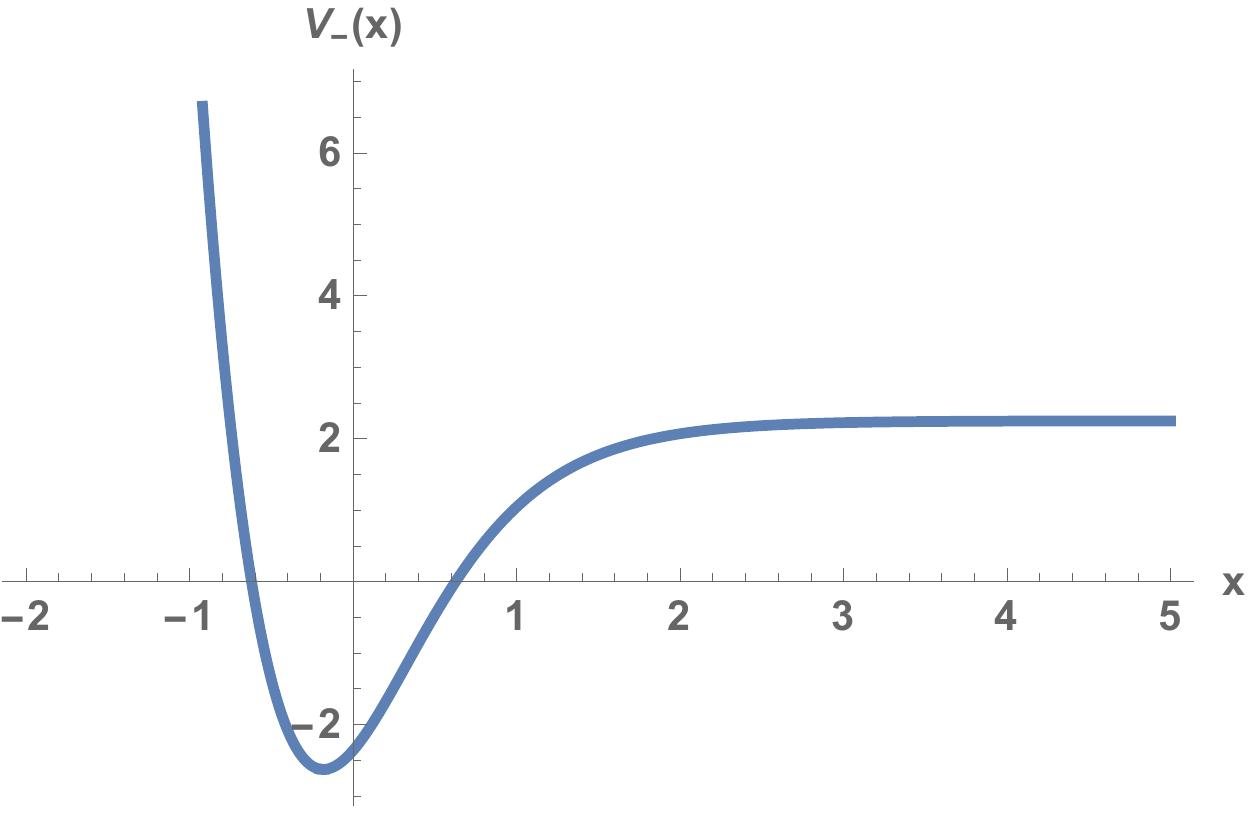}}%
\qquad
\subfloat[]{%
\label{fig:second}%
\includegraphics[height=1.5in]{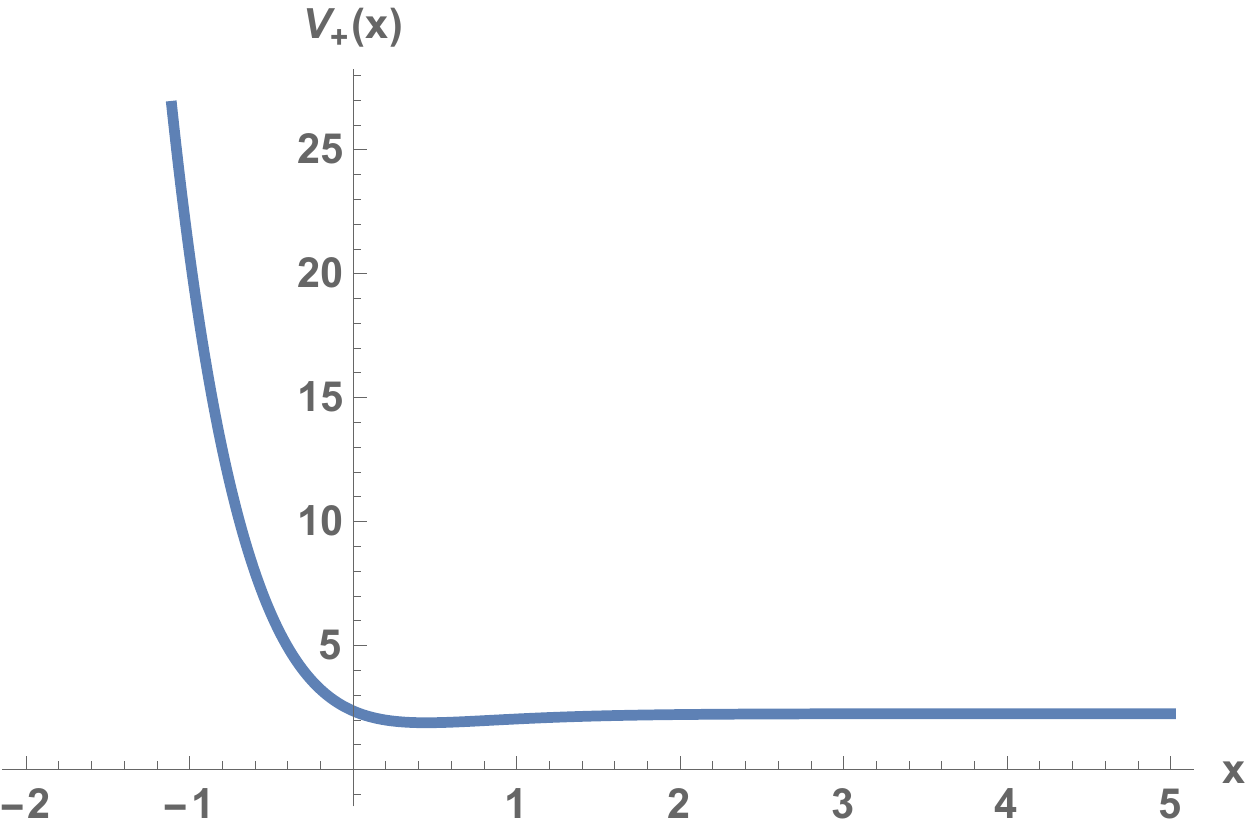}}%
\caption{(a) Minus partner potential for Riemann  Potential II for $A=1/2$. (b) Plus partner potential for Riemann Potential II for $A=1/2$.}
\end{figure}
and finally the minus superpotential that we plot in figure 6.
\begin{equation}{V_ - }(x) =  - 2{e^{ - 2x}}{\operatorname{sech} ^2}({e^{ - x}}) - 2{e^{ - x}}\tanh ({e^{ - x}}) + {\left( {A + 1 - 2\tanh ({e^{ - x}})} \right)^2}\end{equation}
Note the Riemann minus potential II is deeper than Riemann  potential I  but can still only hold one bound state.

\subsection*{Riemann Xi function Potential I}
This potential follows from the integral representation of the Riemann Xi function as:
\begin{equation}\xi (A + ip) = \int_{ - \infty }^\infty  {dx\Phi ({e^{ - \pi {e^{ - 2x}}}}} ){e^{ - (A - 1/2)x}}{e^{ - ipx}}\end{equation}
From this representation one can obtain the Riemann Xi Function prepotential I as :
\begin{equation}
V_0(A,x) = (A-\frac{1}{2})x -\log(\Phi(e^{-\pi e^{-2x}}))
\end{equation}
Then the superpotential can be obtained from:
\begin{equation}W(x) = {V_0}'(x)\end{equation}
and the two partner potentials are:
$${V_ - }(x) = {W^2}(x) - W'(x)$$
\begin{equation}{V_ + }(x) = {W^2}(x) + W'(x)\end{equation}
These are plotted in figure 7 and 8.
\begin{figure}%
\centering
\subfloat[]{%
\label{fig:first}%
\includegraphics[height=1.5in]{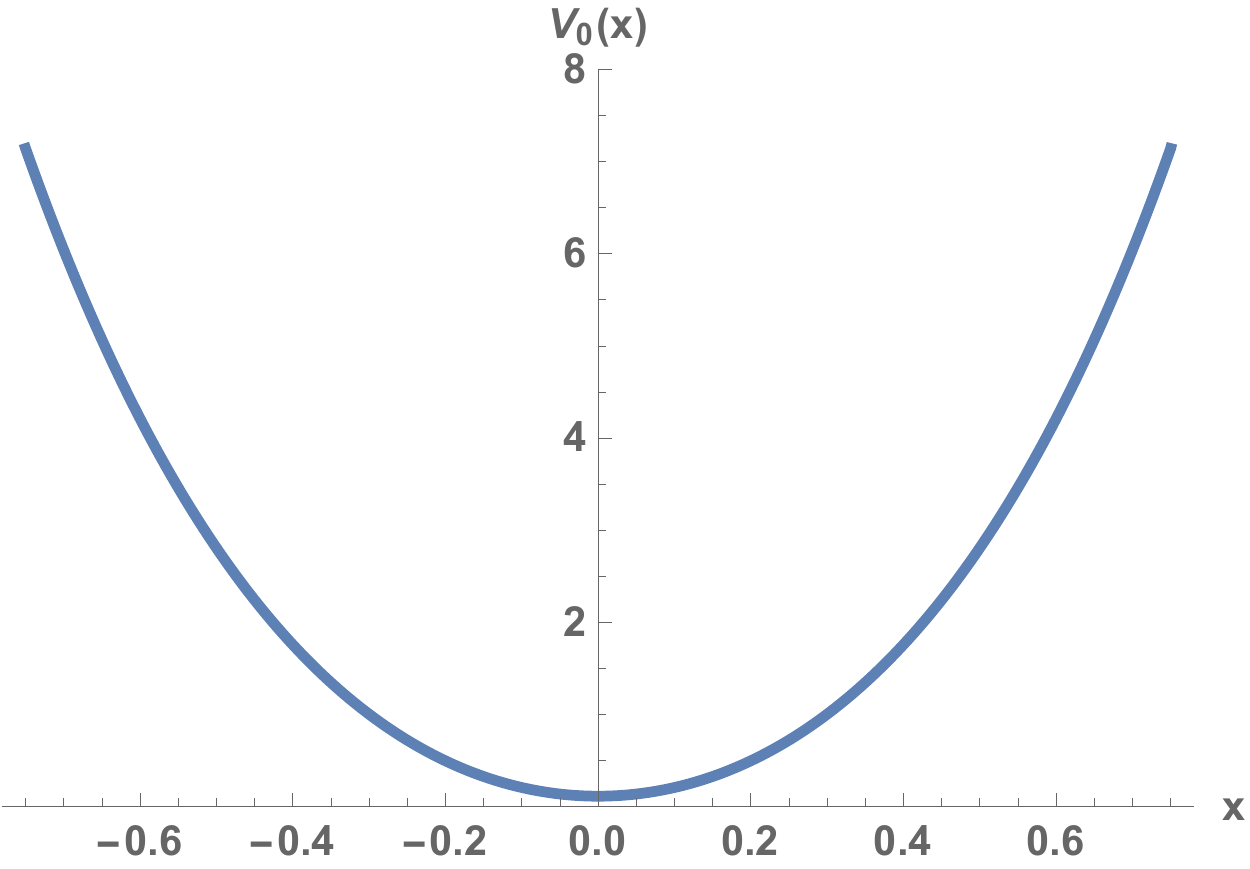}}%
\qquad
\subfloat[]{%
\label{fig:second}%
\includegraphics[height=1.5in]{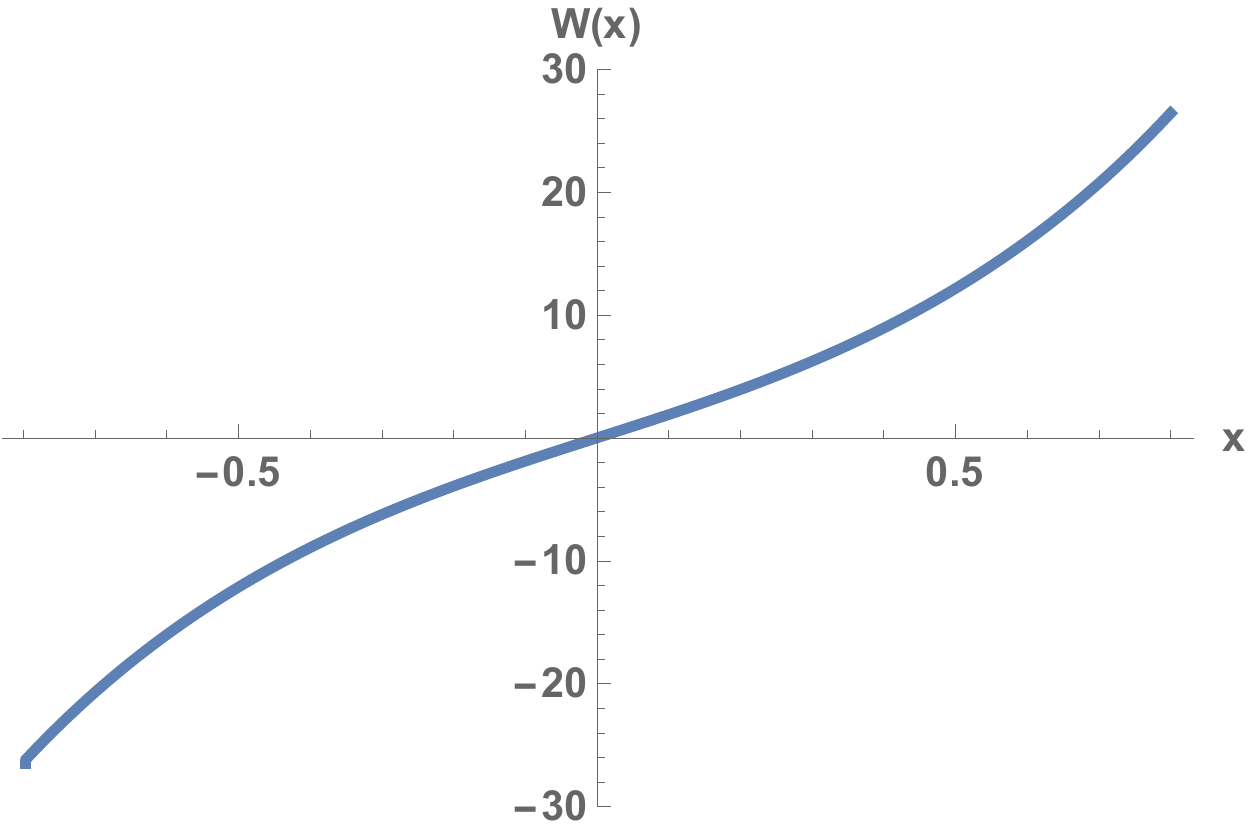}}%
\caption{(a) Prepotential for Riemann Xi Potential I for $A=1/2$. (b) Superpotential for Riemann Xi Pontential I for $A=1/2$.}
\end{figure}
\begin{figure}%
\centering
\subfloat[]{%
\label{fig:first}%
\includegraphics[height=1.5in]{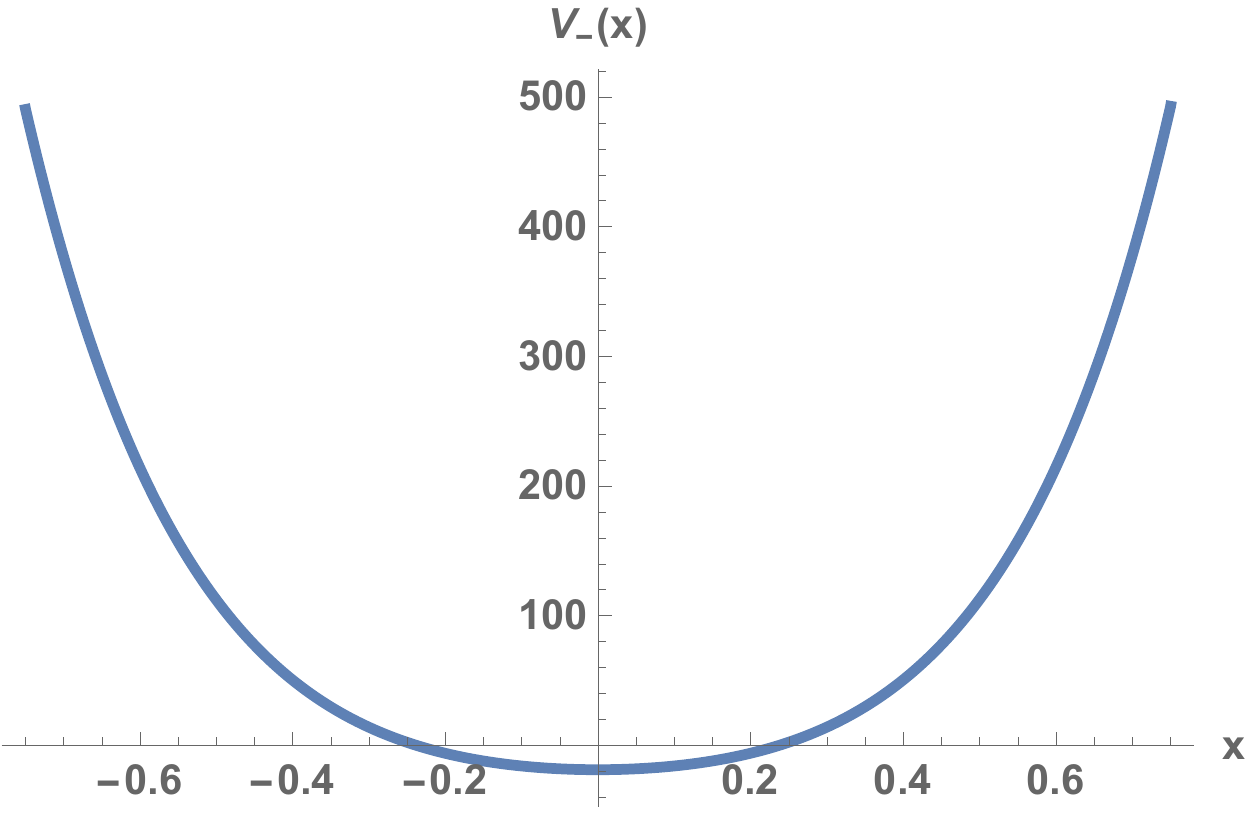}}%
\qquad
\subfloat[]{%
\label{fig:second}%
\includegraphics[height=1.5in]{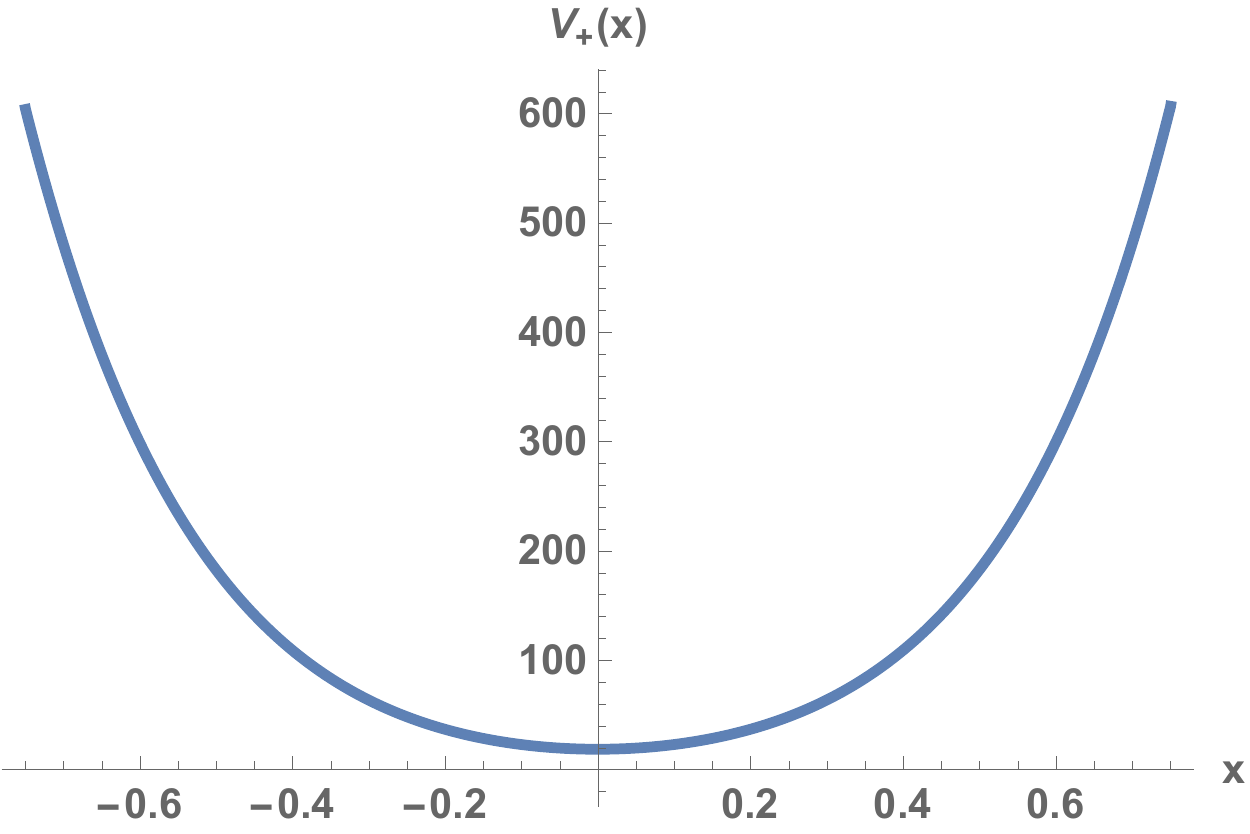}}%
\caption{(a) Minus partner potential for Riemann Xi Potential I for $A=1/2$. (b) Plus partner potential for Riemann Xi Potential I for $A=1/2$.}
\end{figure}

\subsection{Two matrix integrals and Riemann Xi function Potential}
It is interesting that like the Gaussian potential the Riemann Xi prepotential can be used to define a two matrix model defined by:
\begin{equation}Z = \int {d{M_1}d{M_2}{e^{ - tr({V_0}({M_1})) - tr({M_1}{M_2})}}} \end{equation}
This type of model ca be solved by using biorthogonal polynomials which satisfy:
\begin{equation}\int {dadb{e^{ - {V_0}(a) - ab}}} {Q_m}(a){R_n}(b) = {h_m}{\delta _{m,n}}\end{equation}
We can relate the Riemann Xi function potential to a two matrix model in a similar manner to the Gaussian function discussed above.
For the choice  $ {Q_n}(a) = {a^n}$ we find the first ten polynomials  $ {R_n}(b)$ to be
\begin{align}
&R_0(t)=1\nonumber\\
&R_1(t)=t\nonumber\\
&R_2(t)=-2 + t^2\nonumber\\
&R_3(t)=-6 t + t^3\nonumber\\
&R_4(t)=10.3688 - 12 t^2 + t^4\nonumber\\
&R_5(t)=51.844 t - 20 t^3 + t^5\nonumber\\
&R_6(t)=-69.229 + 155.532 t^2 - 30 t^4 + t^6\nonumber\\
&R_7(t)=-484.603 t + 362.908 t^3 - 42 t^5 + t^7\nonumber\\
&R_8(t)=280.027 - 1938.41 t^2 + 725.815 t^4 - 56 t^6 + t^8\nonumber\\
&R_9(t)=2520.24t-5815.24 t^3+1306.47 t^5-72 t^7+t^9\nonumber\\
 \end{align}
Note in computing these polynomials we have scaled the Matrix in $V_0(M_1)$ by $1/\sqrt{9.36345}$ so the coefficient of the quadratic term in the series expansion is normalized to one.

\subsection*{Riemann Xi function Potential II}
This form of the Riemann Xi function potential is somewhat simpler beause we don't have to take derivatives of a theta function to construct it. Defining:
\begin{equation}\Phi_{II}(e^{-\pi e^{-2x}})= e^{-x/2}( {{\theta _4}(0|{e^{ - \pi {e^{ - 2x}}}}) + {\theta _2}(0|{e^{ - \pi {e^{ - 2x}}}}) - {\theta _3}(0|{e^{ - \pi {e^{ - 2x}}}}))}
\end{equation}
we have the integral representation:
\begin{equation}\left( {{2^{1 - s}} + {2^s} - 3} \right)\left( {\frac{2}{{\left( { - 1 + s} \right)s}}} \right)\xi (s) = \int_{ - \infty }^\infty  {\left( {{\theta _4}(0|{e^{ - \pi {e^{ - 2x}}}}) + {\theta _2}(0|{e^{ - \pi {e^{ - 2x}}}}) - {\theta _3}(0|{e^{ - \pi {e^{ - 2x}}}})} \right){e^{ - x s}}} dx\end{equation}
or:
\begin{equation}\left( {{2^{1 - s}} + {2^s} - 3} \right)\left( {\frac{2}{{\left( { - 1 + s} \right)s}}} \right)\xi (s) = \int_{ - \infty }^\infty   \Phi_{II}(e^{-\pi e^{-2x}}){e^{ - x (s-1/2)}} dx\end{equation}
From this representation one can obtain the Riemann Xi Function prepotential II as :
\begin{equation}
V_0(A,x) = (A-\frac{1}{2})x -\log(\Phi_{II}(e^{-\pi e^{-2x}}))
\end{equation}
Then the superpotential can be obtained from:
\begin{equation}W(x) = {V_0}'(x)\end{equation}
and the two partner potentials are:
$${V_ - }(x) = {W^2}(x) - W'(x)$$
\begin{equation}{V_ + }(x) = {W^2}(x) + W'(x)\end{equation}
These are plotted in figure 9 and 10.
\begin{figure}%
\centering
\subfloat[]{%
\label{fig:first}%
\includegraphics[height=1.5in]{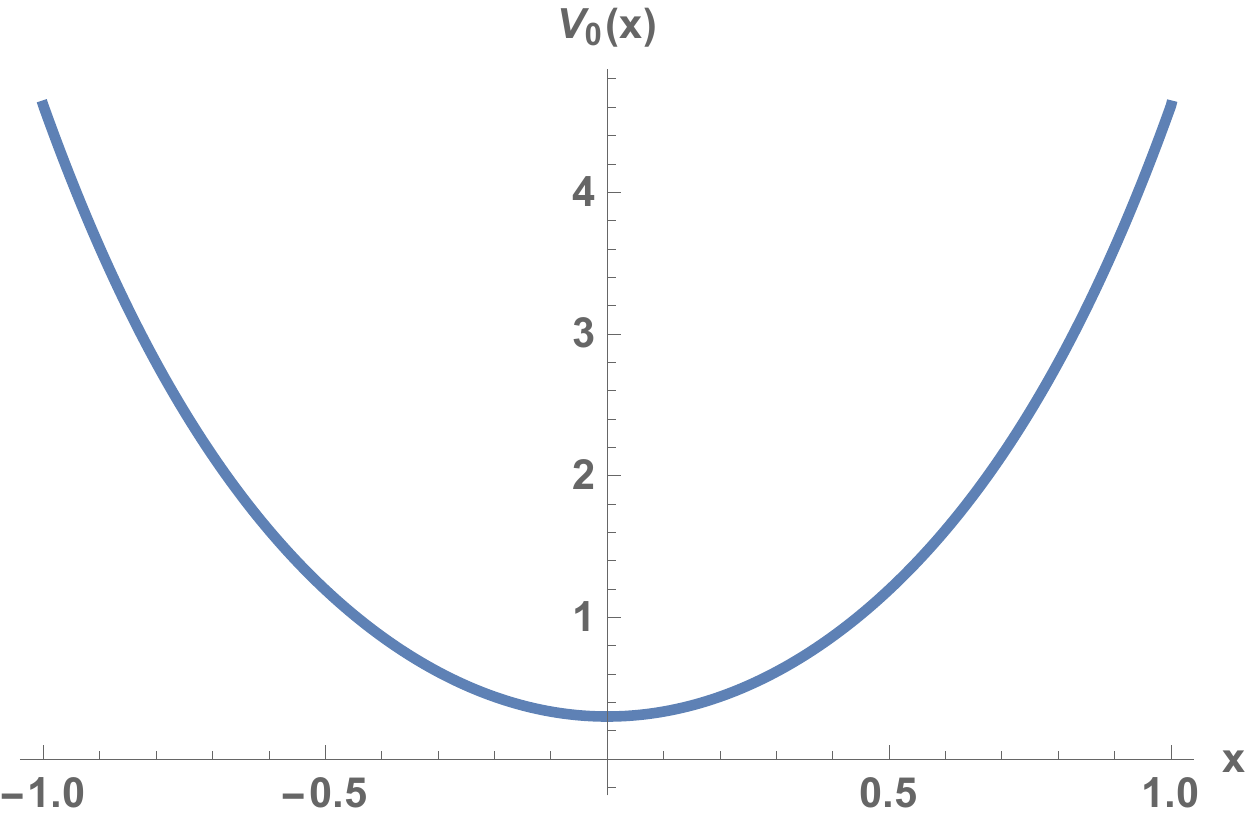}}%
\qquad
\subfloat[]{%
\label{fig:second}%
\includegraphics[height=1.5in]{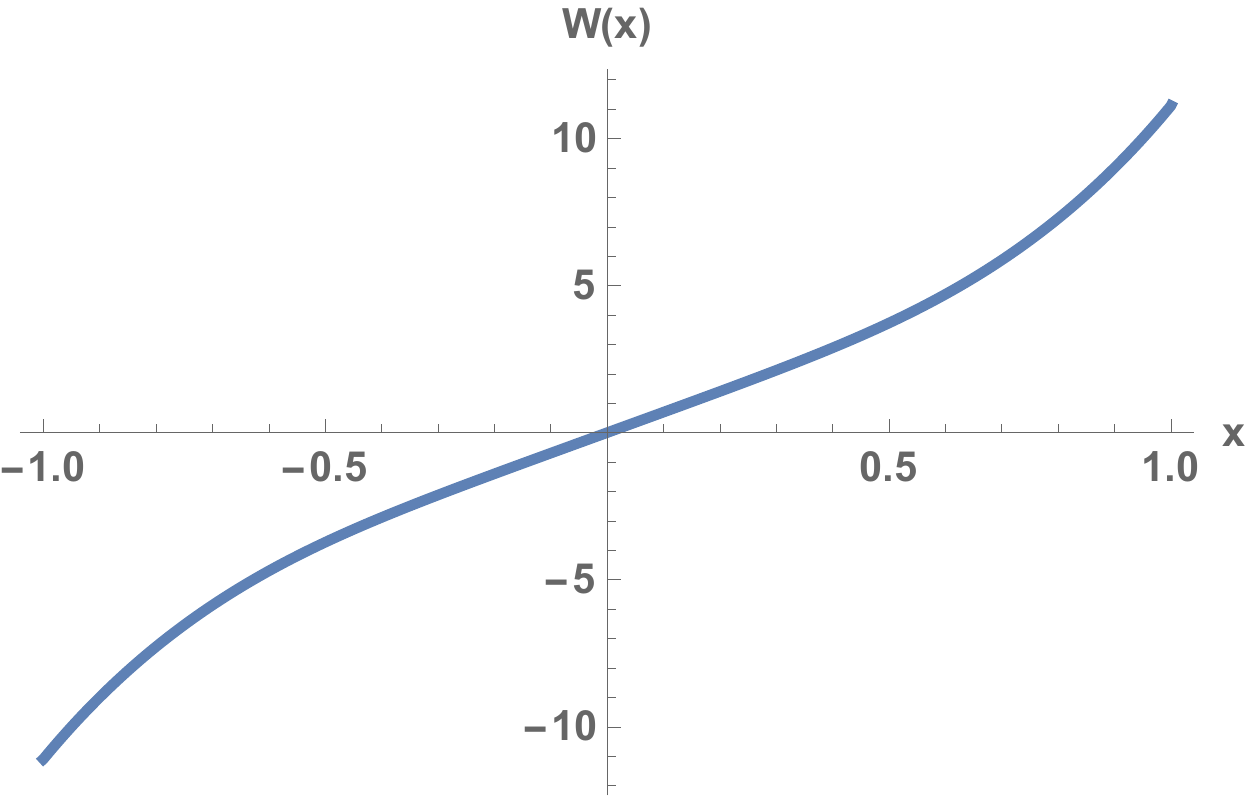}}%
\caption{(a) Prepotential for Riemann Xi Potential II for $A=1/2$. (b) Superpotential for Riemann Xi Pontential II for $A=1/2$.}
\end{figure}
\begin{figure}%
\centering
\subfloat[]{%
\label{fig:first}%
\includegraphics[height=1.5in]{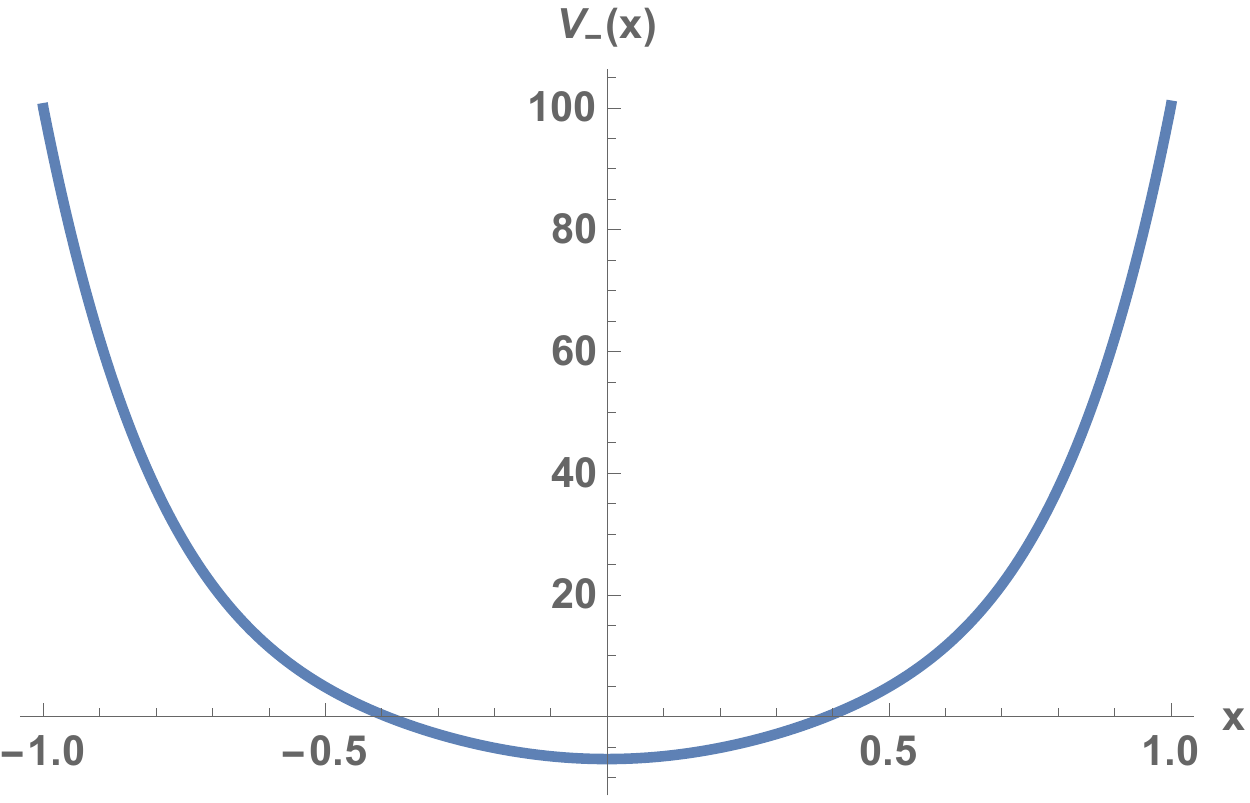}}%
\qquad
\subfloat[]{%
\label{fig:second}%
\includegraphics[height=1.5in]{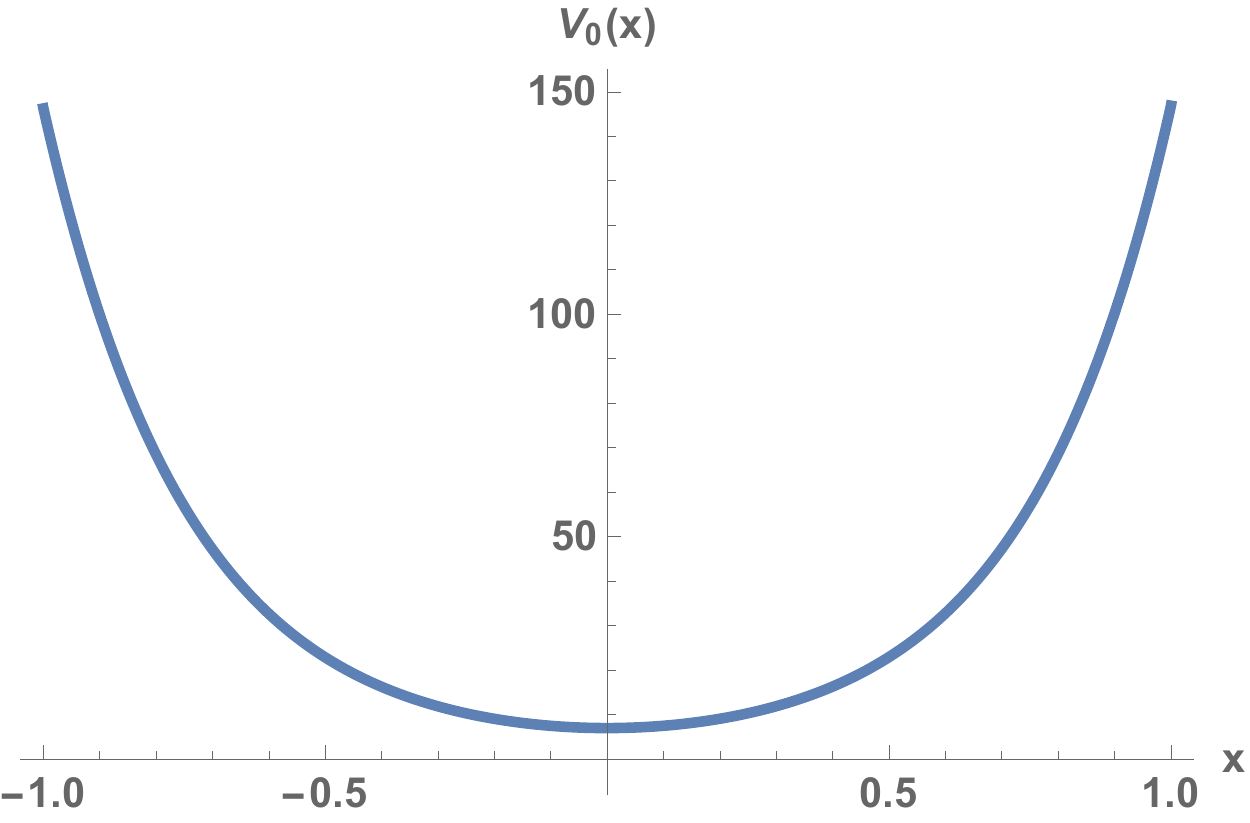}}%
\caption{(a) Minus partner potential for Riemann Xi Potential II for $A=1/2$. (b) Plus partner potential for Riemann Xi Potential II for $A=1/2$.}
\end{figure}


\newpage

\section{ Uncertainty relation for the different representations }

Having obtained various representations of the ground state for the Riemann potential and Riemann Xi potential is both the position and momentum basis one can proceed to make various calculations using those states. In this section we will compute the uncertainty relations obtained for the ground states of these various potentials.

\subsection{Simple Harmonic Oscillator}

For the simple Harmonic Oscillator the ground state in the position basis is:
\begin{equation}{\psi _0}(x) = {\left( {\frac{\omega }{{2\pi }}} \right)^{1/4}}{e^{ - \omega {x^2}/4}}\end{equation}
and in the momentum basis is:
\begin{equation}{{\tilde \psi }_0}(p) = {\left( {\frac{2}{{\pi \omega }}} \right)^{1/4}}{e^{ - {p^2}/\omega }}\end{equation}
The prepotential and superpotential are:
$${V_0}(x) = \frac{1}{4}\omega {x^2}$$
\begin{equation}W(x) = \frac{1}{2}\omega x\end{equation}
and the partner potentials are:
$${V_ - }(x) = \frac{1}{4}{\omega ^2}{x^2} - \frac{1}{2}\omega $$
\begin{equation}{V_ + }(x) = \frac{1}{4}{\omega ^2}{x^2} + \frac{1}{2}\omega \end{equation}
these are plotted in figure 8.

Then we have using the position basis:
$$\left\langle x \right\rangle  = 0$$
$$\left\langle {{x^2}} \right\rangle  = \frac{1}{\omega}$$
\begin{equation}\Delta x = \sqrt {\left\langle {{{\left( {x - \left\langle x \right\rangle } \right)}^2}} \right\rangle }  = \sqrt {\left\langle {{x^2}} \right\rangle  - {{\left\langle x \right\rangle }^2}}  = \frac{1}{\sqrt{\omega}}\end{equation}
and using the momentum basis:
$$\left\langle p \right\rangle  = 0$$
$$\left\langle {{p^2}} \right\rangle  = \frac{\omega}{4}$$
\begin{equation}\Delta p = \sqrt {\left\langle {{{\left( {p - \left\langle p \right\rangle } \right)}^2}} \right\rangle }  = \sqrt {\left\langle {{p^2}} \right\rangle  - {{\left\langle p \right\rangle }^2}}  = \frac{\sqrt{\omega}}{2}\end{equation}
Then we have:
\begin{equation}\Delta p\Delta x = \frac{\sqrt{\omega}}{2}\frac{1}{\sqrt{\omega}}=.5 \geq .5\end{equation}
Which is consistent with the uncertainty relation inequality.

\subsection{Morse Potential}

For the Morse potential the ground state in the position basis is :
\begin{equation}{\psi _0}(x) = \sqrt 2 {e^{ - x/2}}{e^{ - {e^{ - x}}}}\end{equation}
and in the momentum basis:
\begin{equation}{{\tilde \psi }_0}(p) = \frac{1}{{\sqrt \pi  }}\Gamma \left( {\frac{1}{2} + ip} \right)\end{equation}
Then we have using the position basis:
$$\left\langle x \right\rangle  = \gamma + \log(2) = 1.27036$$
$$\left\langle {{x^2}} \right\rangle  = \frac{\pi^2}{6} + (\gamma + \log(2))^2=3.25876$$
\begin{equation}\Delta x = \sqrt {\left\langle {{{\left( {x - \left\langle x \right\rangle } \right)}^2}} \right\rangle }  = \sqrt {\left\langle {{x^2}} \right\rangle  - {{\left\langle x \right\rangle }^2}}  = \frac{\pi}{\sqrt{6}}=1.28255\end{equation}
and using the momentum basis:
$$\left\langle p \right\rangle  = 0$$
$$\left\langle {{p^2}} \right\rangle  = \frac{1}{4}$$
\begin{equation}\Delta p = \sqrt {\left\langle {{{\left( {p - \left\langle p \right\rangle } \right)}^2}} \right\rangle }  = \sqrt {\left\langle {{p^2}} \right\rangle  - {{\left\langle p \right\rangle }^2}}  = \frac{1}{2}\end{equation}
These are plotted in figure 11. Then we have:
\begin{equation}\Delta p\Delta x = \frac{\pi}{2\sqrt{6}}=.641275 \geq .5\end{equation}
Which is consistent with the uncertainty relation inequality.

\begin{figure}%
\centering
\subfloat[]{%
\label{fig:first}%
\includegraphics[height=1.5in]{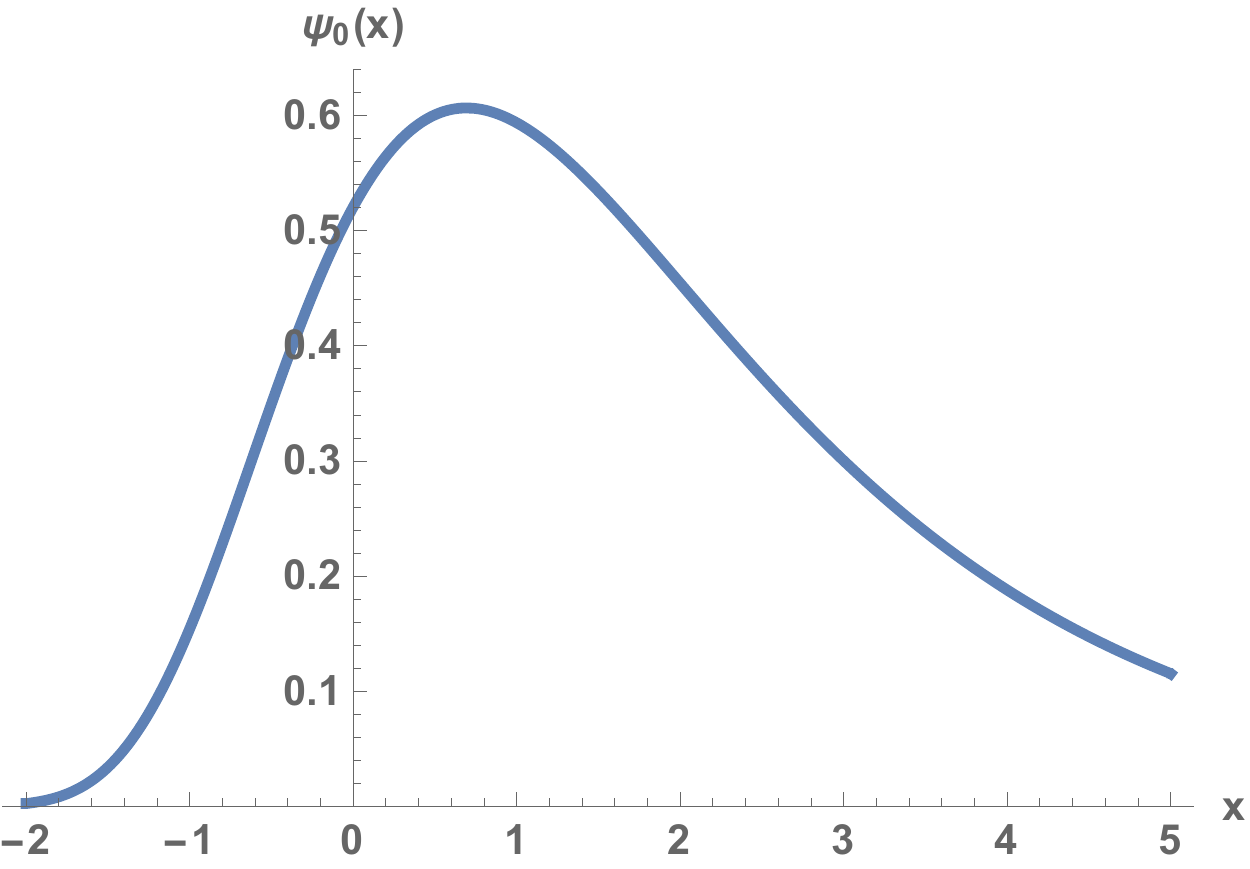}}%
\qquad
\subfloat[]{%
\label{fig:second}%
\includegraphics[height=1.5in]{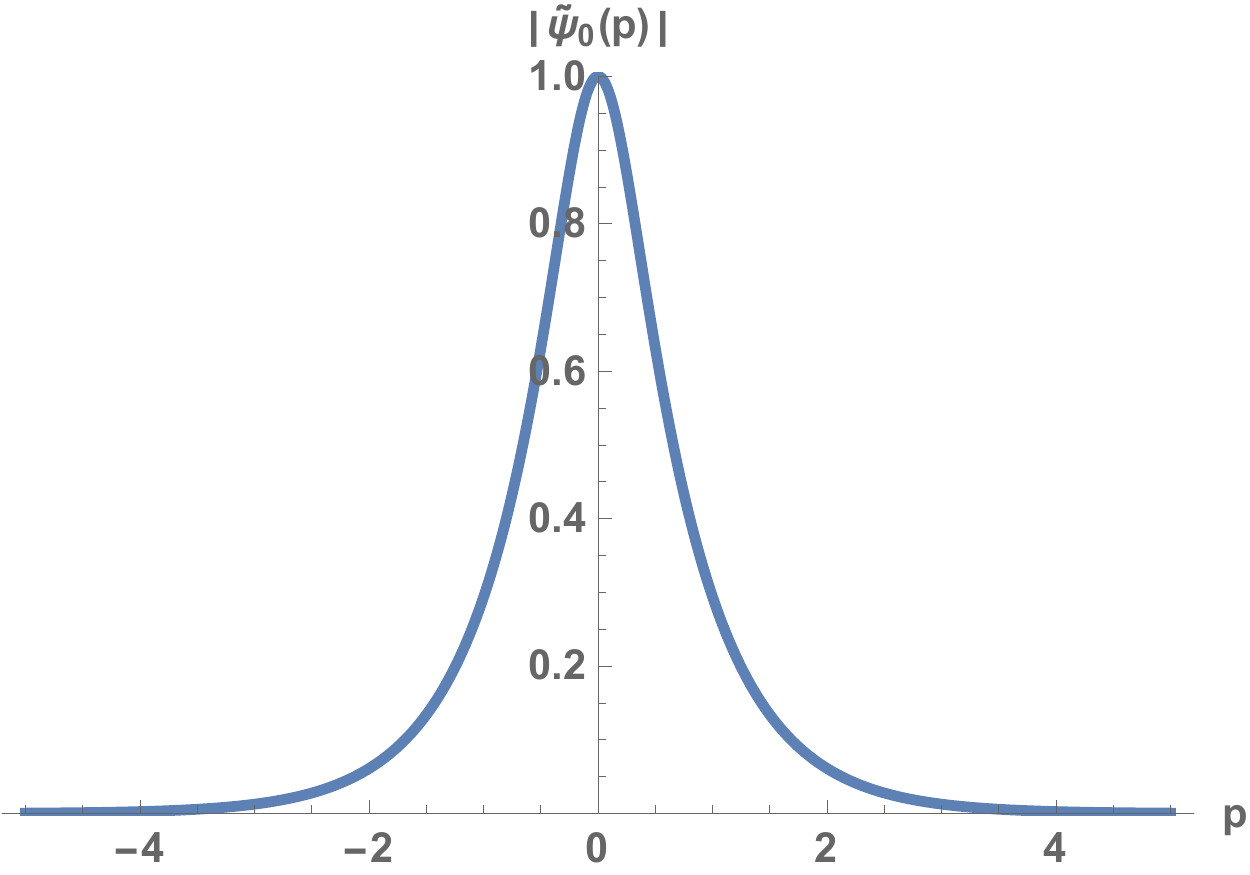}}%
\caption{(a) Magnitude of the ground state wave function in position space for the Morse Potential. (b) Magnitude of the ground State wave function in momentum space for the Morse Potential.}
\end{figure}

\subsection{Riemann Potential I}

For the first representation we have the normalized ground state in the position basis given by:
\begin{equation}{\psi _0}(x) = \frac{1}{{\sqrt { - \frac{1}{2} + \log (2)} }}{e^{ - x/2}}\frac{1}{{{e^{{e^{ - x}}}} + 1}}\end{equation}
The normalized ground state in the momentum basis is given by:
\begin{equation}{{\tilde \psi }_0}(p) = \frac{1}{{\sqrt {2\pi } }}\frac{1}{{\sqrt { - \frac{1}{2} + \log (2)} }}\Gamma \left( {\frac{1}{2} + ip} \right)\eta \left( {\frac{1}{2} + ip} \right)\end{equation}
The we have using the position basis:
$$\left\langle x \right\rangle  = .918522$$
$$\left\langle {{x^2}} \right\rangle  = 2.34964$$
\begin{equation}\Delta x = \sqrt {\left\langle {{{\left( {x - \left\langle x \right\rangle } \right)}^2}} \right\rangle }  = \sqrt {\left\langle {{x^2}} \right\rangle  - {{\left\langle x \right\rangle }^2}}  = 1.22717\end{equation}
and using the momentum basis:
$$\left\langle p \right\rangle  = 0$$
$$\left\langle {{p^2}} \right\rangle  = .306513$$
\begin{equation}\Delta p = \sqrt {\left\langle {{{\left( {p - \left\langle p \right\rangle } \right)}^2}} \right\rangle }  = \sqrt {\left\langle {{p^2}} \right\rangle  - {{\left\langle p \right\rangle }^2}}  = .553637\end{equation}
These are plotted in figure 12 and 13. Then we have:
\begin{equation}\Delta p\Delta x = .67408 \geq .5\end{equation}
Which is consistent with the uncertainty relation inequality.
\begin{figure}%
\centering
\subfloat[]{%
\label{fig:first}%
\includegraphics[height=1.5in]{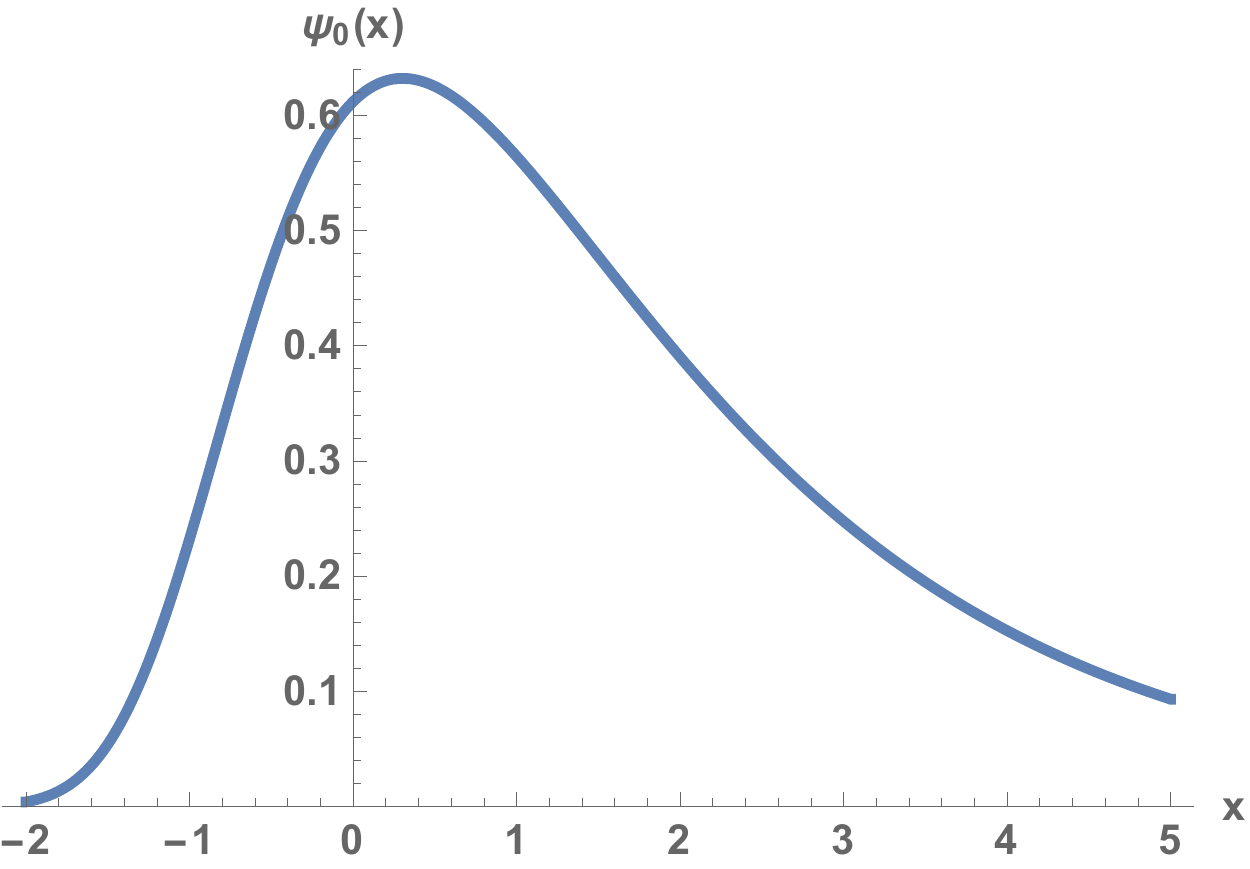}}%
\qquad
\subfloat[]{%
\label{fig:second}%
\includegraphics[height=1.5in]{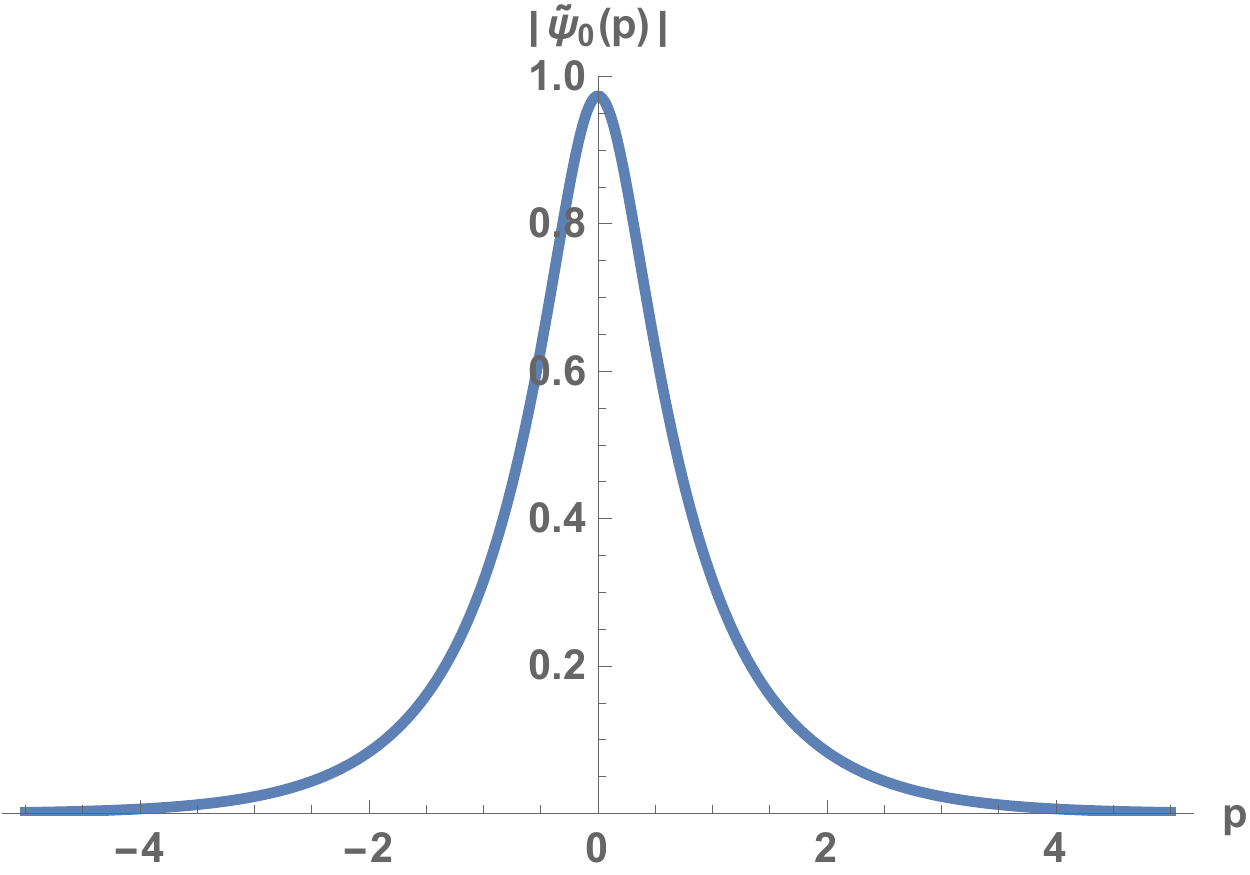}}%
\caption{(a) Magnitude of the ground state wave function in position space for the Riemann Potential I. (b) Magnitude of the ground State wave function in momentum space for the Riemann Potential I.}
\end{figure}
\begin{figure}%
\centering
\subfloat[]{%
\label{fig:first}%
\includegraphics[height=1.5in]{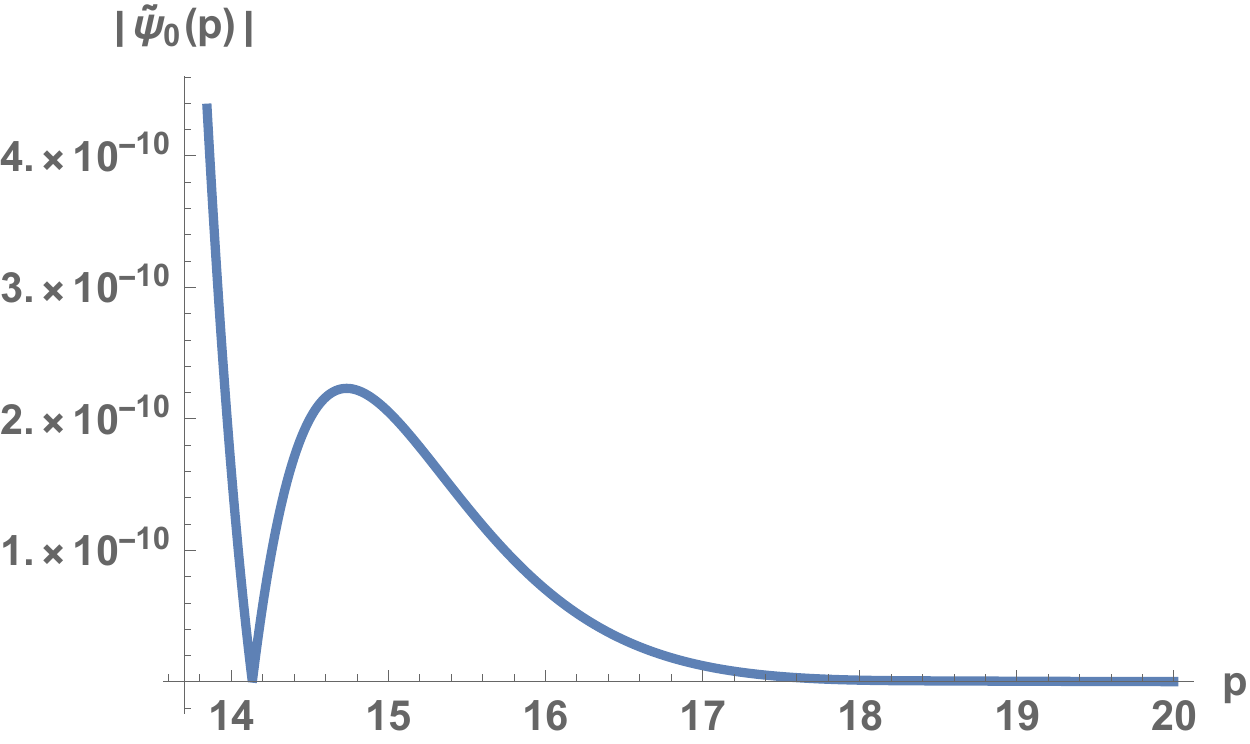}}%
\qquad
\subfloat[]{%
\label{fig:second}%
\includegraphics[height=1.5in]{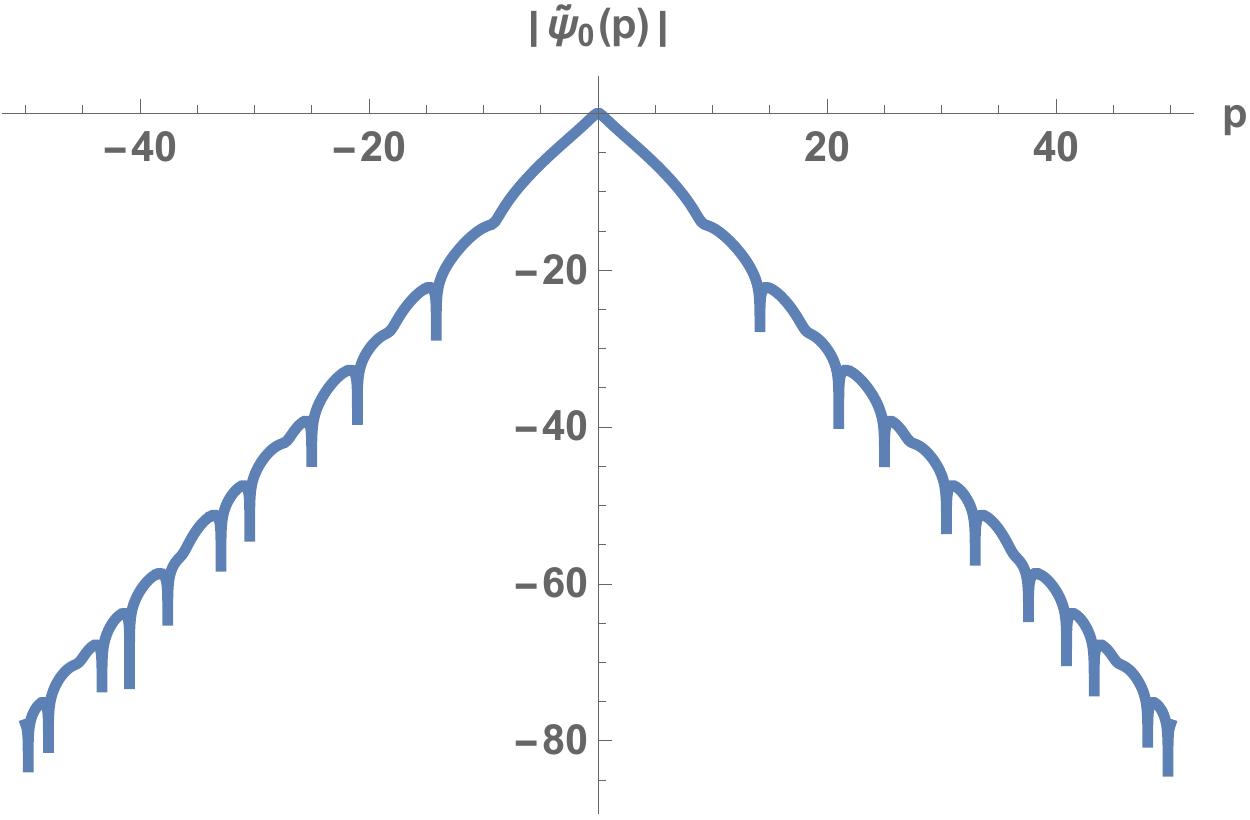}}%
\caption{(a) Closeup of a zero region of the magnitude of the ground state wave function in momentum space of Riemann Potential II. (b) Log of the magnitude of the ground state wave function in momentum space of Riemann Potential I.}
\end{figure}

\subsection{Riemann Potential II}

For the Rieman potential II the  ground state wave function in position space is:
\begin{equation}{\psi _0}(x) = \frac{1}{\sqrt{N_0}}\frac{{{e^{ - x(A + 1)}}}}{{{{\cosh }^2}({e^{ - x}})}}\end{equation}
The ground state wave function in momentum space is:
\begin{equation}{{\tilde \psi }_0}(p) =  \frac{1}{{\sqrt {2\pi } }}\frac{1}{{\sqrt {{N_0}} }}{2^{1 - A - ip}}\left( {A + ip} \right)\Gamma (A + ip)\eta (A + ip)\end{equation}
with ${N_0} = \frac{1}{{18}}\left( { - 6 + {\pi ^2}} \right)$ and $A=1/2$. These are plotted in figures 14 and 15.
\begin{figure}%
\centering
\subfloat[]{%
\label{fig:first}%
\includegraphics[height=1.5in]{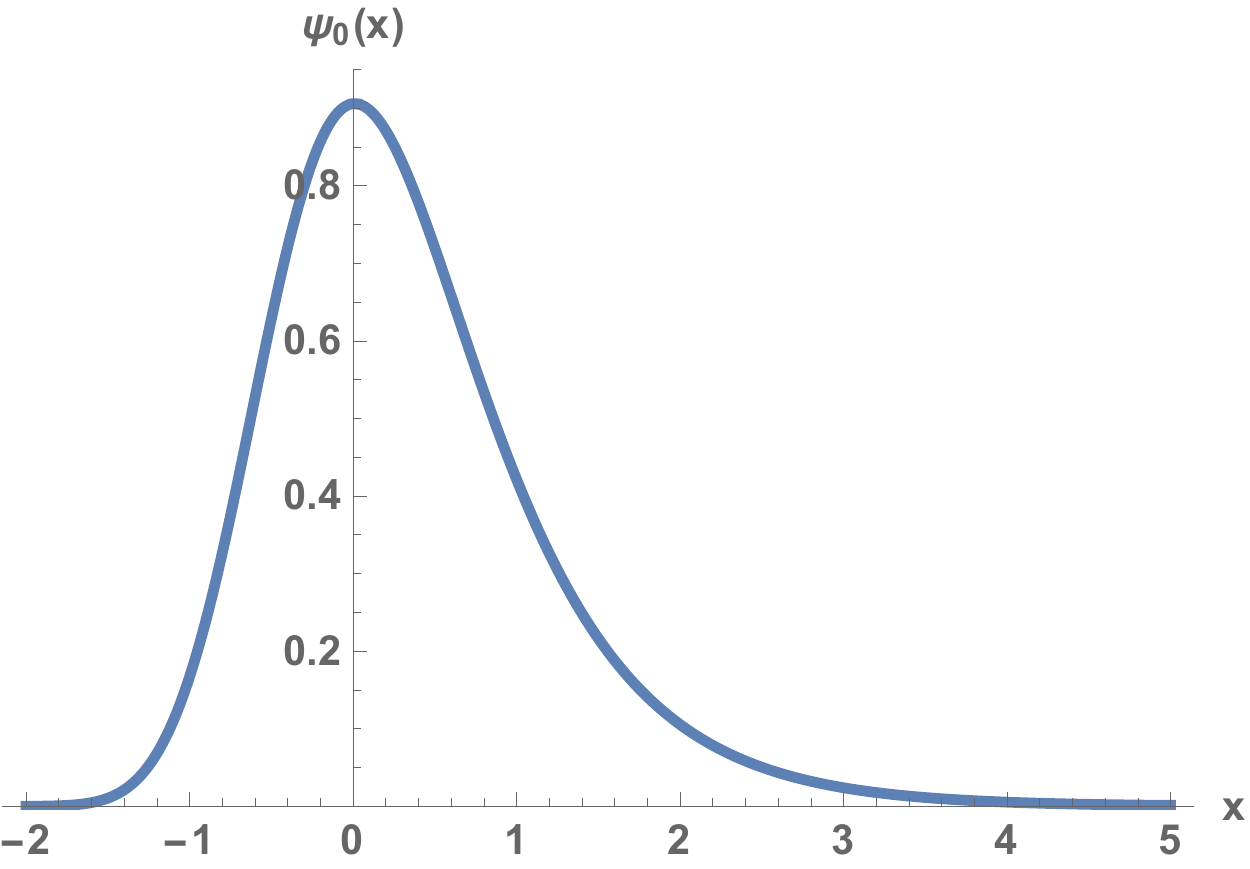}}%
\qquad
\subfloat[]{%
\label{fig:second}%
\includegraphics[height=1.5in]{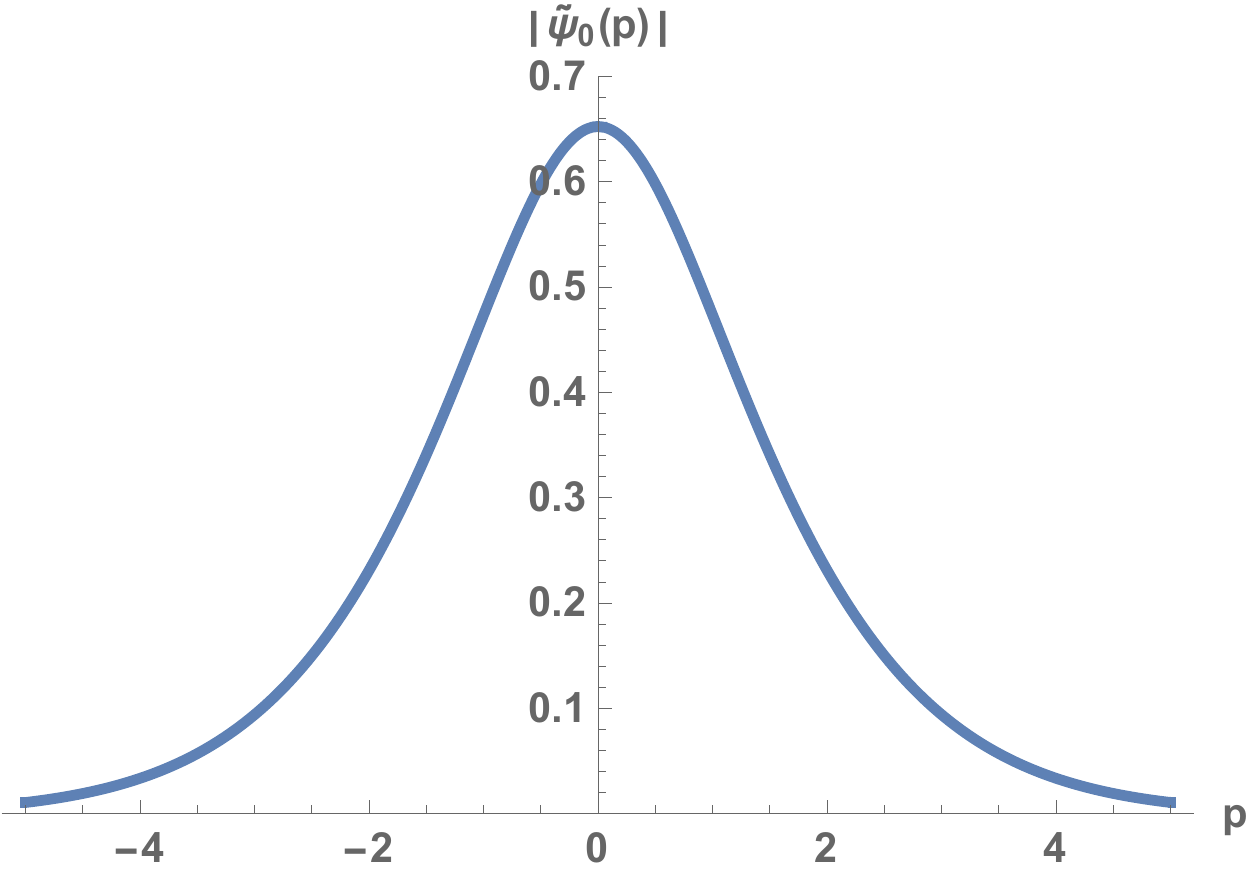}}%
\caption{(a) Magnitude of the ground state wave function in position space for the Riemann Potential II. (b) Magnitude of the ground State wave function in momentum space for the Riemann Potential II.}
\end{figure}
\begin{figure}%
\centering
\subfloat[]{%
\label{fig:first}%
\includegraphics[height=1.5in]{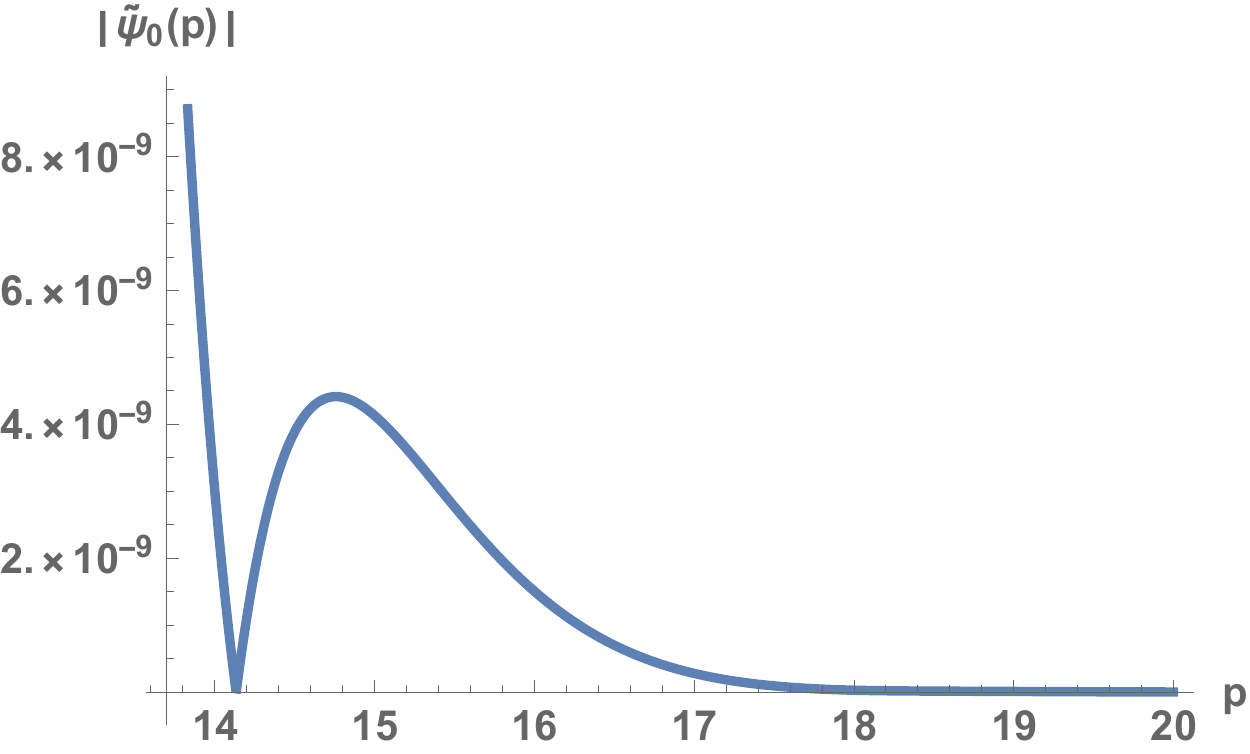}}%
\qquad
\subfloat[]{%
\label{fig:second}%
\includegraphics[height=1.5in]{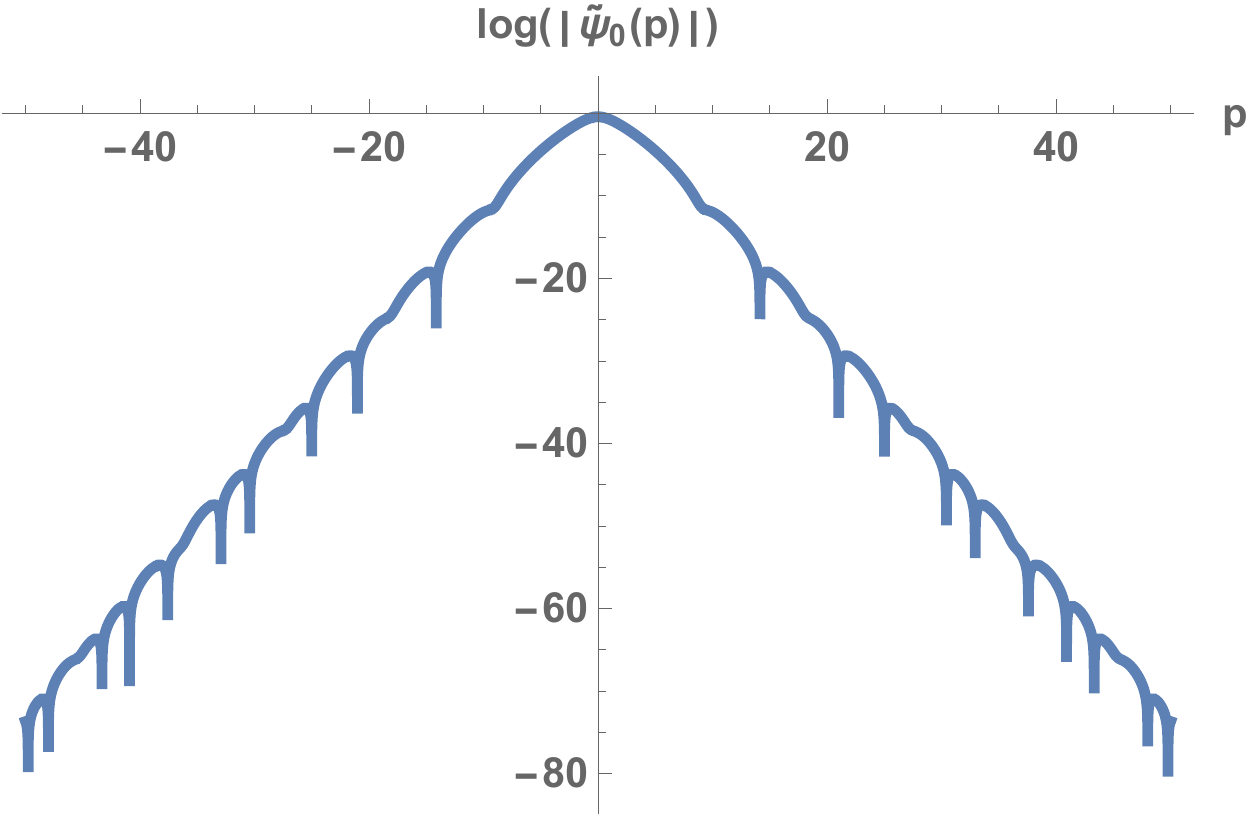}}%
\caption{(a) Closeup of a zero region of the magnitude of the ground state wave function in momentum space of Riemann Potential II. (b) Log of the magnitude of the ground state wave function in momentum space of Riemann Potential II.}
\end{figure}
Then we have:
$$\left\langle x \right\rangle  =.156371$$
$$\left\langle {{x^2}} \right\rangle  = .303422$$
\begin{equation} \Delta x = \sqrt {\left\langle {{{\left( {x - \left\langle x \right\rangle } \right)}^2}} \right\rangle }  = \sqrt {\left\langle {{x^2}} \right\rangle  - {{\left\langle x \right\rangle }^2}}  = .528176\end{equation}
and
$$\left\langle p \right\rangle  = 0$$
$$\left\langle {{p^2}} \right\rangle  = 1.0771$$
\begin{equation}\Delta p = \sqrt {\left\langle {{{\left( {p - \left\langle p \right\rangle } \right)}^2}} \right\rangle }  = \sqrt {\left\langle {{p^2}} \right\rangle  - {{\left\langle p \right\rangle }^2}}  = 1.03783\end{equation}
Then we have:
\begin{equation}\Delta p\Delta x = .548158 \geq .5\end{equation}
which is consistent with the uncertainty relation inequality.

\subsection{ Riemann Xi potential I}

For the first $xi$ representation we have:
\begin{equation}{\psi _0}(x) = \frac{1}{{\sqrt {{N_0}} }}\Phi ({e^{ - \pi {e^{ - 2 x}}}})\end{equation}
and 
\begin{equation}{{\tilde \psi }_0}(p) = \frac{1}{{\sqrt {2\pi } }}\frac{1}{{\sqrt {{N_0}} }} \xi \left( {\frac{1}{2} + ip} \right)\end{equation}
where $ {N_0} = .319752$. These are plotted in figures 16 and 17. Then we have:
$$\left\langle x \right\rangle  = 0$$
$$\left\langle {{x^2}} \right\rangle  = .0245801$$
\begin{equation}\Delta x = \sqrt {\left\langle {{{\left( {x - \left\langle x \right\rangle } \right)}^2}} \right\rangle }  = \sqrt {\left\langle {{x^2}} \right\rangle  - {{\left\langle x \right\rangle }^2}}  = .15678\end{equation}
and
$$\left\langle p \right\rangle  = 0$$
$$\left\langle {{p^2}} \right\rangle  = 10.2076$$
\begin{equation}\Delta p = \sqrt {\left\langle {{{\left( {p - \left\langle p \right\rangle } \right)}^2}} \right\rangle }  = \sqrt {\left\langle {{p^2}} \right\rangle  - {{\left\langle p \right\rangle }^2}}  = 3.19493\end{equation}
Then we have:
\begin{equation}\Delta p\Delta x = .500902 \geq .5 \end{equation}
which is consistent with the uncertainty relation inequality.
\begin{figure}%
\centering
\subfloat[a]{%
\label{fig:first}%
\includegraphics[height=1.5in]{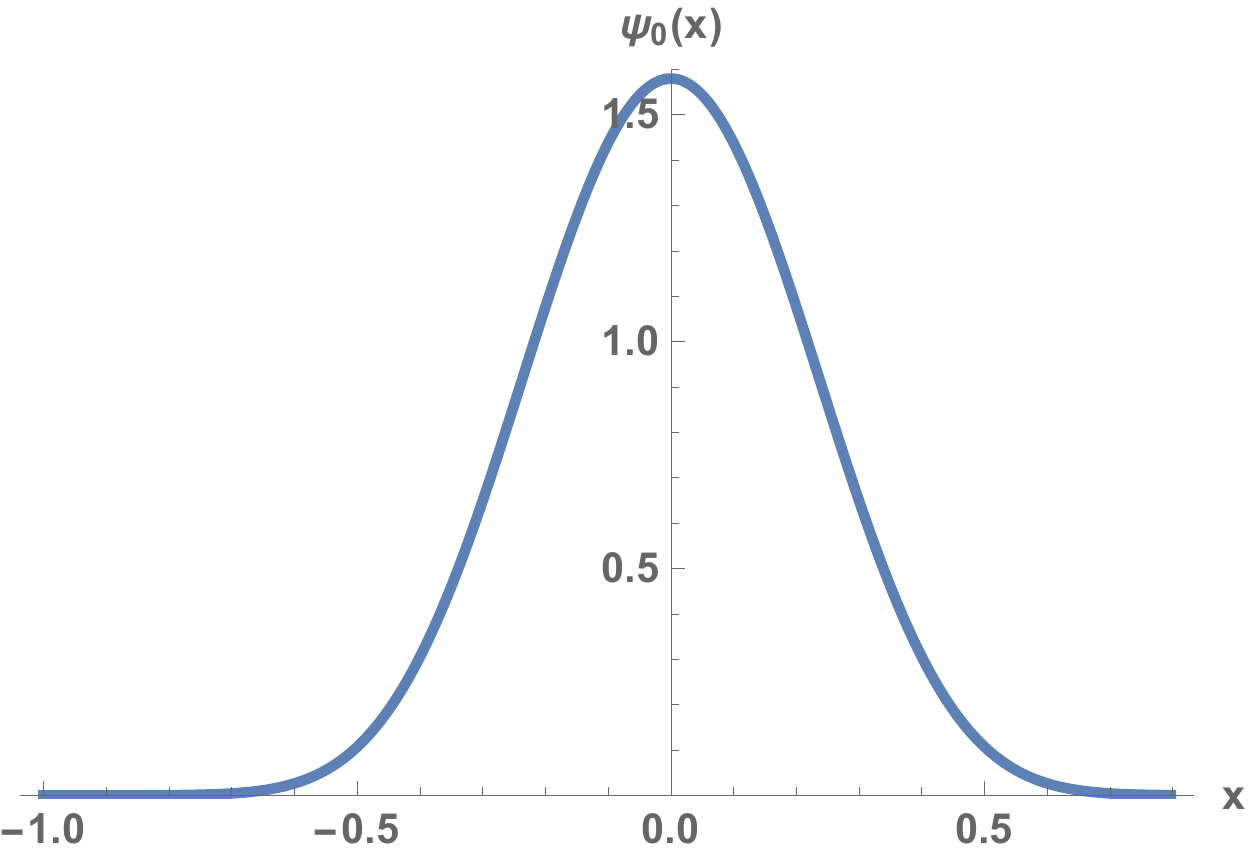}}%
\qquad
\subfloat[b]{%
\label{fig:second}%
\includegraphics[height=1.5in]{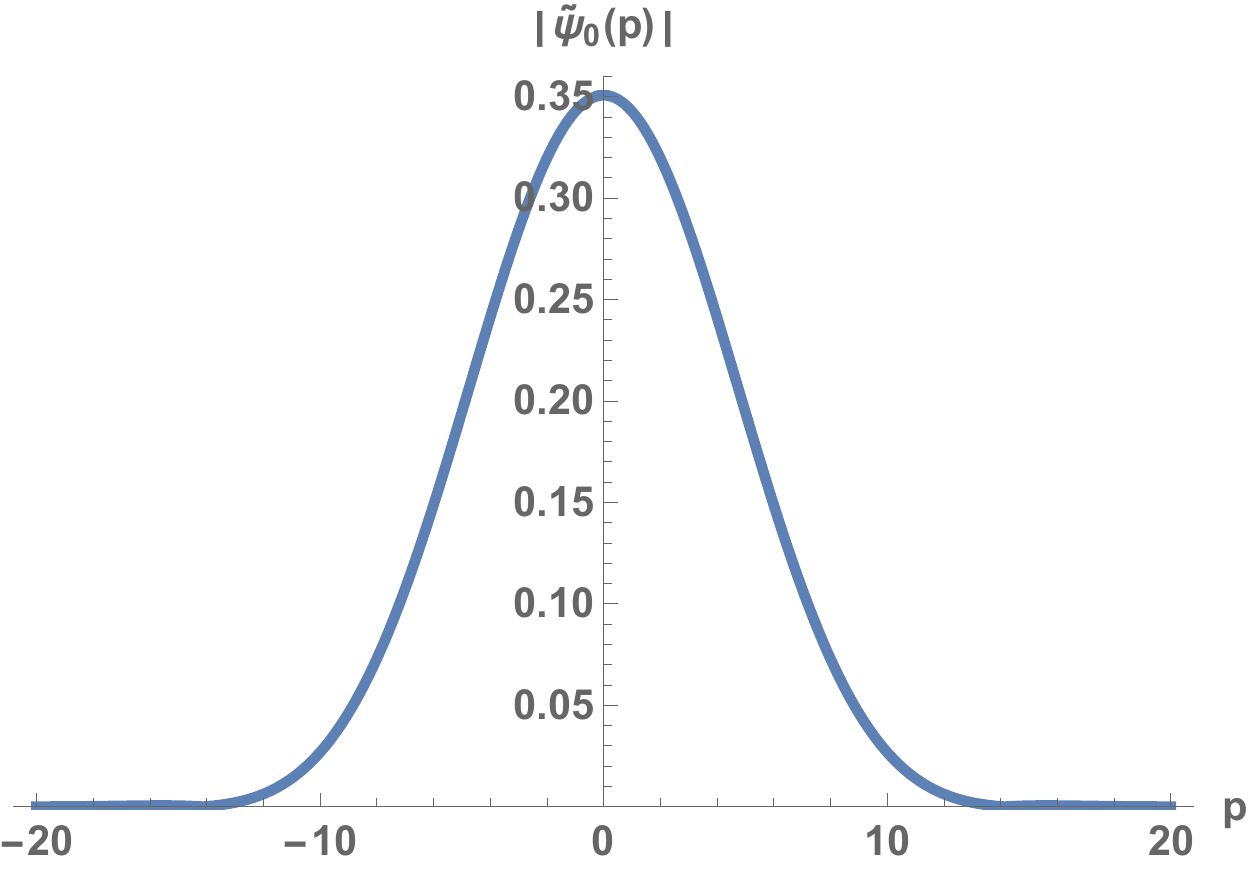}}%
\caption{(a) Magnitude of the ground state wave function in position space for the Riemann Xi function Potential I. (b) Magnitude of the ground State wave function in momentum space for the Riemann Xi Function Potential I.}
\end{figure}

\begin{figure}%
\centering
\subfloat[a]{%
\label{fig:first}%
\includegraphics[height=1.5in]{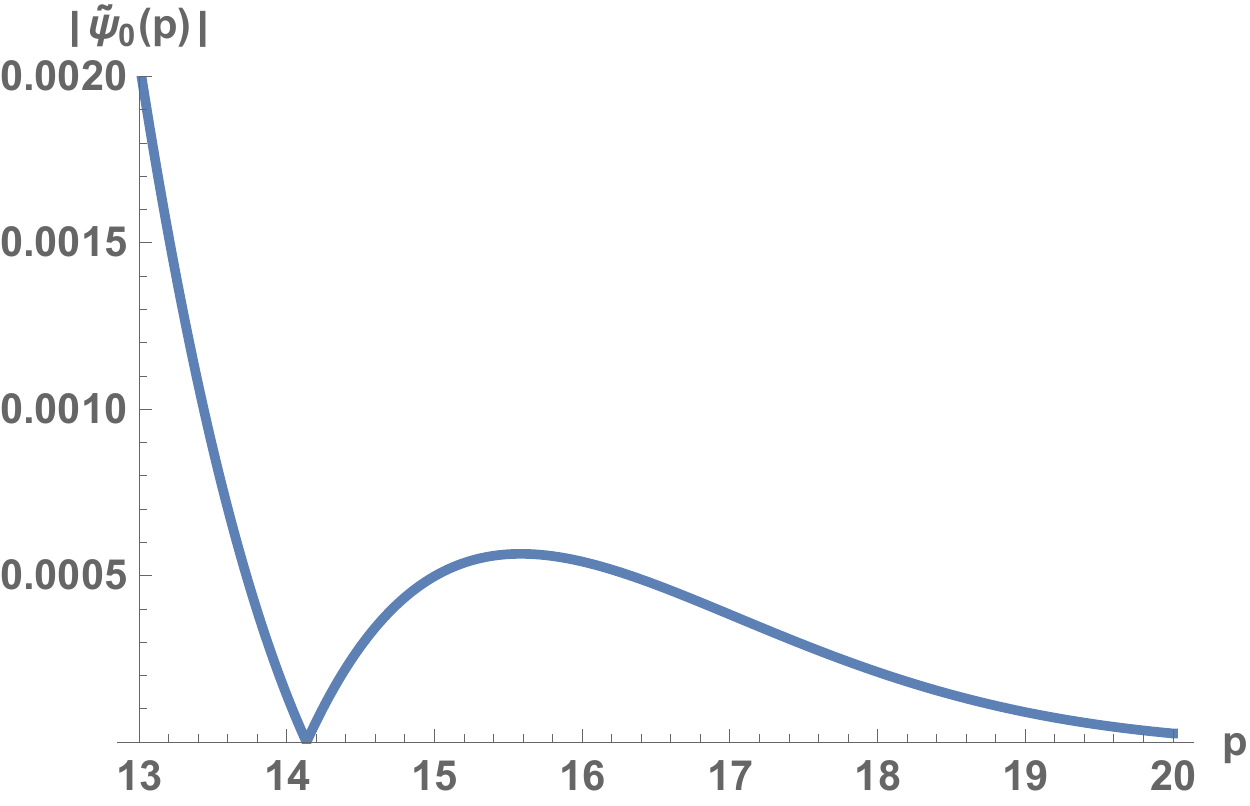}}%
\qquad
\subfloat[b]{%
\label{fig:second}%
\includegraphics[height=1.5in]{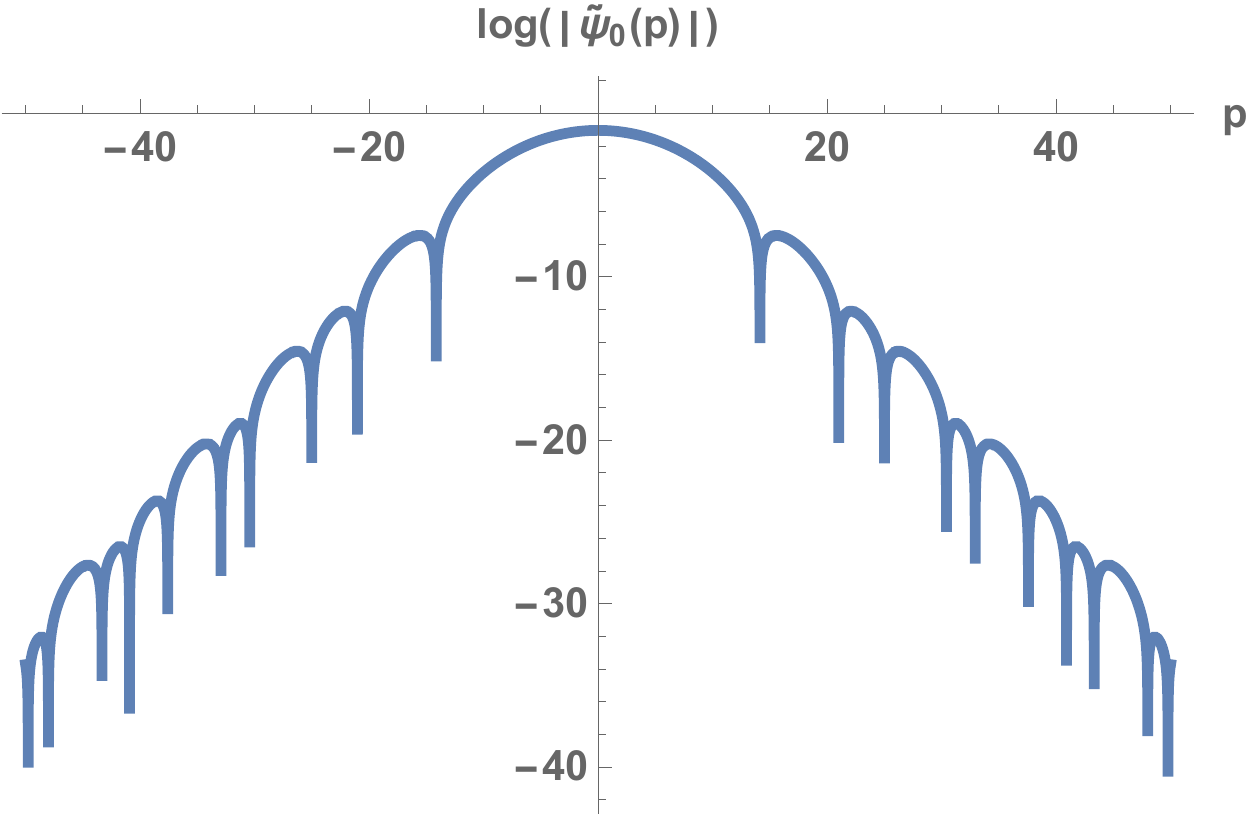}}%
\caption{(a) Closeup of a zero region of the magnitude of the ground state wave function in momentum space of Riemann Xi Potential I. (b) Log of the magnitude of the ground state wave function in momentum space of Riemann Xi Potential I.}
\end{figure}

\subsection{Riemann  Xi potential II }

For the Riemann Xi Potential IIthe position space wave function:
\begin{equation}{\psi _0}(x) = \frac{1}{{\sqrt {{N_0}} }}\left( {{\theta _4}(0|{e^{ - \pi {e^{ - 2x}}}}) + {\theta _2}(0|{e^{ - \pi {e^{ - 2x}}}}) - {\theta _3}(0|{e^{ - \pi {e^{ - 2x}}}})} \right){e^{ - xA}}\end{equation}
and momentum space wave function:
\begin{equation}{{\tilde \psi }_0}(p) = \frac{1}{{\sqrt {2\pi } }}\frac{1}{{\sqrt {{N_0}} }}\left( {{2^{1 - A - ip}} + {2^{A + ip}} - 3} \right)\left( {\frac{2}{{\left( { - 1 + A + ip} \right)(A + ip)}}} \right)\xi (A + ip)\end{equation}
with $N_0=.367016$.

The we have using the position basis for $A=1/2$:
$$\left\langle x \right\rangle  = 0$$
$$\left\langle {{x^2}} \right\rangle  = .0677675$$
\begin{equation}\Delta x = \sqrt {\left\langle {{{\left( {x - \left\langle x \right\rangle } \right)}^2}} \right\rangle }  = \sqrt {\left\langle {{x^2}} \right\rangle  - {{\left\langle x \right\rangle }^2}}  = .260322\end{equation}
and using the momentum basis:
$$\left\langle p \right\rangle  = 0$$
$$\left\langle {{p^2}} \right\rangle  = 3.70515$$
\begin{equation}\Delta p = \sqrt {\left\langle {{{\left( {p - \left\langle p \right\rangle } \right)}^2}} \right\rangle }  = \sqrt {\left\langle {{p^2}} \right\rangle  - {{\left\langle p \right\rangle }^2}}  = 1.92488\end{equation}
Then we have:
\begin{equation}\Delta p\Delta x =  .501088 \geq .5\end{equation}
Which is consistent with the uncertainty relation.

We summarise our results  for all the potentials in table 2. Besides the simple harmonic oscillator Riemann Xi potential II is the closest to saturating the uncertainty relation inequality.
\begin{table}[h]
\centering
\begin{tabular}{|l|l|l|l|l|l|l|l|}
\hline
Potential       & $ \left\langle x \right\rangle $ &  $ \left\langle {{x^2}} \right\rangle $ & $ \Delta x $   & $ \left\langle p \right\rangle $ &  $ \left\langle {{p^2}} \right\rangle $ & $ \Delta p $ & $ \Delta p\Delta x $\\ \hline
SHO   &      $0$          & $ \frac{1}{\omega}$   & $\frac{1}{\sqrt{\omega}}$  &  $0$ & $ \frac{\omega}{4}$ &   $ \frac{\sqrt{\omega}}{2}$ &  $ \frac{1}{2}$ \\ \hline
Morse    & $ 1.27036$ & $ 3.25876$  & $1.28255$   & $0$    & $.25$   & $ .5$  & $ .641275$               \\ \hline
Riemann I    & $ .918522$ & $ 2.34964$  & $ 1.22717$   & $0$    & $.306513$   & $ .553637$  & $ .67408$                     \\ \hline
Riemann II     & $ .156371$ & $ .303422$  & $.528176$   & $0$    & $1.0771$   & $ 1.03783$  & $ .548158$                       \\ \hline
Xi function I  & $ 0$ & $.0245801$  & $.15678$   & $0$    & $10.2076$   & $ 3.19493$  & $ .500902$                    \\ \hline
Xi Function II      & $ 0$ & $ .0677675$  & $.260322$   & $0$    & $3.70515$   & $ 1.92488$  & $ .501088$                    \\ \hline
\end{tabular}
\caption{\label{tab:table-name} Uncertainty relations in position space and momentum space  associated with Simple Harmonic Oscillator, Morse potential, Riemann Zeta function and Riemann Xi Function.}
\end{table}

\section{ Shannon information inequality}

Besides the uncertainty relation we can also compute the Shannon information inequality for the six potentials listed above. Using the definition for the Shannon information entropies in position and momentum basis \cite{Dehesa}\cite{Abdelmonem}\cite{Aydiner}:
$${S_x} =  - {\int_{ - \infty }^\infty  {\left| {{\psi _0}(x)} \right|} ^2}\log \left( {{{\left| {{\psi _0}(x)} \right|}^2}} \right)dx$$
\begin{equation}{S_p} =  - {\int_{ - \infty }^\infty  {\left| {{{\tilde \psi }_0}(p)} \right|} ^2}\log \left( {{{\left| {{{\tilde \psi }_0}(p)} \right|}^2}} \right)dp\end{equation}
these satisfy the inequality:
\begin{equation}{S_x} + {S_p}  \geqslant 1 + \log (\pi)\end{equation}
In table 3 we calculate then quantities for the six potentials. In this case  besides the simple harmonic oscillator Riemann Xi potential I comes closest to satisfying the Shannon information inequality.
\begin{table}[h]
\centering
\begin{tabular}{|l|l|l|l|l|}
\hline
Potential       & $ S_x $ &  $ S_p$ & $ S_x + S_p $   &  $ 1 + \log (\pi )$
  \\ \hline
SHO   &  $  \frac{1}{2} - \frac{1}{2}\log (\omega /2\pi )$           & $   \frac{1}{2} - \frac{1}{2}\log (2/\omega \pi )$    & $ 1 + \log (\pi )$ &  $ 2.14473 $  \\ \hline
Morse    & $ 1.57722$ & $ .693147$  & $2.27036$   & $2.14473$       \\ \hline
Riemann I    & $ 1.5121$ & $ .781932$  & $ 2.29403$   & $2.14473$       \\ \hline
Riemann II     & $ .745831$ & $ 1.44866$  & $ 2.19449$   & $2.14473$      \\ \hline
Xi function I  & $ -.434395$ & $ 2.58012 $  & $ 2.14573$   & $2.14473$            \\ \hline
Xi Function II      & $ .0726135$ & $ 2.07331$  & $ 2.14593$   & $2.14473$          \\ \hline
\end{tabular}
\caption{\label{tab:table-name} Shannon information relations in position space and momentum space  associated with Simple Harmonic Oscillator, Morse potential, Riemann Zeta function and Riemann Xi Function.}
\end{table}

\newpage

\section{Potentials associated with other Dirichlet series}

It is clear that the above methods can be applied to other functions with Dirichlet series. One such function is the Ramanujan Zeta function ${\zeta _{Rj}}(s)$\cite{Rogers}. It's Dirichlet series is:
\begin{equation}{\zeta _{Rj}}(s) = \sum\limits_{n = 1}^\infty  {\frac{{\tau (n)}}{{{n^s}}}} \end{equation}
where the coefficients are defined by the expansion:
\begin{equation}\Delta (iy) = \sum\limits_{n = 1}^\infty  {\tau (n){e^{ - 2\pi yn}}} \end{equation}
where $\Delta(iy)$ is the modular discriminant.

The function can be defined using the integral representation.
\begin{equation}{(2\pi )^{ - (6 + ip)}}\Gamma (6 + ip){\zeta _{Rj}}(6 + ip) = \int_{ - \infty }^\infty  {{e^{ - 6x}}} \Delta (i{e^{ - x}}){e^{ - ipx}}dx\end{equation}
or more generally for arbitrary A off the critical line as:
\begin{equation}{(2\pi )^{ - (A + ip)}}\Gamma (A + ip){\zeta _{Rj}}(A + ip) = \int_{ - \infty }^\infty  {{e^{ - (A - 6)x}}{e^{ - 6x}}} \Delta (i{e^{ - x}}){e^{ - ipx}}dx\end{equation}
Where we have defined the Dedekind eta function as:
\begin{equation}{\rm N}(iy) = {e^{ - \pi y/12}}\prod\limits_{n = 1}^\infty  {\left( {1 - {e^{ - 2\pi yn}}} \right)} \end{equation}
and used the product representation of the derivative of the first Jacobi elliptic function as:
\begin{equation}{2^{ - 1}}{{\theta '}_1}(0,{e^{ - \pi y}}) = {e^{ - \pi y/4}}\prod\limits_{n = 1}^\infty  {{{\left( {1 - {e^{ - 2\pi yn}}} \right)}^3}} \end{equation}
Then the modular discriminat is $\Delta(iy)$ is expressed as:
\begin{equation}\Delta (iy) = {e^{ - 2\pi y}}\prod\limits_{n = 1}^\infty  {{{\left( {1 - {e^{ - 2\pi yn}}} \right)}^{24}}}  = {{\rm N}^{24}}(iy) = {\left( {{2^{ - 1}}{{\theta '}_1}(0,{e^{ - \pi y}})} \right)^8}\end{equation}
By the methods above this yield a prepotential
\begin{equation}
V_0(x) = -\log(2^{-8}{e^{-6 x} (\theta _1^{\prime }(0,e^{-\pi e^{-x}})^8})
\end{equation}
and ground state wave function in position space 
\begin{equation}
\psi_0(x)={e^{ - {V_0}(x)}} =2^{-8}{e^{-6 x} (\theta _1^{\prime }(0,e^{-\pi e^{-x}})^8}
\end{equation}
and in momentum space:
\begin{equation}
{{{\tilde \psi }_0}(p)} =(2 \pi )^{-(6+i p)} \Gamma (6+i p) \zeta_{Rj}(6+i p)
\end{equation}
The superpotential is
\begin{equation}W(x) = {V_0}'(x)\end{equation}
and the two partner potentials are:
$${V_ - }(x) = {W^2}(x) - W'(x)$$
\begin{equation}{V_ + }(x) = {W^2}(x) + W'(x)\end{equation}
These are plotted in figure 18 and 19.
\begin{figure}%
\centering
\subfloat[]{%
\label{fig:first}%
\includegraphics[height=1.5in]{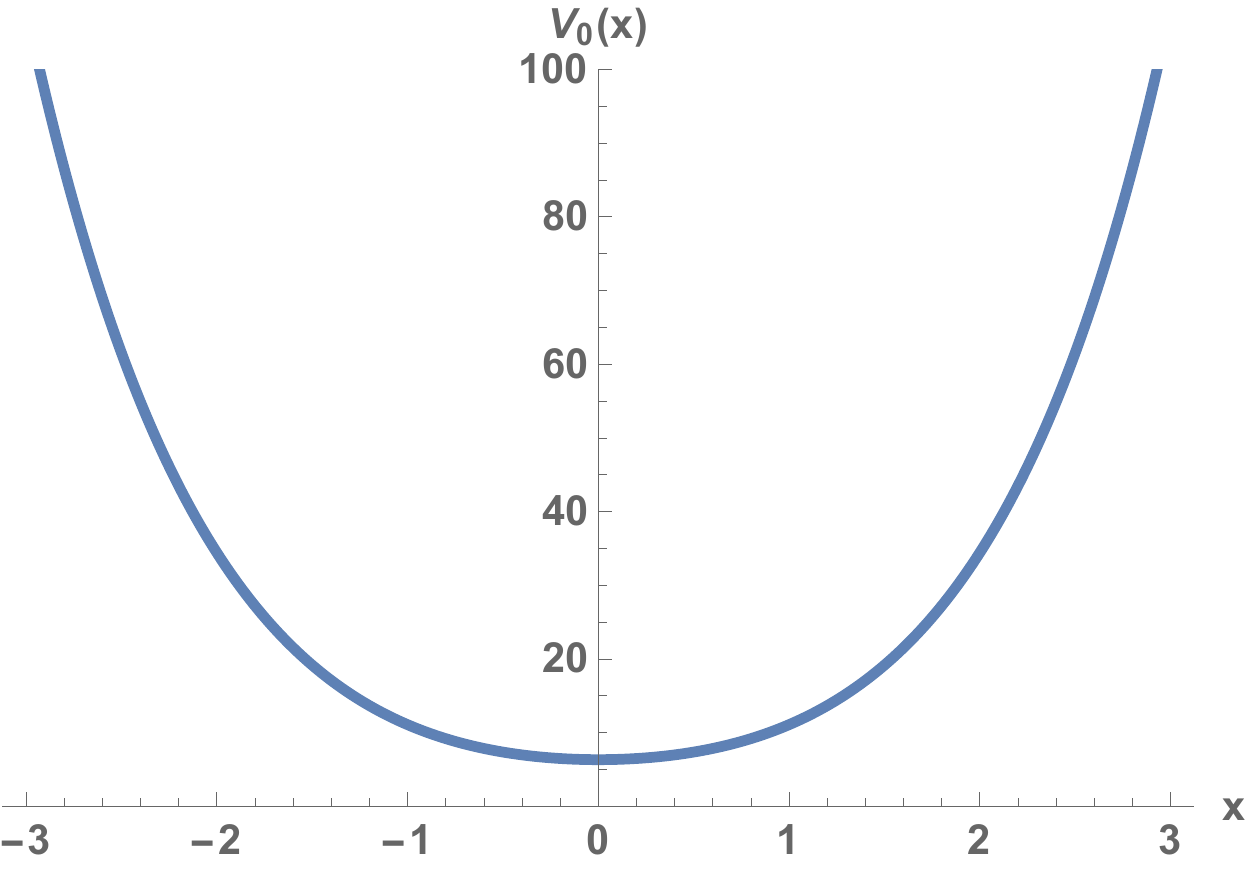}}%
\qquad
\subfloat[]{%
\label{fig:second}%
\includegraphics[height=1.5in]{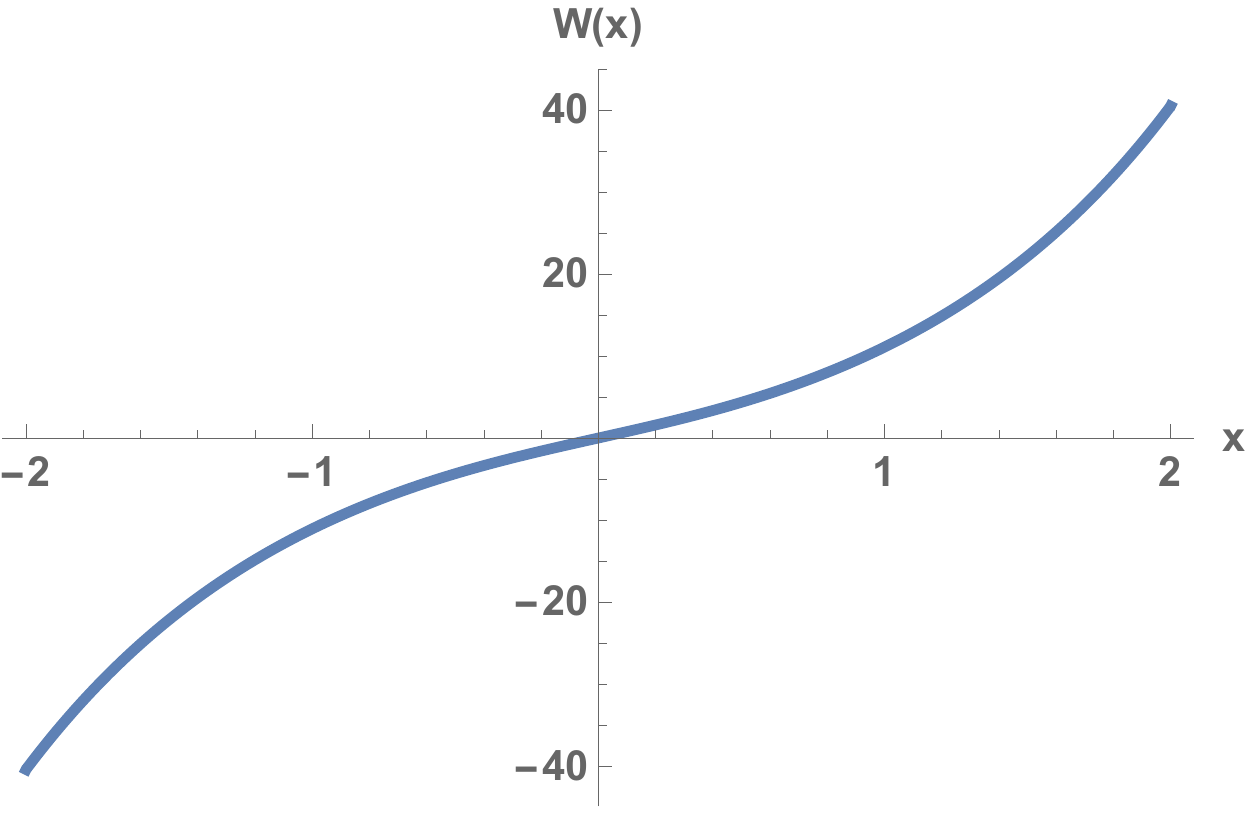}}%
\caption{(a) Prepotential for Ramanujan Zeta Potential for $A=6$. (b) Superpotential for Ramanujan Zeta Potential  for $A=6$.}
\end{figure}
\begin{figure}%
\centering
\subfloat[]{%
\label{fig:first}%
\includegraphics[height=1.5in]{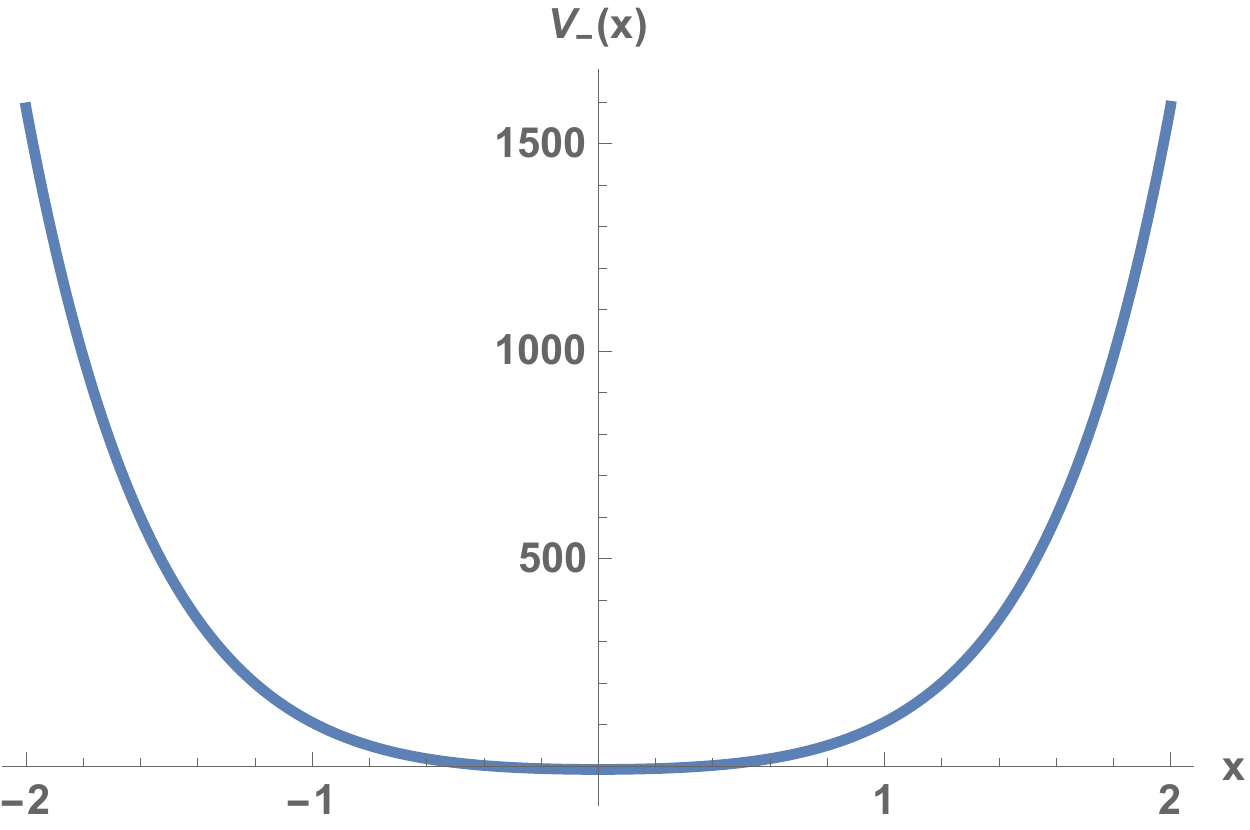}}%
\qquad
\subfloat[]{%
\label{fig:second}%
\includegraphics[height=1.5in]{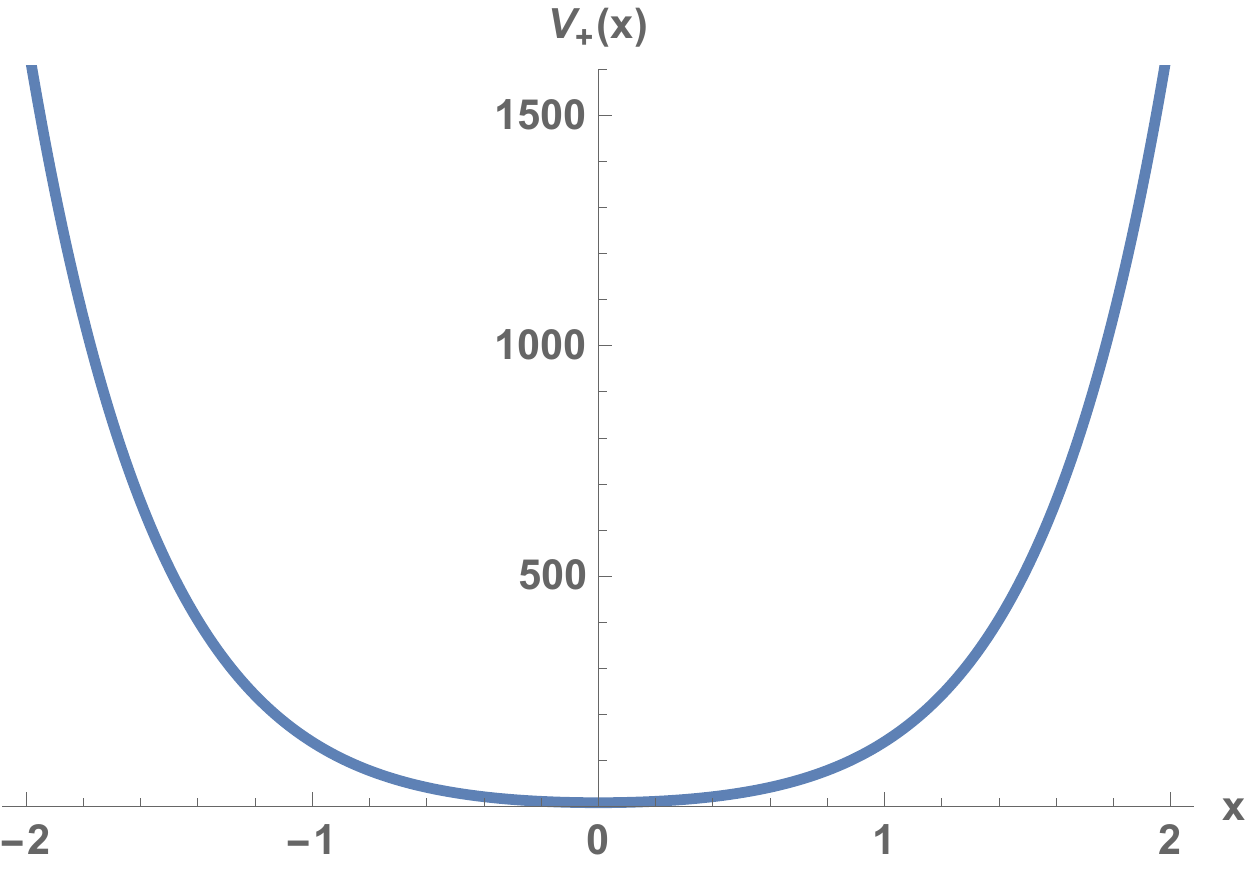}}%
\caption{(a) Minus partner potential for Ramanujan Zeta Potential  for $A=6$. (b) Plus partner potential for Ramanujan Zeta Potential for $A=6$.}
\end{figure}

\section{Series expansion about the minimum for prepotentials}



The prepotential associated with Riemann Xi function yields a simple way to express the Riemann Hypothesis. The prepotential for arbitrary real parameter $A$ is given by:
\begin{equation}V_0(A,x) = -\log(\Phi(e^{-\pi {e^{-2x}}})+(A-\frac{1}{2})x\end{equation}
The series expansion is different about the minimum of the prepotential as one moves from the critical value $A=1/2$. For $A=1/2$ the Riemann Xi potential has the expansion for small $x$
\begin{equation}
V_0(1/2,x) =0.112728 +9.36345 x^2 +5.95896 x^4 -2.09194 x^6 + 3.84 x^8+\ldots
\end{equation}
For $A \ne 1/2$ this is modified. For example for $A=3/4$ the potential about its minimum looks like:
\begin{equation}V_0(3/4,x+.01334675) =0.111059 +9.36982 x^2 + .318029 x^3 + 5.95322 x^4 +\ldots \end{equation}
So we see in this example that moving off the critical line increases the strength of the quadratic term in the Riemann Xi prepotential. The Riemann hypothesis is equivalent to the statement that if the quadratic  term in the prepotential is greater than $9.36982$ than the ground state wave function in momentum space will have zeros in complex momentum space. This can occur by either modifying $A$ from the value $1/2$ or by adding a positive term 
$\lambda x^2$ to the prepotential for the Riemann Xi function.

The series expansion for $V_0(6,x)$ prepotential associated with Ramanujun zeta function is:
\begin{equation}V_0(6,x) = 6.32813 + 0.25 (16.7321 )x^2 + \dots \end{equation}
the first few eigenvalues for the $V_{-}(x)$ partner potential 
\begin{equation}{V_ - }(x) = {W^2}(x) - W'(x)\end{equation}
are:
\begin{equation}\{0,16.8,35.72,56.275,78.21,101.39,125.69,151.04,177.37,204.624,232.76,
261.75,291.55,322.14,353.48\}
\end{equation}
Wheres those of the quadratic function:
\begin{equation}V_{-}(x,quadratic)=0.25 (16.7321)^2 x^2-0.5 (16.7321)\end{equation}
are given by:
\begin{equation}\{0, 16.73, 33.46, 50.2, 66.93, 83.66, 100.39, 
117.125, 133.86, 150.59, 167.32, 184.05, 200.785, 217.52, 
234.25\}\end{equation}
 These plotted in figure 20.
 \begin{figure}%
\centering
\subfloat[]{%
\label{fig:first}%
\includegraphics[height=1.5in]{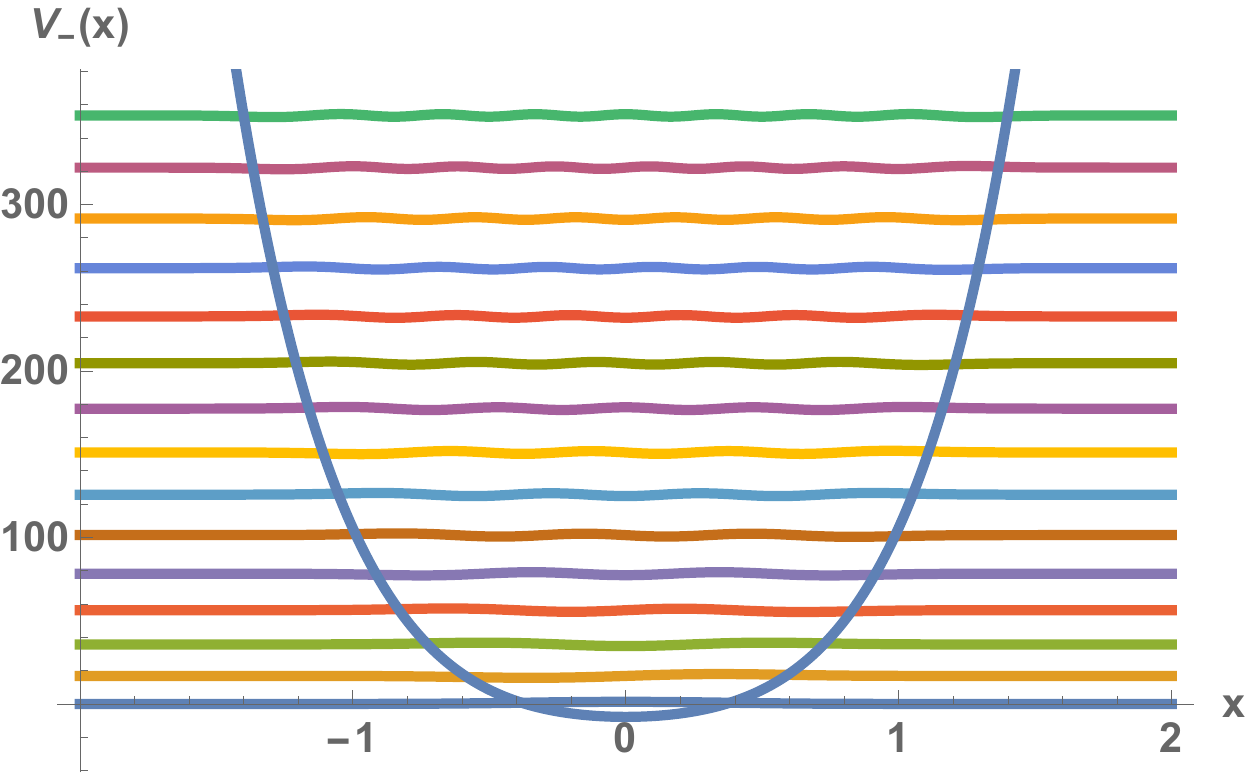}}%
\qquad
\subfloat[]{%
\label{fig:second}%
\includegraphics[height=1.5in]{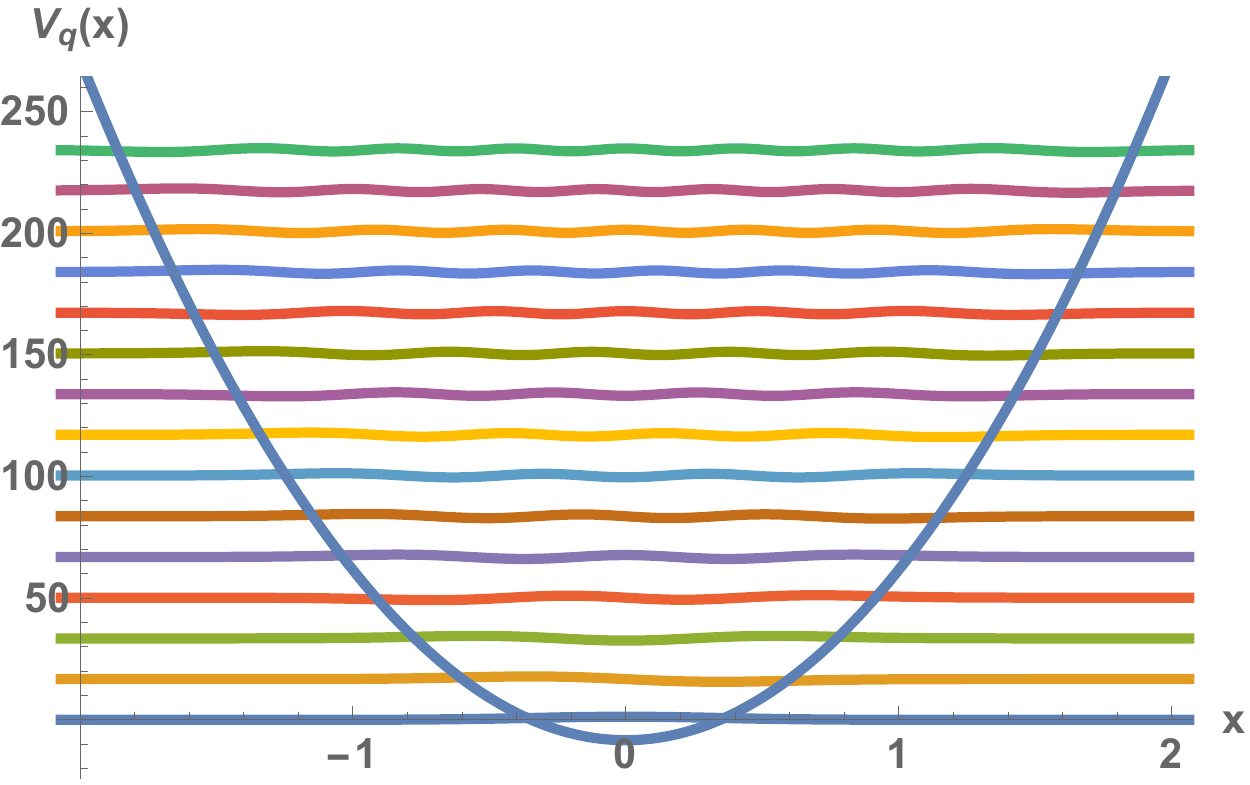}}%
\caption{Comparison for (a) the energy spectrum of the Ramanujan Zeta function potential and (b) a quadratic approximation to the Ramanujan Zeta function potential.}
\end{figure}
 For comparison for the Riemann potential I for $A=5$ the first few eigenvalues are:
 \begin{equation}\{0,9.54345,17.2421,22.4573,24.7907\}\end{equation}
 Whereas for the Morse potential for $A=5$ the first few eigenvalues are
 \begin{equation}\{0,9,16,21,24\}\end{equation}
 These are plotted in figure 21.
 \begin{figure}%
\centering
\subfloat[]{%
\label{fig:first}%
\includegraphics[height=1.5in]{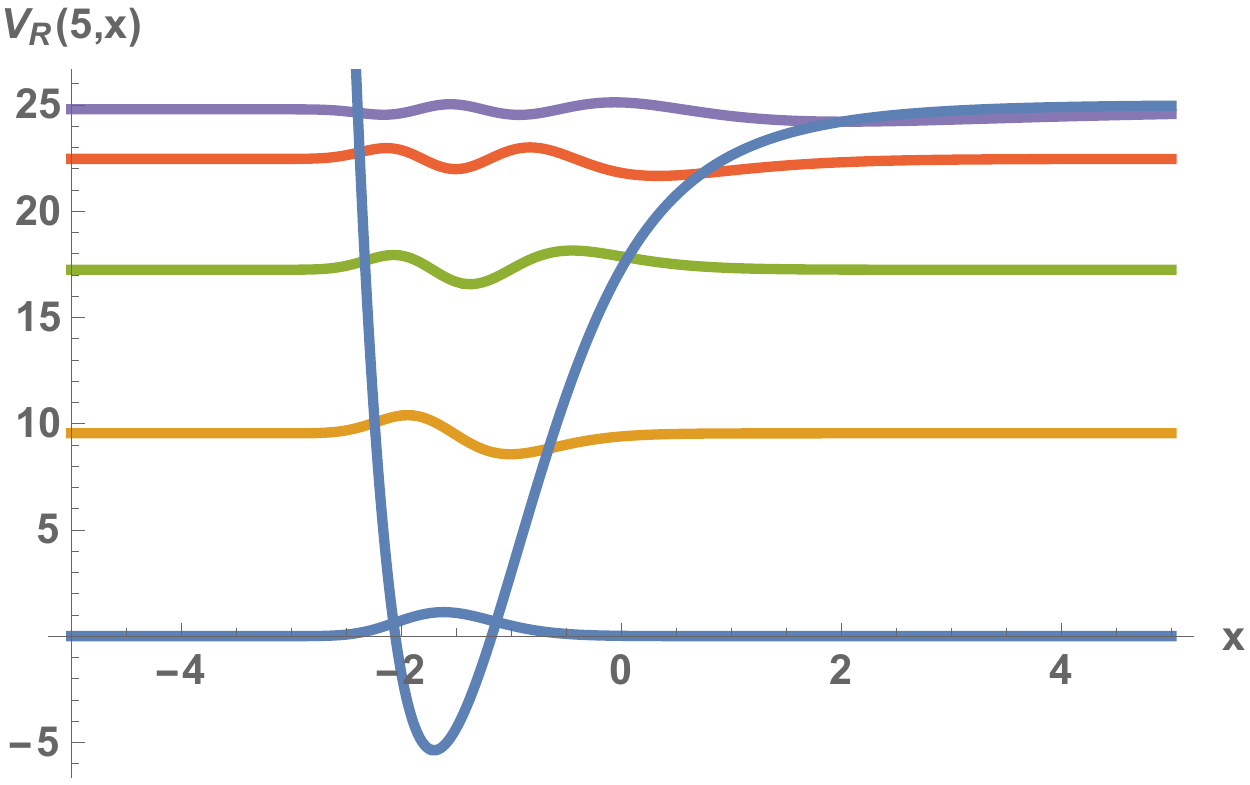}}%
\qquad
\subfloat[]{%
\label{fig:second}%
\includegraphics[height=1.5in]{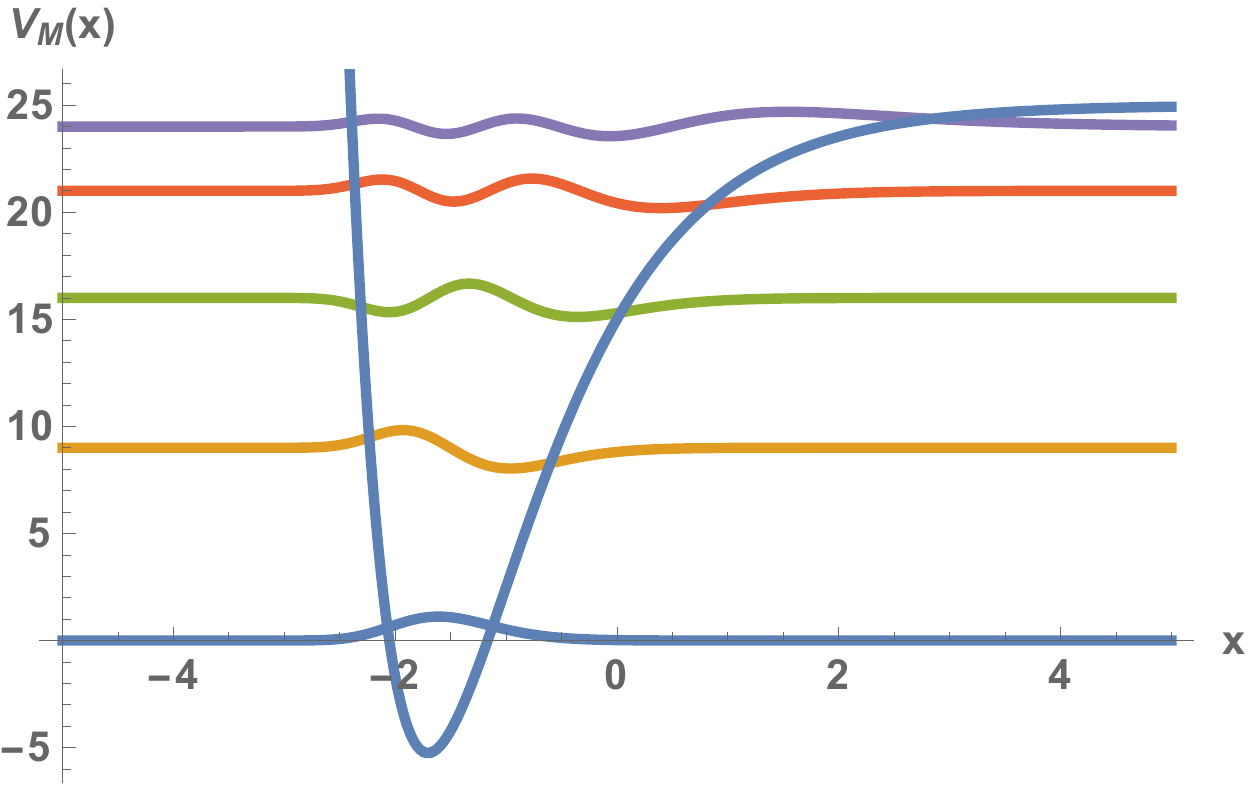}}%
\caption{Comparison for (a) the energy spectrum of the Riemann potential I and $A=5$  (b) and a Morse Potential also with $A=5$.}
\end{figure}
\begin{table}[h]
\centering
\begin{tabular}{|l|l|l|l|l|}
\hline
Potential       & Prepotential \\ \hline
SHO   &  ${V_0}(x) = \frac{1}{4}\omega {x^2}$ \\ \hline
Morse    & ${V_0}(A,x) = A x + {e^{ - x}}$ \\ \hline
Riemann I    & ${V_0}(A,T,x) = Ax + {e^{ - x}} + T\log (1 + {e^{ - {e^{ - x}}/T}})$ \\ \hline
Riemann II     &  ${V_0}(A,x) = (A + 1)x + 2\log (\cosh ({e^{ - x}}))$  \\ \hline
Xi function I  &  $V_0(A,x) = -\log(\Phi(e^{-\pi {e^{-2x}}})+(A-\frac{1}{2})x$\\ \hline
Xi Function II &    $ V_0(A,x) = -\log  \left( {{\theta _4}(0|{e^{ - \pi {e^{ - 2x}}}}) + {\theta _2}(0|{e^{ - \pi {e^{ - 2x}}}}) - {\theta _3}(0|{e^{ - \pi {e^{ - 2x}}}})} \right) + A x$    \\ \hline
Ramanujan Zeta & $V_0(A,x) = -\log(2^{-8}{e^{-6 x} (\theta _1^{\prime }(0,e^{-\pi e^{-x}})^8}) +(A-6)x$ \\ \hline
\end{tabular}
\caption{\label{tab:table-name} Prepotentials associated with Simple Harmonic Oscillator, Morse potential, Riemann Zeta function, Riemann Xi Function and Ramanujan Zeta function.}
\end{table}

\begin{table}[h]
\centering
\begin{tabular}{|l|l|l|l|l|}
\hline
Potential       & Ground State Position Space & Ground State Momentum Space \\ \hline
SHO   &  ${\psi _0}(x) = {\left( {\frac{\omega }{{2\pi }}} \right)^{1/4}}{e^{ - \omega {x^2}/4}}$ & ${{\tilde \psi }_0}(p) = {\left( {\frac{2}{{\pi \omega }}} \right)^{1/4}}{e^{ - {p^2}/\omega }}$  \\ \hline
Morse   & ${\psi _0}(x) = \sqrt 2 {e^{ - x/2}}{e^{ - {e^{ - x}}}}$ & ${{\tilde \psi }_0}(p) = \frac{1}{{\sqrt \pi  }}\Gamma \left( {\frac{1}{2} + ip} \right)$  \\ \hline
Riemann I   & ${\psi _0}(x) = \frac{1}{{\sqrt { - \frac{1}{2} + \log (2)} }}{e^{ - x/2}}\frac{1}{{{e^{{e^{ - x}}}} + 1}}$ & ${{\tilde \psi }_0}(p) = \frac{1}{{\sqrt {2\pi } }}\frac{1}{{\sqrt { - \frac{1}{2} + \log (2)} }}\Gamma \left( {\frac{1}{2} + ip} \right)\eta \left( {\frac{1}{2} + ip} \right)$  \\ \hline
Riemann II  &  ${\psi _0}(x) = \frac{1}{\sqrt{N_0}}\frac{{{e^{ - x(A + 1)}}}}{{{{\cosh }^2}({e^{ - x}})}}$ & ${{\tilde \psi }_0}(p) =  {2^{1 - A - ip}}\left( {A + ip} \right)\Gamma (A + ip)\eta (A + ip)$  \\ \hline
Xi function I  & ${\psi _0}(x) = \frac{1}{{\sqrt {{N_0}} }}\Phi ({e^{ - \pi {e^{ - 2 x}}}})$ & ${{\tilde \psi }_0}(p) = \xi \left( {\frac{1}{2} + ip} \right)$ \\ \hline
Xi Function II & ${\psi _0}(x) = \frac{1}{{\sqrt {{N_0}} }}\Phi_{II} ({e^{ - \pi {e^{ - 2 x}}}})$ & ${{\tilde \psi }_0}(p) = \left( {\frac{2\left( {{2^{1 - A - ip}} + {2^{A + ip}} - 3} \right)}{{\left( { - 1 + A + ip} \right)(A + ip)}}} \right)\xi (A + ip)$   \\ \hline
Ramanujan Zeta & $
\psi_0(x)=2^{-8}{e^{-6 x} (\theta _1^{\prime }(0,e^{-\pi e^{-x}})^8}
$  & $
{{{\tilde \psi }_0}(p)} =(2 \pi )^{-(6+i p)} \Gamma (6+i p) \zeta_{Rj}(6+i p)
$  \\ \hline
\end{tabular}
\caption{\label{tab:table-name} Ground state wave functions in position space and momentum space  associated with Simple Harmonic Oscillator, Morse potential, Riemann Zeta function, Riemann Xi Function and Ramanujan Zeta function.}
\end{table}
\newpage
\section{Conclusion}
In this paper we have examined potentials that give rise to eigenstates which in the momentum representation related to the Riemann Zeta and Xi functions from the point of view of supersymmetric quantum mechanics. We derived matrix models associated with these potentials and discussed their partition functions to the Jacobi matrix. We showed how these potentials are related to the Morse potential with a  deformation. The Riemann and Xi potentials seem to be of the Quasi exactly soluble type with the ground state known exactly but excited states computed numerically. We derived uncertainty and Shannon information relations for the ground state of these potentials. We computed series expansions of these potentials about their minimum and discuss the delicate dependence on the quadratic term as recently investigated by \cite{tau} in the context of the De Bruijn-Newman constant. Finally we showed how these techniques can be used for other potentials with a Dirichlet series such as the Ramanujan Zeta function.

\end{document}